# Median Based Unit Weibull Distribution (MBUW): Do the Higher Order Probability Weighted Moments (PWMs) Add More Information over the Lower Order PWMs in Parameter Estimation

Iman M. Attia *

Imanattiathesis1972@gmail.com ,imanattia1972@gmail.com

*Department of Mathematical Statistics, Faculty of Graduate Studies for Statistical Research, Cairo University, Egypt

**Abstract:** This paper offers an in-depth investigation into the Probability Weighted Moments (PWMs) methodology for estimating parameters of the Median Based Unit Weibull (MBUW) distribution. The author delves into a thorough comparison of the commonly employed first-order PWMs against more advanced higher-order PWMs. The analysis highlights the significant benefits associated with adopting these more sophisticated techniques, particularly in terms of accuracy and reliability in parameter estimation. In addition to this comparative analysis, the author derives the asymptotic distribution of the PWM estimator, which provides a theoretical foundation for the results and enhances the robustness of the conclusions. To further illustrate the practical implications of the findings, the author includes a detailed real data analysis that exemplifies the effectiveness of the proposed methodology. Through these examples, the author emphasizes the relevance of PWMs in real-world applications, demonstrating how this approach can lead to improved parameter estimates when working with the MBUW distribution.

## 1. Introduction

Iman Attia was the first to introduce Median Based Unit Weibull distribution (MBUW)[1]. Given a random variable y that is distributed as Median Based Unit Weibull distribution (MBUW), the PDF, CDF, and quantile functions are as follows:

$$f(y) = \frac{6}{\alpha^{\beta}}\left[1 - y^{\frac{1}{\alpha^{\beta}}}\right] y^{\left(\frac{2}{\alpha^{\beta}}-1\right)} \;,\;\; 0 < y < 1\,, \alpha > 0, \beta > 0 \tag{1}$$

$$F(y) = 3y^{\frac{2}{\alpha^{\beta}}} - 2y^{\frac{3}{\alpha^{\beta}}} \;,\;\; 0 < y < 1\,, \alpha > 0, \beta > 0 \tag{2}$$

$$y = F^{-1}(y) = \left\{-.5\left(cos\left[\frac{cos^{-1}(1-2u)}{3}\right] - \sqrt{3}\,sin\left[\frac{cos^{-1}(1-2u)}{3}\right]\right) + .5\right\}^{\alpha^{\beta}} \tag{3}$$

Various methods are employed to estimate the parameters of a distribution. Maximum Likelihood Estimation (MLE) is popular because it leads to efficient, asymptotically minimum variance estimators, although these estimators may not necessarily be unbiased. The method of moments is straightforward to apply and can serve as starting values for the numerical procedures associated with MLE.

[2] advocated for Probability Weighted Moments (PWMs) and initiated their use in hydrology for small sample sizes, as MLE does not always perform well with limited data. PWMs are a leading alternative to method of moments (MM) and MLE for fitting statistical distributions to data, especially those expressed in inverse form. For instance, if y is a random variable and F represents the cumulative distribution function (CDF) for y, the value of y can be expressed as a function of F; $y = y(F)$. These distributions include the Gumbel, Weibull, and logistic distributions, as well as lesser-known types such as Tukey's symmetric lambda, Thomas' Wakeby, and Mielke's kappa.

[3] advocated for PWMs, demonstrating their superior performance over other estimators through Monte Carlo simulations. For most of these distributions, relatively simple expressions for the parameters can be derived, as noted by [4], including several for which parameter estimates are not easily obtained using MLE or conventional moments These distributions relate to various fields, including hydrology, resource management, and the forecasting of weather parameters such as temperature, precipitation, wind velocity, floods, droughts, and rainfall.

Many researchers have utilized PWMs, including [5] for estimating parameters of the type II extreme value distribution, and[6] who applied the generalized form of PWMs to estimate the two-parameter Weibull distribution. [7] introduced a new class of generalized PWMs for estimating parameters of the Pareto distribution, while [8] implemented partial PWMs for censored data from the generalized extreme value distribution (GEV). Many authors like [9], [10], [11], [12], [13], [14], [15], [16] have applied PWMs to various distributions, mainly focusing on the GEV.

PWMs are more robust than conventional moments in the presence of outliers in the data and yield more efficient estimators from small samples regarding the underlying probability distribution. They are computed as expectations of certain functions of a random variable. A significant advantage over conventional moments is that PWMs are, by definition, linear functions of the ordered data, resulting in reduced sensitivity to sampling variability. Various methods are employed to estimate the parameters of a distribution.

This paper is organized as follows: Section 2 introduces the methodology and the definition of Probability Weighted Moments (PWMs). It explicates the classic PWMs method for parameter estimation. In this section the author derives the PWMs for the Median Biased Unit Weibull (MBUW) and clarifies how to utilize these moments for parameter estimation. The author also illustrates the derivation of the asymptotic variance for different estimators. Section 3 demonstrates Monte Carlo simulation study to validate the premise of the paper. Section 4 explains the results of the simulation study. Section 5 explores the application of this method in real data analysis. Section 6 comprehends the conclusions. Finally, Section 7 suggests the future work.

## 2 Materials and Methods:

## 2.1. Definitions of Probability Weighted Moments (PWMs)

The population PWM, as shown in equations (4-5), is defined as $M_{p,r,s}$, where $p, r, and\ s$ are real numbers.

$$M_{p,r,s} = E(y^p\ [F(y)]^r\ [1 - F(y)]^s) = \int_{-\infty}^{\infty} y^p\ [F(y)]^r\ [1 - F(y)]^s\ f(y)dy \qquad (4)$$

$$M_{p,r,s} = E(y^p\ [F(y)]^r\ [1 - F(y)]^s) = \int_{0}^{1} [y(F)]^p\ [F]^r\ [1 - F]^s\ dF \qquad (5)$$

Where $y(F)$ is a solution to $F(y) = F$ , in other words, the distribution has a closed form quantile function, inverse CDF.

In practice, the value of $p$ is chosen to be 1, and so $M_{1,r,s}$ are used to estimate the parameter. Here, $M_{1,0,0}$ represents the mean. If this mean exists, then $M_{1,r,s}$ exist for any real positive values of r and s. These values are often restricted to small positive integers. Choosing p=1 has the dual advantage of not unduly overweighting sample values while also leading to a class of linear L-moments that exhibit asymptotic normality, [17], [18]. While only small positive integers are needed to estimate the parameters of distributions, using real numbers, regardless of their size, can offer significant advantages. This concept was explored by [19], who extended PWMs into generalized PWMs to include those with $p \neq 1$. PWMs is a generalization of the classic method of moments when $r = s = 0$ , $M_{1,0,0}$ , are the non-central moments of order p. [20] modified the method to accommodate the models without an analytic CDF and quantile function. [2] and [18] strongly advised employing $M_{1,r,s}$ due to the significantly clearer relationships that exist between parameters and moments. This approach simplifies analysis and enhances our understanding of these critical relationships. The empirical estimate of $M_{1,r,s}$ is typically less sensitive to outliers and exhibits good properties, especially when the sample size is small. For convenience, several authors chose to use p=1 and non-negative integer values for r and s. This approach is referred to as the classic PWM method. When p=1, r and s are non-negative, the following equations (6-7) are defined:

$$\alpha_r = M_{1,0,s} = E\left(y\big(1 - F(y)\big)^S\right),\ s = 0,1, \dots .\ \ \&\ r = 0 \qquad (6)$$

$$\beta_s = M_{1,r,0} = E\left(y\big(F(y)\big)^r\right),\ r = 0,1, \dots .\quad \&\quad s = 0 \qquad (7)$$

Both $\alpha_r$ and $\beta_s$ are related by the following equation (8):

$$\alpha_r = \sum_{i=0}^{s} (-1)^i \binom{S}{i} \beta_i \quad and\ \ \beta_s = \sum_{i=0}^{r} (-1)^i \binom{r}{i} \alpha_i \qquad (8)$$

For non-negative integers values of r and s that are as small as possible, both $\alpha_r$ and $\beta_s$ are equivalent. [21] defined the sample unbiased estimators for PWMs $\alpha_r$ and $\beta_s$ as the followings in equation (9):

$$\hat{\alpha}_r = \frac{1}{n}\sum_{i=1}^{n-r}\frac{\binom{n-i}{r}}{\binom{n-1}{r}}y_{i:n} \quad and \quad \hat{\beta}_s = \frac{1}{n}\sum_{i=1+s}^{n}\frac{\binom{i-1}{s}}{\binom{n-1}{s}}y_{i:n} \tag{9}$$

The biased sample estimator for the PWM is defined as

$\widehat{M}_{1,r,s} = \frac{1}{n}\sum_{i=1}^{n} y_{i:n}(p_{i:n})^r\,(1-p_{i:n})^s$ , so it can be defined in equations (10-11) for:

$$\hat{\alpha}_r = M_{1,0,s} = \frac{1}{n}\sum_{i=1}^{n} y_{i:n}(p_{i:n})^0\,(1-p_{i:n})^s \tag{10}$$

$$\hat{\beta}_s = M_{1,r,0} = \frac{1}{n}\sum_{i=1}^{n} y_{i:n}(p_{i:n})^r\,(1-p_{i:n})^0 \tag{11}$$

Where $p_{i:n} = \frac{i-b}{n}$ , $0 \le b \le 1$ $or$ $p_{i:n} = \frac{i-b}{n+1-2b}$ , $-0.5 \le b \le 0.5$

[22] empirically concluded that moderated biased estimates of the PWMs could produce more accurate estimates of upper quantiles. In this paper, $M_{1,0,1}$ and $M_{1,1,0}$ are used as a system of equations to estimate the parameters of the Median Based Unit Weibull (MBUW) distribution. Higher moments $M_{1,0,2}$ and $M_{1,2,0}$ are also used to estimate the parameters and compare the results with lower moments $M_{1,0,1}$ and $M_{1,1,0}$ . Parameter estimation using PWMs is carried out by equating the analytic expression of the population PWMs by the corresponding sample estimates of PWMs and solving the resulting systems of equations in terms of the parameters.

## 2.2. Calculating PWMs for MBUW

### 2.2.1. Calculating $M_{1,0,1}$ : equations (12-20)

Substitute in equation 4 to get: $M_{p,r,s} = M_{1,0,1} = E(y\ [1-F(y)]^s) = \int_0^1 y\ [1-F(y)]^s\ f(y)dy$

$$M_{1,0,1} = \int_0^1 y \left[1-\left(3y^{\frac{2}{\alpha^\beta}} - 2y^{\frac{3}{\alpha^\beta}}\right)\right]^s \frac{6}{\alpha^\beta}\left[1-y^{\frac{1}{\alpha^\beta}}\right]y^{\left(\frac{2}{\alpha^\beta}-1\right)}dy \tag{12}$$

$$= \int_0^1 \left[1-\left(3y^{\frac{2}{\alpha^\beta}} - 2y^{\frac{3}{\alpha^\beta}}\right)\right]^s \frac{6}{\alpha^\beta}\left[1-y^{\frac{1}{\alpha^\beta}}\right]y^{\left(\frac{2}{\alpha^\beta}\right)}dy$$

$$= \int_0^1 \frac{6}{\alpha^\beta}y^{\left(\frac{2}{\alpha^\beta}\right)}\left[1-\left(3y^{\frac{2}{\alpha^\beta}} - 2y^{\frac{3}{\alpha^\beta}}\right)\right]^s dy - \int_0^1 \frac{6}{\alpha^\beta}y^{\left(\frac{3}{\alpha^\beta}\right)}\left[1-\left(3y^{\frac{2}{\alpha^\beta}} - 2y^{\frac{3}{\alpha^\beta}}\right)\right]^s dy \tag{13}$$

$$= \int_0^1 A(y)\,dy - \int_0^1 B(y)dy \tag{14}$$

Using binomial expansion in equation (13):

$$\left[1-\left(3y^{\frac{2}{\alpha^\beta}}-2y^{\frac{3}{\alpha^\beta}}\right)\right]^s = \sum_{i=0}^{s}(-1)^i\binom{s}{i}\left(3y^{\frac{2}{\alpha^\beta}}-2y^{\frac{3}{\alpha^\beta}}\right)^i$$

$$\sum_{i=0}^{s}(-1)^i\binom{s}{i}\ \left(3y^{\frac{2}{\alpha^\beta}}\right)^i\left(1-\frac{2}{3}y^{\frac{1}{\alpha^\beta}}\right)^i =$$

$$\sum_{i=0}^{s}(-1)^i\binom{s}{i}\ \left(3y^{\frac{2}{\alpha^\beta}}\right)^i\sum_{m=0}^{i}(-1)^m\binom{i}{m}\ \left(\frac{2}{3}y^{\frac{1}{\alpha^\beta}}\right)^m =$$

$$\sum_{i=0}^{s}(-1)^i\binom{s}{i}\ 3^i\ y^{\frac{2i}{\alpha^\beta}}\sum_{m=0}^{i}(-1)^m\binom{i}{m}\left(\frac{2}{3}\right)^m\ \ y^{\frac{m}{\alpha^\beta}} \tag{15}$$

Now exchange integration with summation in equations (13) & (14):

$$\int_0^1 A(y)\,dy = \int_0^1 \frac{6}{\alpha^\beta}y^{\left(\frac{2}{\alpha^\beta}\right)}\left[1-\left(3y^{\frac{2}{\alpha^\beta}}-2y^{\frac{3}{\alpha^\beta}}\right)\right]^s dy =$$

$$= \int_0^1 \frac{6}{\alpha^\beta}y^{\left(\frac{2}{\alpha^\beta}\right)}\sum_{i=0}^{s}(-1)^i\binom{s}{i}\ 3^i\ y^{\frac{2i}{\alpha^\beta}}\sum_{m=0}^{i}(-1)^m\binom{i}{m}\left(\frac{2}{3}\right)^m\ \ y^{\frac{m}{\alpha^\beta}}\ dy$$

$$= \tfrac{6}{\alpha^\beta}\textstyle\sum_{i=0}^{s}(-1)^i\binom{s}{i}\ 3^i\ \sum_{m=0}^{i}(-1)^m\binom{i}{m}\left(\frac{2}{3}\right)^m\quad \int_0^1\ y^{\left(\frac{2}{\alpha^\beta}\right)}y^{\frac{2i}{\alpha^\beta}}\,y^{\frac{m}{\alpha^\beta}}\quad dy \tag{16}$$

Where the integral is:

$$\int_0^1 y^{\left(\frac{2}{\alpha^\beta}\right)}y^{\frac{2i}{\alpha^\beta}}\,y^{\frac{m}{\alpha^\beta}}\quad dy\ = \frac{\alpha^\beta}{2+2i+m+\alpha^\beta}$$

$$so\!:\int_0^1 A(y)\,dy = \quad\int_0^1 \frac{6}{\alpha^\beta}y^{\left(\frac{2}{\alpha^\beta}\right)}\left[1-\left(3y^{\frac{2}{\alpha^\beta}}-2y^{\frac{3}{\alpha^\beta}}\right)\right]^s dy =$$

$$\frac{6}{\alpha^\beta}\sum_{i=0}^{s}(-1)^i\binom{s}{i}\ 3^i\ \sum_{m=0}^{i}(-1)^m\binom{i}{m}\left(\frac{2}{3}\right)^m\quad\left(\frac{\alpha^\beta}{2+2i+m+\alpha^\beta}\right) \tag{17}$$

The same steps follow for:

$$\int_0^1 B(y)dy = \int_0^1 \frac{6}{\alpha^\beta}y^{\left(\frac{3}{\alpha^\beta}\right)}\left[1-\left(3y^{\frac{2}{\alpha^\beta}}-2y^{\frac{3}{\alpha^\beta}}\right)\right]^s dy =$$

$$\frac{6}{\alpha^{\beta}}\sum_{i=0}^{s}(-1)^{i}\binom{s}{i}\ 3^{i}\ \sum_{m=0}^{i}(-1)^{m}\binom{i}{m}\left(\frac{2}{3}\right)^{m}\ \left(\frac{\alpha^{\beta}}{3+2i+m+\alpha^{\beta}}\right) \qquad (18)$$

Substitute s=1 in equations (17) and (18):

$$\int_{0}^{1} A(y)\,dy =$$

$$\frac{6}{\alpha^{\beta}}\left\{(-1)^{0}\binom{1}{0}(3)^{0}\left[(-1)^{0}\binom{0}{0}\left(\frac{2}{3}\right)^{0}\left(\frac{\alpha^{\beta}}{2+2(0)+0+\alpha^{\beta}}\right)+(-1)^{1}\binom{0}{1}\left(\frac{2}{3}\right)^{1}\left(\frac{\alpha^{\beta}}{2+2(0)+1+\alpha^{\beta}}\right)\right]\right.$$

$$\left.+(-1)^{1}\binom{1}{1}(3)^{1}\left[(-1)^{0}\binom{1}{0}\left(\frac{2}{3}\right)^{0}\left(\frac{\alpha^{\beta}}{2+2(1)+0+\alpha^{\beta}}\right)+(-1)^{1}\binom{1}{1}\left(\frac{2}{3}\right)^{1}\left(\frac{\alpha^{\beta}}{2+2(1)+1+\alpha^{\beta}}\right)\right]\right\}$$

$$=\frac{6}{\alpha^{\beta}}\left\{\frac{\alpha^{\beta}}{2+\alpha^{\beta}}-\frac{3\alpha^{\beta}}{4+\alpha^{\beta}}+\frac{2\alpha^{\beta}}{5+\alpha^{\beta}}\right\}=6\left\{\frac{1}{2+\alpha^{\beta}}-\frac{3}{4+\alpha^{\beta}}+\frac{2}{5+\alpha^{\beta}}\right\} \qquad (19)$$

For

$$\int_{0}^{1} B(y)\,dy =$$

$$\frac{6}{\alpha^{\beta}}\left\{(-1)^{0}\binom{1}{0}(3)^{0}\left[(-1)^{0}\binom{0}{0}\left(\frac{2}{3}\right)^{0}\left(\frac{\alpha^{\beta}}{3+2(0)+0+\alpha^{\beta}}\right)+(-1)^{1}\binom{0}{1}\left(\frac{2}{3}\right)^{1}\left(\frac{\alpha^{\beta}}{3+2(0)+1+\alpha^{\beta}}\right)\right]\right.$$

$$\left.+(-1)^{1}\binom{1}{1}(3)^{1}\left[(-1)^{0}\binom{1}{0}\left(\frac{2}{3}\right)^{0}\left(\frac{\alpha^{\beta}}{3+2(1)+0+\alpha^{\beta}}\right)+(-1)^{1}\binom{1}{1}\left(\frac{2}{3}\right)^{1}\left(\frac{\alpha^{\beta}}{3+2(1)+1+\alpha^{\beta}}\right)\right]\right\}$$

$$=\frac{6}{\alpha^{\beta}}\left\{\frac{\alpha^{\beta}}{3+\alpha^{\beta}}-\frac{3\alpha^{\beta}}{5+\alpha^{\beta}}+\frac{2\alpha^{\beta}}{6+\alpha^{\beta}}\right\}=6\left\{\frac{1}{3+\alpha^{\beta}}-\frac{3}{5+\alpha^{\beta}}+\frac{2}{6+\alpha^{\beta}}\right\} \qquad (20)$$

Substitute equations (19) and (20) into equation (14)

$$M_{1,0,1}=\frac{360+108\alpha^{\beta}}{1044\alpha^{\beta}+580\alpha^{2\beta}+155\alpha^{3\beta}+20\alpha^{4\beta}+\alpha^{5\beta}+720}$$

### 2.2.2. Calculating $M_{1,1,0}$ equations (21-26)

$$M_{1,1,0}=E(y^{p}\,[F(y)]^{r}\,)=\int_{0}^{1} y^{p}\,[F(y)]^{r}\ \ f(y)dy$$

$$\int_{0}^{1} y\ \left[3y^{\frac{2}{\alpha^{\beta}}}-2y^{\frac{3}{\alpha^{\beta}}}\right]^{r}\ \frac{6}{\alpha^{\beta}}\left[1-y^{\frac{1}{\alpha^{\beta}}}\right]y^{\left(\frac{2}{\alpha^{\beta}}-1\right)}dy \qquad (21)$$

$$=\int_{0}^{1}\frac{6}{\alpha^{\beta}}y^{\left(\frac{2}{\alpha^{\beta}}\right)}\left[3y^{\frac{2}{\alpha^{\beta}}}-2y^{\frac{3}{\alpha^{\beta}}}\right]^{r}dy-\int_{0}^{1}\frac{6}{\alpha^{\beta}}y^{\left(\frac{3}{\alpha^{\beta}}\right)}\left[3y^{\frac{2}{\alpha^{\beta}}}-2y^{\frac{3}{\alpha^{\beta}}}\right]^{r}dy=\int_{0}^{1} C(y)\,dy-\int_{0}^{1} D(y)dy \qquad (22)$$

$$\int_0^1 C(y)\,dy = \int_0^1 \frac{6}{\alpha^\beta} y^{\left(\frac{2}{\alpha^\beta}\right)} \left[3y^{\frac{2}{\alpha^\beta}} - 2y^{\frac{3}{\alpha^\beta}}\right]^r dy = \frac{6}{\alpha^\beta}\int_0^1 y^{\left(\frac{2}{\alpha^\beta}\right)} \left(3y^{\frac{2}{\alpha^\beta}}\right)^r \left[1 - \frac{2}{3}y^{\frac{1}{\alpha^\beta}}\right]^r dy$$

Using binomial expansion: $\left[1 - \frac{2}{3}y^{\frac{1}{\alpha^\beta}}\right]^r = \sum_{i=0}^{r}(-1)^i\binom{r}{i}\left(\frac{2}{3}y^{\frac{1}{\alpha^\beta}}\right)^i$

$$So \int_0^1 C(y)\,dy = \frac{6}{\alpha^\beta}\int_0^1 y^{\left(\frac{2}{\alpha^\beta}\right)}\left(3y^{\frac{2}{\alpha^\beta}}\right)^r \sum_{i=0}^{r}(-1)^i\binom{r}{i}\left(\frac{2}{3}y^{\frac{1}{\alpha^\beta}}\right)^i dy \tag{23}$$

Exchange the integral and the sum $\frac{6}{\alpha^\beta}\sum_{i=0}^{r} 3^r\ (-1)^i\binom{r}{i}\left(\frac{2}{3}\right)^i \int_0^1 y^{\left(\frac{2}{\alpha^\beta}\right)} y^{\frac{2r}{\alpha^\beta}}\, y^{\frac{i}{\alpha^\beta}}\, dy =$

$$\int_0^1 C(y)\,dy = \frac{6}{\alpha^\beta}\sum_{i=0}^{r} 3^r\ (-1)^i\binom{r}{i}\left(\frac{2}{3}\right)^i\left(\frac{\alpha^\beta}{2+2r+i+\alpha^\beta}\right) \tag{24}$$

The same steps are followed and give:

$$\int_0^1 D(y)\,dy = \frac{6}{\alpha^\beta}\sum_{i=0}^{r} 3^r\ (-1)^i\binom{r}{i}\left(\frac{2}{3}\right)^i\left(\frac{\alpha^\beta}{3+2r+i+\alpha^\beta}\right) \tag{25}$$

Substitute equation (24) and equation (25) into equation (22) and let r=1

$$M_{1,1,0} = \frac{6}{\alpha^\beta}\sum_{i=0}^{r} 3^r\ (-1)^i\binom{r}{i}\left(\frac{2}{3}\right)^i\left[\left(\frac{\alpha^\beta}{2+2r+i+\alpha^\beta}\right) - \left(\frac{\alpha^\beta}{3+2r+i+\alpha^\beta}\right)\right]$$

$$= \frac{6}{\alpha^\beta}\ \{3^1\ (-1)^0\binom{1}{0}\left(\frac{2}{3}\right)^0\left[\left(\frac{\alpha^\beta}{2+2(1)+(0)+\alpha^\beta}\right) - \left(\frac{\alpha^\beta}{3+2(1)+(0)+\alpha^\beta}\right)\right]$$

$$+ 3^1\ (-1)^1\binom{1}{1}\left(\frac{2}{3}\right)^1\left[\left(\frac{\alpha^\beta}{2+2(1)+(1)+\alpha^\beta}\right) - \left(\frac{\alpha^\beta}{3+2(1)+(1)+\alpha^\beta}\right)\right]\}$$

$$= \frac{6}{\alpha^\beta}\left\{\frac{3\alpha^\beta}{4+\alpha^\beta} - \frac{2\alpha^\beta}{5+\alpha^\beta} - \frac{3\alpha^\beta}{5+\alpha^\beta} + \frac{2\alpha^\beta}{6+\alpha^\beta}\right\} = \frac{18}{4+\alpha^\beta} - \frac{30}{5+\alpha^\beta} + \frac{12}{6+\alpha^\beta}$$

$$M_{1,1,0} = \frac{6(10+\alpha^\beta)}{(4+\alpha^\beta)(5+\alpha^\beta)(6+\alpha^\beta)} = \frac{60+6\alpha^\beta}{74\alpha^\beta + 15\alpha^{2\beta} + \alpha^{3\beta} + 120} \tag{26}$$

To sum up, PWMs method for estimating the parameters using $M_{1,0,1}$ and $M_{1,1,0}$ :

Step 1: Calculate the population PWMs for the order p=1

$$M_{1,0,1} = \frac{360 + 108\alpha^\beta}{1044\alpha^\beta + 580\alpha^{2\beta} + 155\alpha^{3\beta} + 20\alpha^{4\beta} + \alpha^{5\beta} + 720} \tag{27}$$

$$M_{1,1,0} = \frac{60 + 6\alpha^\beta}{74\alpha^\beta + 15\alpha^{2\beta} + \alpha^{3\beta} + 120} \tag{28}$$

Step 2: Calculate the estimated sample PWMs whether the unbiased or biased estimators and equate these estimators with the corresponding population PWMs.

$$\hat{\alpha}_{r=1} = \frac{1}{n}\sum_{i=1}^{n-1} \frac{(n-i)}{(n-1)}\, y_{i:n} = M_{1,0,1}, \qquad or \quad \hat{\alpha}_r = \frac{1}{n}\sum_{i=1}^{n} y_{i:n}(p_{i:n})^0\,(1-p_{i:n})^s = M_{1,0,1}$$

$$\hat{\beta}_{s=1} = \frac{1}{n}\sum_{i=1+s}^{n} \frac{(i-1)}{(n-1)}\, y_{i:n} = M_{1,1,0}\ , \qquad or \quad \hat{\beta}_s = \frac{1}{n}\sum_{i=1}^{n} y_{i:n}(p_{i:n})^1\,(1-p_{i:n})^0 = M_{1,1,0}$$

Step 3: The above equations construct a system of equations to be solved numerically. In this paper, the author used the Levenberg-Marquardt (LM) algorithm. The objective functions to be minimized are equations (27-28). Differentiate the previous equations (27-28) with respect to alpha and beta.

$$\text{The Jacobian matrix is } \begin{bmatrix} \frac{\partial}{\partial \alpha} M_101 & \frac{\partial}{\partial \beta} M_101 \\ \frac{\partial}{\partial \alpha} M_110 & \frac{\partial}{\partial \beta} M_110 \end{bmatrix}$$

$$\left(y - f\left(\boldsymbol{\theta}^{(a)}\right)\right) = \begin{bmatrix} \hat{\alpha}_r - \dfrac{360 + 108\alpha^{\beta}}{1044\alpha^{\beta} + 580\alpha^{2\beta} + 155\alpha^{3\beta} + 20\alpha^{4\beta} + \alpha^{5\beta} + 720} \\ \hat{\beta}_s - \dfrac{60 + 6\alpha^{\beta}}{74\alpha^{\beta} + 15\alpha^{2\beta} + \alpha^{3\beta} + 120} \end{bmatrix}, \&\ \boldsymbol{\theta}^{(a)} = \begin{bmatrix} \alpha^{(a)} \\ \beta^{(a)} \end{bmatrix}$$

Apply LM algorithm: $\theta^{(a+1)} = \theta^{(a)} + \left[J'\left(\theta^{(a)}\right)J\left(\theta^{(a)}\right) + \lambda^{(a)} I^{(a)}\right]^{-1}\left[J'\left(\theta^{(a)}\right)\right]\left(y - f\left(\theta^{(a)}\right)\right)$.

Where the parameters used in the first iteration are the initial guess, then they are updated according to the sum of squares of errors. The LM algorithm is an iterative algorithm. $J'\left(\theta^{(a)}\right)$ is the Jacobian function which is the first derivative of the objective function evaluated at the initial guess $\theta^{(a)}$ and $\lambda^{(a)}$ is a damping factor that adjusts the step size in each iteration direction. The starting value of this factor is usually 0.001 and according to the sum square of errors (SSE) in each iteration this damping factor is adjusted:

$SSE^{(a+1)} \geq SSE^{(a)}$ *so update*: $\lambda_{updated} = 10 * \lambda_{old}$

$SSE^{(a+1)} < SSE^{(a)}$ *so update*: $\lambda_{updated} = \frac{1}{10} * \lambda_{old}$ , $f\left(\theta^{(a)}\right)$ is the objective function (population PWMs, $M_{1,0,1}$ & $M_{1,1,0}$) evaluated at the initial guess, and $y$ is the sample estimates of population PWMs , the $M_{1,0,1}$ and $M_{1,1,0}$. Steps of the LM algorithm:

1- Start with the initial guess of parameters (alpha and beta).
2- Substitute these values in the objective function and the Jacobian.
3- Choose the damping factor, say lambda=0.001
4- Substitute in the equation (LM equation) to get the new parameters.
5- Calculate the SSE at these parameters and compare this SSE value with the previous one when using the initial parameters to adjust for the damping factor.
6- Update the damping factor accordingly as previously explained.

7- Start new iteration with the new parameters and the new updated damping factor, i.e, apply the previous steps many times till convergence is achieved or a pre-specified number of iterations is accomplished.

The value of this quantity: $\left[J'(\theta^{(a)})J(\theta^{(a)}) + \lambda^{(a)}I^{(a)}\right]^{-1}$ can be considered a good approximation to the variance–covariance matrix of the estimated parameters. Standard errors for the estimated parameters are the square root of the diagonal of the elements in this matrix.

### 2.2.3. Calculating $M_{1,0,2}$ equation (29)

Substitute s=2 in equation (13) and solve to get equation (29):

$$\frac{6}{\alpha^{\beta}}\{(-1)^0\binom{2}{0}(3)^0\left[(-1)^0\binom{0}{0}\left(\frac{2}{3}\right)^0\left(\frac{\alpha^{\beta}}{2+2(0)+0+\alpha^{\beta}}\right)+(-1)^1\binom{0}{1}\left(\frac{2}{3}\right)^1\left(\frac{\alpha^{\beta}}{2+2(0)+1+\alpha^{\beta}}\right)+(-1)^2\binom{0}{2}\left(\frac{2}{3}\right)^2\left(\frac{\alpha^{\beta}}{2+2(0)+2+\alpha^{\beta}}\right)\right]$$

$$+(-1)^1\binom{2}{1}(3)^1\left[(-1)^0\binom{1}{0}\left(\frac{2}{3}\right)^0\left(\frac{\alpha^{\beta}}{2+2(1)+0+\alpha^{\beta}}\right)+(-1)^1\binom{1}{1}\left(\frac{2}{3}\right)^1\left(\frac{\alpha^{\beta}}{2+2(1)+1+\alpha^{\beta}}\right)+(-1)^2\binom{1}{2}\left(\frac{2}{3}\right)^2\left(\frac{\alpha^{\beta}}{2+2(1)+2+\alpha^{\beta}}\right)\right]$$

$$+(-1)^2\binom{2}{2}(3)^2\left[(-1)^0\binom{2}{0}\left(\frac{2}{3}\right)^0\left(\frac{\alpha^{\beta}}{2+2(2)+0+\alpha^{\beta}}\right)+(-1)^1\binom{2}{1}\left(\frac{2}{3}\right)^1\left(\frac{\alpha^{\beta}}{2+2(2)+1+\alpha^{\beta}}\right)+(-1)^2\binom{2}{2}\left(\frac{2}{3}\right)^2\left(\frac{\alpha^{\beta}}{2+2(2)+2+\alpha^{\beta}}\right)\right]\}$$

$$-\frac{6}{\alpha^{\beta}}\{(-1)^0\binom{2}{0}(3)^0\left[(-1)^0\binom{0}{0}\left(\frac{2}{3}\right)^0\left(\frac{\alpha^{\beta}}{3+2(0)+0+\alpha^{\beta}}\right)+(-1)^1\binom{0}{1}\left(\frac{2}{3}\right)^1\left(\frac{\alpha^{\beta}}{3+2(0)+1+\alpha^{\beta}}\right)+(-1)^2\binom{0}{2}\left(\frac{2}{3}\right)^2\left(\frac{\alpha^{\beta}}{3+2(0)+2+\alpha^{\beta}}\right)\right]$$

$$+(-1)^1\binom{2}{1}(3)^1\left[(-1)^0\binom{1}{0}\left(\frac{2}{3}\right)^0\left(\frac{\alpha^{\beta}}{3+2(1)+0+\alpha^{\beta}}\right)+(-1)^1\binom{1}{1}\left(\frac{2}{3}\right)^1\left(\frac{\alpha^{\beta}}{3+2(1)+1+\alpha^{\beta}}\right)+(-1)^2\binom{1}{2}\left(\frac{2}{3}\right)^2\left(\frac{\alpha^{\beta}}{3+2(1)+2+\alpha^{\beta}}\right)\right]$$

$$+(-1)^2\binom{2}{2}(3)^2\left[(-1)^0\binom{2}{0}\left(\frac{2}{3}\right)^0\left(\frac{\alpha^{\beta}}{3+2(2)+0+\alpha^{\beta}}\right)+(-1)^1\binom{2}{1}\left(\frac{2}{3}\right)^1\left(\frac{\alpha^{\beta}}{3+2(2)+1+\alpha^{\beta}}\right)+(-1)^2\binom{2}{2}\left(\frac{2}{3}\right)^2\left(\frac{\alpha^{\beta}}{3+2(2)+2+\alpha^{\beta}}\right)\right]\}$$

$$M_{1,0,2}=\frac{6}{\alpha^{\beta}}\left\{\frac{\alpha^{\beta}}{2+\alpha^{\beta}}-\frac{6\alpha^{\beta}}{4+\alpha^{\beta}}+\frac{4\alpha^{\beta}}{5+\alpha^{\beta}}+\frac{9\alpha^{\beta}}{6+\alpha^{\beta}}-\frac{12\,\alpha^{\beta}}{7+\alpha^{\beta}}+\frac{4\alpha^{\beta}}{8+\alpha^{\beta}}\right\}$$

$$-\frac{6}{\alpha^\beta}\left\{\frac{\alpha^\beta}{3+\alpha^\beta}-\frac{6\alpha^\beta}{5+\alpha^\beta}+\frac{4\alpha^\beta}{6+\alpha^\beta}+\frac{9\alpha^\beta}{7+\alpha^\beta}-\frac{12\,\alpha^\beta}{8+\alpha^\beta}+\frac{4\alpha^\beta}{9+\alpha^\beta}\right\}$$

$$M_{1,0,2}=6\left\{\frac{1}{2+\alpha^\beta}-\frac{6}{4+\alpha^\beta}+\frac{4}{5+\alpha^\beta}+\frac{9}{6+\alpha^\beta}-\frac{12}{7+\alpha^\beta}+\frac{4}{8+\alpha^\beta}\right\}$$

$$-6\left\{\frac{1}{3+\alpha^\beta}-\frac{6}{5+\alpha^\beta}+\frac{4}{6+\alpha^\beta}+\frac{9}{7+\alpha^\beta}-\frac{12}{8+\alpha^\beta}+\frac{4}{9+\alpha^\beta}\right\}$$

$$M_{1,0,2}=\frac{K}{L}\quad where: \tag{29}$$

$$K=58320\ \alpha^\beta+6480\,\alpha^{2\beta}+120960$$

$$L=663696\,\alpha^\beta+509004\,\alpha^{2\beta}+214676\,\alpha^{3\beta}+54649\alpha^{4\beta}+8624\,\alpha^{5\beta}+826\,\alpha^{6\beta}+44\,\alpha^{7\beta}+\alpha^{8\beta}+362880$$

### 2.2.4. Calculating $M_{1,2,0}$ equation (30)

Substitute r=2 in equation (21) and solve to get equation (30):

$$M_{1,2,0}=\frac{6}{\alpha^\beta}\sum_{i=0}^{2}3^r\ (-1)^i\binom{r}{i}\left(\frac{2}{3}\right)^i\left[\left(\frac{\alpha^\beta}{2+2r+i+\alpha^\beta}\right)-\left(\frac{\alpha^\beta}{3+2r+i+\alpha^\beta}\right)\right]$$

$$=\frac{6}{\alpha^\beta}\ \{3^2\ (-1)^0\binom{2}{0}\left(\frac{2}{3}\right)^0\left[\left(\frac{\alpha^\beta}{2+2(2)+(0)+\alpha^\beta}\right)-\left(\frac{\alpha^\beta}{3+2(2)+(0)+\alpha^\beta}\right)\right]$$

$$+3^2\ (-1)^1\binom{2}{1}\left(\frac{2}{3}\right)^1\left[\left(\frac{\alpha^\beta}{2+2(2)+(1)+\alpha^\beta}\right)-\left(\frac{\alpha^\beta}{3+2(2)+(1)+\alpha^\beta}\right)\right]$$

$$+3^2\ (-1)^2\binom{2}{2}\left(\frac{2}{3}\right)^2\left[\left(\frac{\alpha^\beta}{2+2(2)+(2)+\alpha^\beta}\right)-\left(\frac{\alpha^\beta}{3+2(2)+(2)+\alpha^\beta}\right)\right]\}$$

$$M_{1,2,0}=\frac{6}{\alpha^\beta}\left\{\frac{9\alpha^\beta}{6+\alpha^\beta}-\frac{12\alpha^\beta}{7+\alpha^\beta}+\frac{4\alpha^\beta}{8+\alpha^\beta}-\frac{9\alpha^\beta}{7+\alpha^\beta}+\frac{12\alpha^\beta}{8+\alpha^\beta}-\frac{4\alpha^\beta}{9+\alpha^\beta}\right\}$$

$$M_{1,2,0}=6\left\{\frac{9}{6+\alpha^\beta}-\frac{21}{7+\alpha^\beta}+\frac{16}{8+\alpha^\beta}-\frac{4\alpha^\beta}{9+\alpha^\beta}\right\}$$

$$M_{1,2,0}=\frac{150\,\alpha^\beta+6\,\alpha^{2\beta}+1008}{1650\,\alpha^\beta+335\,\alpha^{2\beta}+30\,\alpha^{3\beta}+\alpha^{4\beta}+3024} \tag{30}$$

In summary, PWMs method for estimating the parameters using $M_{1,0,2}$ and $M_{1,2,0}$ can be explained in the following steps: the first step is to calculate the population PWMs for the order p=1, using equations (29-30). The second step is to calculate the estimated sample PWMs, whether the unbiased or biased estimators, and equate these estimators with the corresponding population PWMs.

$$\hat{\alpha}_{r=2} = \frac{1}{n}\sum_{i=1}^{n-r}\frac{(n-i)(n-i-1)}{(n-1)(n-2)}\, y_{i:n} = M_{1,0,2} \quad or \quad \hat{\alpha}_r = \frac{1}{n}\sum_{i=1}^{n} y_{i:n}(p_{i:n})^0\,(1-p_{i:n})^2 = M_{1,0,2}$$

$$\hat{\beta}_{s=2} = \frac{1}{n}\sum_{i=1+s}^{n}\frac{(i-1)(i-2)}{(n-1)(n-2)}\, y_{i:n} = M_{1,2,0} \quad or \quad \hat{\beta}_s = \frac{1}{n}\sum_{i=1}^{n} y_{i:n}(p_{i:n})^2\,(1-p_{i:n})^0 = M_{1,2,0}$$

The last step is to construct a system of the above equations to solve it numerically. Using the Levenberg-Marquardt (LM) algorithm to minimize equations (29) and (30). Differentiate the previous equations (29-30) with respect to alpha and beta

$$\text{The Jacobian matrix is } \begin{bmatrix} \frac{\partial}{\partial\alpha}M_102 & \frac{\partial}{\partial\beta}M_102 \\ \frac{\partial}{\partial\alpha}M_120 & \frac{\partial}{\partial\beta}M_120 \end{bmatrix}$$

$$\left(y - f(\boldsymbol{\theta}^{(a)})\right) = \begin{bmatrix} \hat{\alpha}_r - \dfrac{K}{L} \\ \hat{\beta}_s - \dfrac{150\,\alpha^{\beta} + 6\,\alpha^{2\beta} + 1008}{1650\,\alpha^{\beta} + 335\,\alpha^{2\beta} + 30\,\alpha^{3\beta} + \alpha^{4\beta} + 3024} \end{bmatrix}, \& \ \boldsymbol{\theta}^{(a)} = \begin{bmatrix} \alpha^{(a)} \\ \beta^{(a)} \end{bmatrix}$$

Apply the LM algorithm : $\theta^{(a+1)} = \theta^{(a)} + \left[J'(\theta^{(a)})J(\theta^{(a)}) + \lambda^{(a)}I^{(a)}\right]^{-1}\left[J'(\theta^{(a)})\right]\left(y - f(\theta^{(a)})\right)$.

See Appendix A for the Jacobian matrices for both $M_{1,0,1}$ , $M_{1,1,0}$ and $M_{1,0,2}$ , $M_{1,2,0}$

## 2.3. Asymptotic Distribution of PWM Estimators:

For optimal evaluations of estimators derived from various Probability Weighted Moments (PWMs) method, it is essential to leverage analytical expressions for their variances, as established by asymptotic theory. This approach not only streamlines the process but also eliminates the need for extensive and potentially flawed computer simulations. Unfortunately, in numerous cases, deriving straightforward expressions for the asymptotic variance of moments and parameter estimators proves to be a significant challenge. Moreover, as highlighted by Hosking1986 and further explored by [4], the reliability of asymptotic variance expressions diminishes, particularly when working with small sample sizes.

In his 1986 study, Hosking offered critical insights into the asymptotic variances of generalized Pareto parameters estimated through classical PWMs. These asymptotic variances can be misleading, significantly misrepresenting true sampling variances when the sample size (n) is below 50. The asymptotic distribution of sample PWMs manifests as a linear combination of order statistics. [23] proved that the vector of $\hat{b} = (\hat{b}_1, \hat{b}_2)$ has asymptotically a multivariate Normal Distribution with mean $\hat{b} = (\hat{b}_1, \hat{b}_2)$ and covariance matrix $n^{-1}V$ . Supporting this, [24] employed the delta method to show that the covariance matrix can be articulated in terms of the variance of the parameter being estimated, hence, using the delta method, the covariance matrix has the expression of $n^{-1}\,GVG^T$, where V is the variance of the parameter. The quantity $\left[J'(\theta^{(a)})J(\theta^{(a)}) + \lambda^{(a)}I^{(a)}\right]^{-1}$ derived from applying the LM algorithm for parameter estimation serves as a reliable approximation of the variance–covariance matrix for the estimated parameters, known as the V matrix. Moreover, the G matrix, integral to the LM algorithm, functions as the Jacobian matrix, reflecting the relationships between the parameters being estimated. Harnessing these methods can vastly improve the accuracy and reliability of estimators, ultimately leading to more credible and impactful results in statistical analysis.

Using Taylor series expansion of the $g_1(\theta) = M_{101}$ and $g_2(\theta) = M_{110}$ around the $\theta_0$ gives

$g(\hat{\theta}) = g(\theta_0) + (\hat{\theta} - \theta_0)g'(\theta_0)$ . Therefore, the delta method can be applied. Taking the variance on both sides yields $var[g(\hat{\theta})] = [g'(\theta_0)]var(\hat{\theta})[g'(\theta_0)]^T$ where $g'(\theta_0) = \frac{\partial g(\theta_0)}{\partial \theta_0}\Big|_{\theta_0=\hat{\theta}}$ which is the Jacobian matrix in LM algorithm. Hence, $var(\hat{\theta}) = \left[[g'(\theta_0)]^T \left(var[g(\hat{\theta})]\right)^{-1}[g'(\theta_0)]\right]^{-1}$ and $var[g(\hat{\theta})] = \begin{bmatrix} var(M_{101}) & cov(M_{101}, M_{110}) \\ cov(M_{101}, M_{110}) & var(M_{110}) \end{bmatrix}$. This matrix, $[g'(\theta_0)]^T \left(var[g(\hat{\theta})]\right)^{-1}[g'(\theta_0)]$ may have a high condition number that inflates the $var(\hat{\theta})$ .

To ameliorate this, an appropriate regularization factor $(\xi)$ can be added to the matrix like this $\left[[g'(\theta_0)]^T \left(var[g(\hat{\theta})]\right)^{-1}[g'(\theta_0)] + \xi I\right]^{-1}$, where $\xi = \frac{\lambda_{max} - C^* \lambda_{min}}{C^* - 1}$ such that $\lambda_{max}$ is the maximum eigenvalue of the matrix $[g'(\theta_0)]^T \left(var[g(\hat{\theta})]\right)^{-1}[g'(\theta_0)]$ and $\lambda_{min}$ is the minimum eigenvalue for the same matrix. While the $C^*$ is the desired condition number. The desired value of this number is usually from 5 up to 10. Other types of variance is the value of this quantity $\left[J'(\theta^{(a)})J(\theta^{(a)}) + \lambda^{(a)} I^{(a)}\right]^{-1}$ which can be considered a good approximation to the variance-covariance matrix of the estimated parameters $\hat{\theta}$ . The delta method is applied to derive variance of the function of the estimated parameters.

The same discussion is applied to $M_{1,0,2}, M_{1,2,0}$. See appendix B for the variance and covariance of the lower and higher order PWMs.

# 3. Monte Carlo Simulation Study

## 3.1. Simulation Study for the Parameters

A simulation was conducted by generating 1000 replicates (N) of the MBUW random variable $y$, each replicate with different sample sizes (n). The author chosen suitable pairs of $\alpha$ and $\beta$ parameters respectively as follows (0.5, 0.6), (0.5, 1.3), (1.1, 0.6), and (1.1, 1.6) . The sample sizes were 20, 50, 100, and 500. For each sample, the parameters were estimated using the PWMs method (the low and high orders), maximum likelihood (MLE) and the Method of Moment (MOM, the first and the second raw moments). For the MOM, $m_1'(y) = \mu_1'$ and , $m_2'(y) = \mu_2'$ where $m_1'(y) = \frac{1}{n}\sum_i^n y_i = \bar{y}$ and $m_2'(y) = \frac{1}{n}\sum_i^n y_i^2$. $\mu_1' = \frac{6}{(2+\alpha^\beta)(3+\alpha^\beta)}$ and $\mu_2' = \frac{6}{(2+2\alpha^\beta)(3+2\alpha^\beta)}$. For each method, where $\hat{\theta} = \begin{bmatrix} \hat{\alpha} \\ \hat{\beta} \end{bmatrix}$, the statistical indices used for the comparisons are:

1. $Average\ Absolute\ Bias\ (AAB) = \frac{1}{N}\sum_{j=1}^N \left|\hat{\boldsymbol{\theta}} - \boldsymbol{\theta}\right|$
2. $Mean\ Square\ Error\ (MSE) = \frac{1}{N}\sum_{j=1}^N \left(\hat{\boldsymbol{\theta}} - \boldsymbol{\theta}\right)^2$

3. $Mean\ Relative\ Error\ (MRE) = \frac{1}{N}\sum_{j=1}^{N}\frac{|\hat{\boldsymbol{\theta}}-\boldsymbol{\theta}|}{\theta}$

$The\ empirical\ variance\ for\ each\ parameter = MSE - bias^2.$

The 2.5th and the 97.5th quantiles of the sampling distribution for each parameter are also recorded. The number of the valid samples, which shows GoF, is recorded. These results are illuminated in Tables (1-4). Each table shows the results of the comparisons between the methods for each sample size. The method yielding the estimator with the least bias is colored. Tables (5-8) show the empirical variance-covariance matrices obtained from the sampling distribution of the parameters. These last four tables demonstrate the Pearson correlation coefficient between the parameters, the condition number and eigenvalues of the empirical covariance matrix. Table (9) expounds the variance decay rate for each simulation validating the asymptotic consistency of the estimator $(\hat{\theta}_n)$ as n increases. In asymptotic theory, for a consistent estimator $\hat{\theta}_n$ of a parameter $\theta$, the variance typically behaves like $var(\hat{\theta}_n) \approx \frac{C}{n}$ for large n, where C is some constant. This means that the variance decays inversely with sample size, it decays as 1/n. Taking log of both sides of $var(\hat{\theta}_n) \approx \frac{C}{n}$ gives $\log[var(\hat{\theta}_n)] \approx \log(C) - \log n$ so the slope in the log-log plot of the variance vs. sample size should be approximately (-1). For each simulation the last two values of the variance of each parameter at n=500 and n=100 will be substituted in this log equation, in other words, the slope is calculated as

$\frac{\log[var(\hat{\theta}_{n=500})]-\log[var(\hat{\theta}_{n=100})]}{\log 500-\log 10} \approx -1$ . This is calculated for $(\hat{\alpha})$ & $(\hat{\beta})$ . Tables (10-13) illustrate the computational time for each simulation and the relative efficiency of each method vs. other methods. The most efficient method (the one with the least MSE among other methods in the row, for $(\hat{\alpha})$ & $(\hat{\beta})$) are colored. The computational time is recorded in seconds per replicate and also the total time of all replicates is recorded. The relative efficiency of estimator A obtained from method A to estimator B gained from method B is defined as $RE(\ A\ vs. B) = \frac{\widehat{MSE}_B}{\widehat{MSE}_A}$, this means that the estimator A is more efficient than the estimator B.

## 3.2. Simulation Study for the Function of the Parameters (Validation of the Delta Method)

Another simulation was conducted to validate the asymptotic normality of the delta method as follows: for sample size, n=500, alpha=0.5 , beta=1.3, and N=1000

Step 1: for each replicate the $g_1(\hat{\theta}) = \widehat{M}_{101}$ and $g_2(\hat{\theta}) = \widehat{M}_{110}$ .

Step 2: $g_1(\theta_0) = M_{101}$ and $g_2(\theta_0) = M_{110}$ where $\theta_0 = [\alpha = 0.5, \quad \beta = 1.3]$ .

Step 3: $error_{101} = \widehat{M}_{101} - M_{101}$ and $error_{110} = \widehat{M}_{110} - M_{110}$

Step 4: calculate the variance-covariance between $error_{101}$ and $error_{110}$, say it is $(cov_{emp})$. This is the empirical covariance matrix (2by2 matrix).

Step 5: obtain the theoretical covariance matrix ($cov_{theo}$) using the delta method applied on each replicate then take the mean, let's call this $mean_cov_{theo}$

Step 6: apply the singular value decomposition (SVD) on the error matrix $error = [error_{101} \quad error_{110}]$. The error matrix is N by 2 matrix.

$[error_{101} \quad error_{110}] = U_1 * S_1 * V_1^T$, where $S_1$ is the large non-zero singular value, and $V_1$ is the corresponding principal vector, $\lambda_{empirical} = S_1^2/(N-1)$. Let's call $s_{emp} = [error_{101} \quad error_{110}] * V_1$, is the projection of the error matrix in the direction of the principle vector. Let's call $Z_{empirical} = s_{emp}/\sqrt{\lambda_{empirical}}$ , this standardized empirical Z should be distributed as standard normal. These Zs' are the standardized errors after projection along the direction of maximum variability.

Step 7: get the trace of the $cov_{emp}$ and the trace of the $Mean_cov_{theo}$ . Let's call $scalar = trace(cov_{emp.})/trace(mean_cov_{theo})$. Let's call $cov_{theo_scaled} = scalar * mean_cov_{theo}$ .

Step 8: apply the singular value decomposition (SVD) on $mean_cov_{theo}$, so the $mean_{cov_{theo}} = U_2 * S_2 * V_2^T$ where $S_2$ is the large non-zero singular value, and $V_2$ is the corresponding principal vector, Let's call $s_{theo} = [error_{101} \quad error_{110}] * V_2$, is the projection of the error matrix in the direction of the maximum variability of the theoretical covariance matrix.

Step 9: calculate $\lambda_{theo_scaled} = scalar * V_2 * mean_cov_{theo} * V_2^T$ . Then calculate $Z_{theo_scaled} = s_{theo}/\sqrt{\lambda_{theo_scaled}}$. This standardized theoretical Z should be distributed as standard normal.

The intuition behind this algorithm is that the the theoretical covariance matrix (obtained from the delta method) should approximate the empirical covariance matrix from the replicates. The empirical and the theoretical standardized error can be defined respectively as

$Z_{emp} = \Sigma_{emp}^{-1/2}\left(g(\hat{\theta}) - g(\theta_0)\right)$ and $Z_{theo_scaled} = \Sigma_{theo_scaled}^{-1/2}\left(g(\hat{\theta}) - g(\theta_0)\right)$ and to hold for asymptotic normality these Zs' should be distributed as standard normal. A consistent observation across simulations (for all pairs of parameter and at different sample sizes) was that both covariance matrices (empirical and theoretical) shared nearly identical eigenvectors, as revealed by the singular value decomposition(SVD). This indicates that the empirical and theoretical covariance matrices capture a similar dependence structure between the two parameters. However, notable differences were observed in their singular values, particularly for large sample size. The theoretical matrix was a full rank for small sample size (n=20) but it is rank-one as n increases (n=500). This degeneration reflects the increasing linear dependency between the two parameter estimators as the sample size grows, implying that one parameter becomes an almost deterministic function of the other in large samples. To adjust for the overall discrepancy between the empirical and the theoretical covariance, a scalar rescaling factor was applied to the estimated theoretical matrix. This factor is $trace(cov_{emp.})/trace(mean_cov_{theo})$ , where the trace of covariance matrix represents the total marginal variance. Multiplying the theoretical covariance matrix by this scalar aligns the overall variance magnitude of the theoretical covariance with that of the empirical one while preserving its correlation structure. In other words, the rescaling equalizes the total variability but does not alter the orientation (the dependence structure, the eigenvectors) between the parameters. After this adjustment,; the standardized vector, the theoretical Z, exhibited approximately zero mean and unit standard deviation, confirming that the scaling effectively normalized the parameter estimators (function of parameters) .Hence, the asymptotic normality of the delta method was thus validated: as n increased, the distribution of the standardized estimators (PWMs are functions of the parameters) approached the standard normal, and the theoretical covariance structure aligned closely with the empirical results, despite the asymptotic rank deficiency inherent in the parameter dependence. Figure (3) illustrates the histogram, the fitted standard normal curves, and the QQ plot for the Z empirical and the Z theoretical (scaled).

## 4. Results and Discussion of the Simulation Study

Tables (1-4) show that at various sample sizes where ($\alpha = 0.5$ & $\beta = 0.6$) the estimated ($\hat{\alpha}$) obtained by MLE has the least Average Absolute Bias (AAB) and the least Mean Square Error (MSE) among the methods while the estimated $(\hat{\beta})$ obtained by $M_{1,0,2}, M_{1,2,0}$ has the least AAB and MSE among the methods. This is also true at simulation runs using the pair ($\alpha = 0.5$ & $\beta = 1.3$) and the pair ($\alpha = 1.1$ & $\beta = 1.6$) . However, at true values of the pair ($\alpha = 1.1$ & $\beta = 0.6$ ) , the higher order PWMs ($M_{1,0,2}, M_{1,2,0}$) produce ($\hat{\alpha}$) and $(\hat{\beta})$ with the least AAB and MSE among the methods. The method yielding an estimate of each parameter with the least bias and least MSE is colored in each table. The tables record the number of valid samples which pass the GoF in each run of the simulation. Method of Moments (MOM) give non-identifiable estimates of the parameters at ($\alpha = 1.1$ & $\beta = 0.6$ ) and ($\alpha = 1.1$ & $\beta = 1.6$) especially at small sample size; n=20 and n=50. Moreover, at large sample sizes; n=100 and n=500 the number of valid samples gained by (MOM) are less than the number of valid samples obtained by other methods at the same values of the parameters pair. MOMs almost give the highest AAB among other methods at different sample sizes and at different pairs of parameters. MOMs almost yield the highest MSE among other methods across different sample sizes and across different pairs of the simulated parameters. The AAB of the higher order PWMs is approximately equal to that obtained from the lower order PWMs especially at ($\alpha = 0.5$ & $\beta = 1.3$) and at almost all sample sizes. The standard error of each of the estimated parameter is recorded beside the estimated value of the parameter. It is also obvious that the AAB and the MSE decrease as the sample size increases. This is almost true for all methods and for both estimated parameters. And this supports that the methods yield consistent estimators for the parameters.

Table(1): results of simulation study using sample size 20 with different values of $\alpha$ & $\beta$ .

| | Statistical indices | MLE | MOM | PWM M101,M110 | PWM M102,M120 |
|---|---|---|---|---|---|
| $n=20$ $\alpha=0.5$ $\beta=0.6$ | $Mean(\alpha)$ | 0.5055,(0.0842) | 0.4943,(0.1113) | 0.4975,(0.1007) | 0.4978,(0.0985) |
| | $Mean(\beta)$ | 0.5963,(0.1077) | 0.6298,(0.0812) | 0.6014,(0.0581) | 0.6013,(0.0569) |
| | $AAB(\alpha)$ | 0.0647 | 0.0878 | 0.0798 | 0.078 |
| | $AAB(\beta)$ | 0.0794 | 0.0648 | 0.0461 | 0.045 |
| | $MSE(\alpha)$ | 0.0071 | 0.0124 | 0.0101 | 0.0096 |
| | $MSE(\beta)$ | 0.0116 | 0.0075 | 0.0034 | 0.0032 |
| | $MRE(\alpha)$ | 0.1293 | 0.1756 | 0.1596 | 0.1559 |
| | $MRE(\beta)$ | 0.1324 | 0.1079 | 0.0768 | 0.0751 |
| | $Quantile(\alpha)$ | (0.3653,0.7004) | (0.2838,0.7266) | (0.3008,0.7042) | (0.2961,0.6909) |
| | $Quantile(\beta)$ | (0.3366,0.7596) | (0.4767,0.8019) | (0.4821,0.7151) | (0.4897,0.7177) |
| **Number of valid samples** | | **1000 of 1000** | **987 out of 1000** | **999 out of 1000** | **999 out of 1000** |
| $n=20$ $\alpha=0.5$ $\beta=1.3$ | $Mean(\alpha)$ | 0.5000 (0.0372) | 0.4441 (0.0648) | 0.4955 (0.0580) | 0.4954 (0.0578) |
| | $Mean(\beta)$ | 1.3210,0.0963 | 1.1365 (0.0724) | 1.3012 (0.0155) | 1.3012 (0.0154) |
| | $AAB(\alpha)$ | 0.02297 | 0.0706 | 0.0467 | 0.0468 |
| | $AAB(\beta)$ | 0.0800 | 0.1637 | 0.0124 | 0.0125 |
| | $MSE(\alpha)$ | 0.0014 | 0.0073 | 0.0034 | 0.0034 |
| | $MSE(\beta)$ | 0.0097 | 0.0320 | 2.4062e-4 | 2.3897e-4 |
| | $MRE(\alpha)$ | 0.0593 | 0.1412 | 0.0934 | 0.0936 |
| | $MRE(\beta)$ | 0.0615 | 0.1259 | 0.0096 | 0.0096 |
| | $Quantile(\alpha)$ | (0.4281,0.5786) | (0.2679,0.6069) | (0.3815,0.6117) | (0.3837,0.6118) |
| | $Quantile(\beta)$ | (1.111,1.5046) | (1.0330,1.3073) | (1.2702,1.3316) | (1.2702,1.3310) |
| **Number of valid samples** | | **1000 of 1000** | **1000 of 1000** | **1000 of 1000** | **1000 of 1000** |
| $n=20$ $\alpha=1.1$ $\beta=0.6$ | $Mean(\alpha)$ | 1.1056 (0.2651) | 1.1772 (0.3668) | 1.0920 (0.2883) | 1.0942 (0.2586) |
| | $Mean(\beta)$ | 0.6509 (0.0407) | 0.4717 (0.2795) | 0.5986 (0.0504) | 0.5990 (0.0452) |
| | $AAB(\alpha)$ | 0.2153 | 0.2933 | 0.2287 | 0.2052 |
| | $AAB(\beta)$ | 0.0518 | 0.2534 | 0.0400 | 0.0359 |
| | $MSE(\alpha)$ | 0.0702 | 0.1371 | 0.0831 | 0.0668 |
| | $MSE(\beta)$ | 0.0043 | 0.0926 | 0.0025 | 0.0020 |
| | $MRE(\alpha)$ | 0.1958 | 0.2666 | 0.2079 | 0.1865 |
| | $MRE(\beta)$ | 0.0864 | 0.4224 | 0.0666 | 0.0598 |
| | $Quantile(\alpha)$ | (0.6824,1.7008) | (0.5143.1.8809) | (0.5476,1.6682) | (0.6037,1.5896) |
| | $Quantile(\beta)$ | (0.5945,0.7521) | (0.0074,0.8971) | (0.5035,0.6993) | (0.5133,0.6856) |
| **Number of valid samples** | | **1000 of 1000** | **40 out of 1000** | **1000 of 1000** | **1000 of 1000** |
| $n=20$ $\alpha=1.1$ $\beta=1.6$ | $Mean(\alpha)$ | 1.0928 (0.1050) | | 1.0979 (0.0172) | 1.0983 (0.1139) |
| | $Mean(\beta)$ | 1.6595 (0.0546) | | 1.5999 (0.0011) | 1.5999 (0.0075) |
| | $AAB(\alpha)$ | 0.086 | | 0.0959 | 0.0927 |
| | $AAB(\beta)$ | 0.0646 | | 0.0063 | 0.0061 |
| | $MSE(\alpha)$ | 0.0111 | | 0.0138 | 0.0130 |
| | $MSE(\beta)$ | 0.0065 | | 5.9109e-5 | 5.5617e-5 |
| | $MRE(\alpha)$ | 0.0782 | | 0.0128 | 0.0843 |
| | $MRE(\beta)$ | 0.0403 | | 5.7592e-4 | 0.0038 |
| | $Quantile(\alpha)$ | (0.8940,1.2883) | | (0.8661,1.13024) | (8.700,1.3045) |
| | $Quantile(\beta)$ | (1.5711,1.7822) | | (1.5847,1.6133) | (1.5849,1.6134) |
| **Number of valid samples** | | **1000 of 1000** | | **1000 of 1000** | **1000 of 1000** |

Table(2): results of simulation study using sample size of 50 with different values of $\alpha$ & $\beta$.

| | Statistical indices | MLE | MOM | PWM M101,M110 | PWM M102,M120 |
|---|---|---|---|---|---|
| $n = 50$ $\alpha = 0.5$ $\beta = 0.6$ | $Mean(\alpha)$ | 0.5052 (0.0533) | 0.4988 (0.074) | 0.5002 (0.0666) | 0.5001 (0.0645) |
| | $Mean(\beta)$ | 0.6012 (0.0652) | 0.6135 (0.0417) | 0.5999 (0.0385) | 0.5999 (0.0373) |
| | $AAB(\alpha)$ | 0.04427 | 0.0585 | 0.0540 | 0.0522 |
| | $AAB(\beta)$ | 0.0520 | 0.0348 | 0.0312 | 0.0301 |
| | $MSE(\alpha)$ | 0.0029 | 0.0055 | 0.0044 | 0.0042 |
| | $MSE(\beta)$ | 0.0042 | 0.0019 | 0.0015 | 0.0014 |
| | $MRE(\alpha)$ | 0.0854 | 0.1170 | 0.1080 | 0.1043 |
| | $MRE(\beta)$ | 0.0866 | 0.0579 | 0.0520 | 0.0502 |
| | $Quantile(\alpha)$ | (0.4097,0.6114) | (0.3495,0.6400) | (0.3693,0.6242) | (0.3718,0.6231) |
| | $Quantile(\beta)$ | (0.4535,0.7133) | (0.5302,0.6930) | (0.5283,0.6755) | (0.5289,0.6740) |
| **Number of valid samples** | | **999 out of 1000** | **999 out of 1000** | **1000 of 1000** | **999 out of 1000** |
| $n = 50$ $\alpha = 0.5$ $\beta = 1.3$ | $Mean(\alpha)$ | 0.5049 (0.0237) | 0.4699 (0.041) | 0.5015 (0.0374) | 0.5016 (0.0375) |
| | $Mean(\beta)$ | 1.3136 (0.0652) | 1.1947 (0.0546) | 1.2996 (0.0100) | 1.2996 (0.0100) |
| | $AAB(\alpha)$ | 0.0192 | 0.0415 | 0.0301 | 0.0302 |
| | $AAB(\beta)$ | 0.0537 | 0.1057 | 0.0080 | 0.0081 |
| | $MSE(\alpha)$ | 5.8625e-4 | 0.0026 | 0.0014 | 0.0014 |
| | $MSE(\beta)$ | 0.0044 | 0.0141 | 9.9379-5 | 1.0005e-4 |
| | $MRE(\alpha)$ | 0.0384 | 0.0830 | 0.0602 | 0.0605 |
| | $MRE(\beta)$ | 0.0413 | 0.0813 | 0.0062 | 0.0062 |
| | $Quantile(\alpha)$ | (0.4540,0.5486) | (0.3916,0.5492) | (0.4313,0.5756) | (0.4286,0.5769) |
| | $Quantile(\beta)$ | (1.1705,1.4177) | (1.1100,1.3013) | (1.2798,1.3183) | (1.2795,1.3190) |
| **Number of valid samples** | | **1000 of 1000** | **1000 of 1000** | **1000 of 1000** | **1000 of 1000** |
| $n = 50$ $\alpha = 1.1$ $\beta = 0.6$ | $Mean(\alpha)$ | 1.1094 (0.1773) | | 1.1040 (0.1804 | 1.1034 (0.1610) |
| | $Mean(\beta)$ | 0.6336 (0.0282) | | 0.6007 (0.0315) | 0.6006 (0.0281) |
| | $AAB(\alpha)$ | 0.1439 | | 0.1444 | 0.1292 |
| | $AAB(\beta)$ | 0.0351 | | 0.0252 | 0.0226 |
| | $MSE(\alpha)$ | 0.0315 | | 0.0325 | 0.0259 |
| | $MSE(\beta)$ | 0.0019 | | 9.9281e-4 | 7.9139e-4 |
| | $MRE(\alpha)$ | 0.1308 | | 0.1312 | 0.1175 |
| | $MRE(\beta)$ | 0.0584 | | 0.0420 | 0.0376 |
| | $Quantile(\alpha)$ | (0.8060,1.4720) | | (0.7622,1.4710) | (0.7980,1.4223) |
| | $Quantile(\beta)$ | (0.5924,0.6997) | | (0.5410,0.6648) | (0.5472,0.6563) |
| **Number of valid samples** | | **1000 of 1000** | | **1000 of 1000** | **1000 of 1000** |
| $n = 50$ $\alpha = 1.1$ $\beta = 1.6$ | $Mean(\alpha)$ | 1.0989 (0.0687) | | 1.1014 (0.0747) | 1.1012 (0.0726) |
| | $Mean(\beta)$ | 1.6399 (0.0394) | | 1.6001 (0.0049) | 1.6001 (0.0048) |
| | $AAB(\alpha)$ | 0.0557 | | 0.0597 | 0.0581 |
| | $AAB(\beta)$ | 0.0458 | | 0.0039 | 0.0038 |
| | $MSE(\alpha)$ | 0.0047 | | 0.0056 | 0.0053 |
| | $MSE(\beta)$ | 0.0031 | | 2.3924e-5 | 2.2608e-5 |
| | $MRE(\alpha)$ | 0.0506 | | 0.0543 | 0.0528 |
| | $MRE(\beta)$ | 0.0286 | | 0.0024 | 0.0024 |
| | $Quantile(\alpha)$ | (0.9697,1.2345) | | (0.9606,1.2525) | (0.9610,1.2454) |
| | $Quantile(\beta)$ | (1.5658,1.7226) | | (1.5909,1.6100) | (1.5909,1.6095) |
| **Number of valid samples** | | **1000 of 1000** | | **1000 of 1000** | **1000 of 1000** |

Table(3) : results of simulation study using sample size of 100 with different values of $\alpha$ & $\beta$.

| | Statistical indices | MLE | MOM | PWM M101,M110 | PWM M102,M120 |
|---|---|---|---|---|---|
| $n = 100$ $\alpha = 0.5$ $\beta = 0.6$ | $Mean(\alpha)$ | 0.5043 (0.0353) | 0.5027 (0.0484) | 0.5005 (0.0449) | 0.5001 (0.0434) |
| | $Mean(\beta)$ | 0.6042 (0.0433) | 0.6106 ( 0.0274) | 0.5997 (0.0259) | 0.5999 (0.0251) |
| | $AAB(\alpha)$ | 0.0282 | 0.0383 | 0.0359 | 0.0347 |
| | $AAB(\beta)$ | 0.0349 | 0.0232 | 0.0207 | 0.0200 |
| | $MSE(\alpha)$ | 0.0013 | 0.0023 | 0.002 | 0.0019 |
| | $MSE(\beta)$ | 0.0019 | 8.6160e-4 | 6.7061e-4 | 6.2760e-4 |
| | $MRE(\alpha)$ | 0.0564 | 0.0767 | 0.0718 | 0.0694 |
| | $MRE(\beta)$ | 0.0581 | 0.0387 | 0.0345 | 0.0334 |
| | $Quantile(\alpha)$ | (0.4332,0.5752) | (0.4059,0.5999) | (0.4118,0.5884) | (0.4110,0.5851) |
| | $Quantile(\beta)$ | (0.5107,0.6776) | (0.5577,0.6669) | (0.5490,0.6509) | (0.5509,0.6514) |
| **Number of valid samples** | | **1000 of 1000** | **1000 of 1000** | **1000 out of 1000** | **1000 of 1000** |
| $n = 100$ $\alpha = 0.5$ $\beta = 1.3$ | $Mean(\alpha)$ | 0.5048 (0.0164) | 0.4845 (0.0277) | 0.5000 (0.0256) | 0.5002 (0.0253) |
| | $Mean(\beta)$ | 1.3183 (0.0441) | 1.247 (0.0330) | 1.3000 (0.0068) | 1.2999 (0.0067) |
| | $AAB(\alpha)$ | 0.0137 | 0.0254 | 0.0207 | 0.0204 |
| | $AAB(\beta)$ | 0.0379 | 0.0534 | 0.0055 | 0.0054 |
| | $MSE(\alpha)$ | 2.9281e-4 | 0.001 | 6.5674e-4 | 6.4032e-4 |
| | $MSE(\beta)$ | 0.0023 | 0.0039 | 4.6676e-5 | 4.5509e-5 |
| | $MRE(\alpha)$ | 0.0273 | 0.0508 | 0.0414 | 0.0407 |
| | $MRE(\beta)$ | 0.0292 | 0.0411 | 0.0042 | 0.0042 |
| | $Quantile(\alpha)$ | (0.4731,0.5354) | (0.4322,0.5397) | (0.4503,0.5508) | (0.4507,0.5494) |
| | $Quantile(\beta)$ | (1.2219,1.3884) | (1.1916,1.3021) | (1.2865,1.3132) | (1.2868,1.3131) |
| **Number of valid samples** | | **1000 of 1000** | **1000 of 1000** | **1000 of 1000** | **1000 of 1000** |
| $n = 100$ $\alpha = 1.1$ $\beta = 0.6$ | $Mean(\alpha)$ | 1.1025 (0.122) | 1.2234 (0.4310) | 1.1006 (0.1254) | 1.1014 (0.1114) |
| | $Mean(\beta)$ | 0.6232 (0.0216) | 0.2910 (0.2066) | 0.6001 (0.0219) | 0.6002 (0.0195) |
| | $AAB(\alpha)$ | 0.0974 | 0.3368 | 0.0999 | 0.0888 |
| | $AAB(\beta)$ | 0.0248 | 0.3267 | 0.0175 | 0.0155 |
| | $MSE(\alpha)$ | 0.0149 | 0.2005 | 0.0157 | 0.0124 |
| | $MSE(\beta)$ | 0.0010 | 0.1381 | 4.7959e-4 | 3.7833e-4 |
| | $MRE(\alpha)$ | 0.0885 | 0.3062 | 0.0908 | 0.0807 |
| | $MRE(\beta)$ | 0.0413 | 0.5445 | 0.0291 | 0.0259 |
| | $Quantile(\alpha)$ | (0.8780,1.3561) | (0.2606,2.1869) | (0.8532,1.3501) | (0.8829,1.3176) |
| | $Quantile(\beta)$ | (0.5907,0.6774) | (0.0076,0.7279) | (0.5569,0.6437) | (0.5621,0.6380) |
| **Number of valid samples** | | **1000 of 1000** | **409 out of 1000** | **1000 of 1000** | **1000 of 1000** |
| $n = 100$ $\alpha = 1.1$ $\beta = 1.6$ | $Mean(\alpha)$ | 1.0989 (0.0474) | 1.3354 (0.1711) | 1.1010 (0.0505) | 1.1008 (0.0492) |
| | $Mean(\beta)$ | 1.6365 (0.0318) | 0.6937 (0.3946) | 1.6001 (0.0033) | 1.6001 (0.0032) |
| | $AAB(\alpha)$ | 0.0379 | 0.2446 | 0.0409 | 0.0398 |
| | $AAB(\beta)$ | 0.0405 | 0.9064 | 0.0027 | 0.0026 |
| | $MSE(\alpha)$ | 0.0022 | 0.0847 | 0.0026 | 0.0024 |
| | $MSE(\beta)$ | 0.0023 | 0.9770 | 1.0953e-5 | 1.0394e-5 |
| | $MRE(\alpha)$ | 0.0345 | 0.2224 | 0.0372 | 0.0362 |
| | $MRE(\beta)$ | 0.0253 | 0.5665 | 0.0017 | 0.0016 |
| | $Quantile(\alpha)$ | (1.0089,1.1856) | (1.0331,1.6236) | (1.0036,1.1972) | (1.0056,1.1941) |
| | $Quantile(\beta)$ | (1.5761,1.6994) | (0.2126,1.5451) | (1.5937,1.6064) | (1.5938,1.6062) |
| **Number of valid samples** | | **1000 of 1000** | **997 out of 1000** | **1000 of 1000** | **1000 of 1000** |

Table(4): results of simulation study using sample size of 500 with different values of $\alpha$ & $\beta$.

| | | MLE | MOM | PWM M101,M110 | PWM M102,M120 |
|---|---|---|---|---|---|
| $n = 500$ $\alpha = 0.5$ $\beta = 0.6$ | $Mean(\alpha)$ | 0.5044 (0.0162) | 0.5011 (0.0214) | 0.4988 (0.0202) | 0.4989 (0.0197) |
| | $Mean(\beta)$ | 0.6097 (0.0201) | 0.6066 (0.0123) | 0.6007 (0.0117) | 0.6006 (0.0114) |
| | $AAB(\alpha)$ | 0.0131 | 0.0171 | 0.0161 | 0.0156 |
| | $AAB(\beta)$ | 0.0177 | 0.0110 | 0.0093 | 0.0090 |
| | $MSE(\alpha)$ | 2.8252e-4 | 4.5718e-4 | 4.1113e-4 | 3.8840e-4 |
| | $MSE(\beta)$ | 4.9920e-4 | 1.9447e-4 | 1.3717e-4 | 1.2959e-4 |
| | $MRE(\alpha)$ | 0.0261 | 0.0341 | 0.0322 | 0.0312 |
| | $MRE(\beta)$ | 0.0294 | 0.0184 | 0.0155 | 0.0150 |
| | $Quantile(\alpha)$ | (0.4731,0.5360) | (0.4590,0.5431) | (0.4581,0.5388) | (0.4578,0.5380) |
| | $Quantile(\beta)$ | (0.5659,0.6457) | (0.5819,0.6303) | (0.5776,0.6242) | (0.5781,0.6244) |
| **Number of valid samples** | | **1000 of 1000** | **1000 of 1000** | **1000 of 1000** | **1000 of 1000** |
| $n = 500$ $\alpha = 0.5$ $\beta = 1.3$ | $Mean(\alpha)$ | 0.5051 (0.0073) | 0.4971 (0.0118) | 0.5000 (0.0113) | 0.4999 (0.0113) |
| | $Mean(\beta)$ | 1.3195 (0.0242) | 1.2894 (0.0068) | 1.3000 (0.003) | 1.3 (0.0030) |
| | $AAB(\alpha)$ | 0.0068 | 0.0098 | 0.0092 | 0.0092 |
| | $AAB(\beta)$ | 0.0237 | 0.0108 | 0.0024 | 0.0024 |
| | $MSE(\alpha)$ | 7.9362e-5 | 1.4693e-4 | 1.2707e-4 | 1.2682e-4 |
| | $MSE(\beta)$ | 9.6674e-4 | 1.5855e-4 | 9.0313e-6 | 9.0134e-6 |
| | $MRE(\alpha)$ | 0.0137 | 0.0197 | 0.0183 | 0.0183 |
| | $MRE(\beta)$ | 0.0182 | 0.0083 | 0.0019 | 0.0019 |
| | $Quantile(\alpha)$ | (0.4932,0.5216) | (0.4753,0.5205) | (0.4795,0.5223) | (0.4793,0.5219) |
| | $Quantile(\beta)$ | (1.2824,1.3668) | (1.2766,1.3018) | (1.2940,1.3055) | (1.2942,1.3055) |
| **Number of valid samples** | | **1000 of 1000** | **1000 of 1000** | **1000 of 1000** | **1000 of 1000** |
| $n = 500$ $\alpha = 1.1$ $\beta = 0.6$ | $Mean(\alpha)$ | 1.0978 (0.0556) | 1.2494 ( 0.0868) | 1.1001 (0.0548) | 1.1 (0.0488) |
| | $Mean(\beta)$ | 0.6137 (0.0126) | 0.2754 (0.1023) | 0.6000 (0.0096) | 0.6 (0.0085) |
| | $AAB(\alpha)$ | 0.0453 | 0.1532 | 0.0444 | 0.0396 |
| | $AAB(\beta)$ | 0.0155 | 0.3247 | 0.0077 | 0.0069 |
| | $MSE(\alpha)$ | 0.0031 | 0.0299 | 0.0030 | 0.0024 |
| | $MSE(\beta)$ | 3.4713e-4 | 0.1158 | 9.1734e-5 | 7.2737e-5 |
| | $MRE(\alpha)$ | 0.0412 | 0.1393 | 0.0403 | 0.0360 |
| | $MRE(\beta)$ | 0.0258 | 0.5412 | 0.0129 | 0.0115 |
| | $Quantile(\alpha)$ | (0.9962,1.2096) | (1.075, 1.4236) | (0.9952,1.2103) | (1.0074,1.1972) |
| | $Quantile(\beta)$ | (0.5904,0.6388) | (0.0953, 0.5332) | (0.5817,0.6193) | (0.5838,0.6170) |
| **Number of valid samples** | | **1000 of 1000** | **675 out of 1000** | **1000 of 1000** | **1000 of 1000** |
| $n = 500$ $\alpha = 1.1$ $\beta = 1.6$ | $Mean(\alpha)$ | 1.0976 (0.0213) | 1.1300 (0.0454) | 1.1001 (0.0227) | 1.1 (0.0220) |
| | $Mean(\beta)$ | 1.6352 (0.0279) | 1.3005 (0.23550 | 1.6000 (0.0015) | 1.6 (0.0014) |
| | $AAB(\alpha)$ | 0.0171 | 0.0388 | 0.0183 | 0.0178 |
| | $AAB(\beta)$ | 0.0384 | 0.2995 | 0.0012 | 0.0012 |
| | $MSE(\alpha)$ | 4.5729e-4 | 0.0030 | 5.1549e-4 | 4.8480e-4 |
| | $MSE(\beta)$ | 0.0020 | 0.1451 | 2.2133e-6 | 2.0816e-6 |
| | $MRE(\alpha)$ | 0.0155 | 0.0353 | 0.0167 | 0.0162 |
| | $MRE(\beta)$ | 0.0240 | 0.1872 | 7.5140e-4 | 7.3023e-4 |
| | $Quantile(\alpha)$ | (1.0584,1.1414) | (1.0631,1.248) | (1.0564,1.1454) | (1.0584,1.1441) |
| | $Quantile(\beta)$ | (1.5789,1.6842) | (0.6479,1.600) | (1.5971,1.6030) | (1.5973, 1.6029) |
| **Number of valid samples** | | **1000 of 1000** | **1000 of 1000** | **1000 of 1000** | **1000 of 1000** |

Table(5): analysis of the sampling distribution of the estimated $(\hat{\alpha})$ & $(\hat{\beta})$ using PWMs $M_{1,0,1}, M_{1,1,0}$

| n | | PWM: M101,M110 $\alpha = 0.5$ & $\beta = 0.6$ | | PWM: M101,M110 $\alpha = 0.5$ & $\beta = 1.3$ | | PWM: M101,M110 $\alpha = 1.1$ & $\beta = 0.6$ | | PWM: M101,M110 $\alpha = 1.1$ & $\beta = 1.6$ | |
|---|---|---|---|---|---|---|---|---|---|
| 500 | Var_cov | 4.1004e-4 | -2.368e-4 | 1.272e-4 | -3.391e-5 | 0.003 | 5.255e-4 | 5.1600e-4 | 3.3811e-5 |
| | | -2.368e-4 | 1.3681e-4 | -3.391e-5 | 9.0404e-6 | 5.255e-4 | 9.1825e-5 | 3.3811e-5 | 2.2155e-6 |
| | Corr | -1 | | -1 | | 1 | | 1 | |
| | Cond num. Eigen.val | Cond. num=1.395e+16 Eigen_val=-1.3553e-20, 5.4685e-4 | | Cond_num=9.0086e+16 Eigen_val=1.6941e-21, 1.3624e-4 | | Cond_num=2.2935e+17 Eigen_val=0, 0.0031 | | Cond_num=6.692e+17 Eigen_val=0, 5.1822e-4 | |
| 100 | Var_cov | 0.0020 | -0.0012 | 6.574e-4 | -1.7526e-4 | 0.0157 | 0.0027 | 0.0026 | 1.6725e-4 |
| | | -0.0012 | 6.7119e-4 | 1.7526e-4 | 4.6723e-5 | 0.0027 | 4.8006e-4 | 1.6725e-4 | 1.0959e-5 |
| | Corr | -1 | | -1 | | 1 | | 1 | |
| | Cond num. Eigen.val | Cond. num=1.632e+16 Eigen_val=1.0842e-19, 0.0027 | | Cond. num=5.8349e+16 Eigen_val=0, 7.0412e-4 | | Cond. num=1.2587e+17 Eigen_val=0.0162, 2.1684e-19 | | Cond. num=2.0801e+18 Eigen_val=0.0026, 3.3881e-21 | |
| 50 | Var_cov | 0.0044 | -0.0026 | 0.0014 | -3.7257e-4 | 0.0325 | 0.0057 | 0.0056 | 3.6534e-4 |
| | | -0.0026 | 0.0015 | -3.7257e-4 | 9.9325e-5 | 0.0057 | 9.9332e-4 | 3.6534e-4 | 2.3939e-5 |
| | Corr | -1 | | -1 | | 1 | | 1 | |
| | Cond num. Eigen.val | Cond. num=2.548e+16 Eigen_val=0, 0.0059 | | Cond. num=5.654e+16 Eigen_val=0.0015, 1.3551e-20 | | Cond. num=5.2012e+16 Eigen_val=0.0335, 4.3368e-19 | | Cond. num=6.3331e+17 Eigen_val=0.0056, -6.7763e-21 | |
| 20 | Var_cov | 0.0101 | -0.0059 | 0.0034 | -8.9798e-4 | 0.0831 | 0.0145 | 0.0138 | 9.0267e-4 |
| | | -0.0059 | 0.0034 | -8.9798e-4 | 2.3940e-4 | 0.0145 | 0.0025 | 9.0267e-4 | 5.9148e-5 |
| | Corr | -1 | | -1 | | 1 | | 1 | |
| | Cond num. Eigen.val | Cond.num=2.7699e+16 Eigen.val=0.0135, 4.3368e-19 | | Cond.num=1.0764e+17 Eigen.val=0.0036, 5.4210e-20 | | Cond.num=9.2362e+16 Eigen.val=0.0856, 4.3368e-20 | | Cond.num=1.328e+18 Eigen.val=0.0138, -1.3553e-20 | |

Table 5 shows the variance-covariance matrix for each estimated parameter $(\hat{\alpha})$ and $(\hat{\beta})$ obtained by the lower order PWMs $(M_{1,0,1}, M_{1,1,0})$. These estimated parameters are perfectly correlated as seen in all simulation across different sample sizes and across different pairs. These matrices show high condition number and one eigenvalue that is nearly zero. This may lead to identifiability problem and inflate the estimated variance which impairs the confidence interval construction. The lower order PWMs are derived from the order statistics so correlation is inherent in the method. The question is does the distribution contribute to this high dependency? To answer this question; the correlation between the estimated parameters obtained from deploying other methods were investigated using the sampling distribution gained from the simulation. Table 6 depicts the sampling distribution obtained from applying the higher order PWMs $(M_{1,0,2}, M_{1,2,0})$ for parameters estimation. The same outcomes were observed. But this may be due to the PWMs inheriting the high dependency from the order statistics. But as the same results were also returned from MLE and MOMs as shown in Tables (7-8), this is a strong evidence that the distribution is the main root cause. Although, the method can contribute, but when every reasonable method shows the same correlation patterns, so the distribution itself is likely to cause this coupling. The Weibull distribution and any derived distribution from the Weibull, like the one between our hands (MBUW), the parameters control related features of the sample data (mean, spread, and tail) by giving information about some combinations of the parameters more strongly than about each parameter separately. Whether the dependency relation is linear, nonlinear, monotonic, or non-onotonic, the method can influence this. Figure (1) shows the relationship between the estimated $(\hat{\alpha})$ and estimated $(\hat{\beta})$ obtained from the simulation using the lower order PWMs at n=20. Table (6-8) show the sampling distribution of the estimated parameters obtained by the simulation study. The empirical variance-covariance matrices and the Pearson correlation coefficient are recorded. The condition number and the eigenvalues of the covariance matrix for each simulation are also shown. Other figures for the different shapes of the dependency are seen in Appendix C

Table (6): analysis of the sampling distribution of the estimated $(\hat{\alpha})$ & $(\hat{\beta})$ using PWMs $M_{1,0,2}, M_{1,2,0}$

| n | | PWM: M102,M120 $\alpha = 0.5$ & $\beta = 0.6$ | | PWM: M102,M120 $\alpha = 0.5$ & $\beta = 1.3$ | | PWM: M102,M120 $\alpha = 1.1$ & $\beta = 0.6$ | | PWM: M102,M120 $\alpha = 1.1$ & $\beta = 1.6$ | |
|---|---|---|---|---|---|---|---|---|---|
| 500 | Var_cov | 3.8765e-4 | -2.239e-4 | 1.2694e-4 | -3.384e-5 | 0.0024 | 4.1669e-4 | 4.8529e-4 | 3.1799e-5 |
| | | -2.239e-4 | 1.2934e-4 | -3.384e-5 | 9.0221e-6 | 4.1669e-4 | 7.281e-5 | 3.1799e-5 | 2.0836e-6 |
| | Corr | -1 | | -1 | | 1 | | 1 | |
| | Cond num. Eigen.val | Cond.num=4.1394e+17 Eigen.val=-4.0658e-20, 5.1700e-4 | | Cond.num=2.7053e+17 Eigen.val=3.3881e-21, 1.3596e-4 | | Cond.num=1.3574e+17 Eigen.val=1.3553e-20, 0.0025 | | Cond.num=7.5123e+17 Eigen.val=1.2705e-21, 4.8727e-4 | |
| 100 | Var_cov | 0.0019 | -0.0011 | 6.4091e-4 | -1.7086e-4 | 0.0124 | 0.0022 | 0.0024 | 1.5875e-4 |
| | | -0.0011 | 6.2822e-4 | -1.7086e-4 | 4.5552e-5 | 0.0022 | 3.7865e-4 | 1.5875e-4 | 1.0402e-5 |
| | Corr | -1 | | -1 | | 1 | | 1 | |
| | Cond num. Eigen.val | Cond.num=1.4424e+16 Eigen.val=1.0842e-19, 0.0025 | | Cond.num=1.0285e+17 Eigen.val=-1.3553e-20, 6.8647e-4 | | Cond.num=1.2979e+17 Eigen.val=-2.1684e-19, 0.0128 | | Cond.num=8.5287e+17 Eigen.val=-5.0822e-21, 0.0024 | |
| 50 | Var_cov | 0.0042 | -0.0024 | 0.0014 | -3.749e-4 | 0.0259 | 0.0045 | 0.0053 | 3.4528e-4 |
| | | -0.0024 | 0.0014 | -3.749e-4 | 9.9962e-5 | 0.0045 | 7.9183e-4 | 3.4528e-4 | 2.2625e-5 |
| | Corr | -1 | | -1 | | 1 | | 1 | |
| | Cond num. Eigen.val | Cond.num=1.8731e+16 Eigen.val=0, 0.0055 | | Cond.num=4.3350e+16 Eigen.val=-4.0658e-20, 0.0015 | | Cond.num=2.0248e+17 Eigen.val=1.0842e-19, 0.0267 | | Cond.num=4.8561e+17 Eigen.val=1.3553e-20, 0.0053 | |
| 20 | Var_cov | 0.0096 | -0.0056 | 0.0033 | -8.917e-4 | 0.0668 | 0.0117 | 0.0130 | 8.4944e-4 |
| | | -0.0056 | 0.0032 | -8.917e-4 | 2.3772e-4 | 0.0117 | 0.0020 | 8.4944e-4 | 5.5660e-5 |
| | Corr | -1 | | -1 | | 1 | | 1 | |
| | Cond num. Eigen.val | Cond.num=2.0154e+16 Eigen.val=8.6736e-19, 0.0128 | | Cond.num=3.6958e+17 Eigen.val=0.0036, 0 | | Cond.num=1.1993e+17 Eigen.val=8.6736e-19, 0.0689 | | Cond.num=6.3129e+17 Eigen.val=-2.7105e-20, 0.0130 | |

The importance to recognize that the cause of the correlation is the distribution itself is the application of the practical remedies to lessen this correlation and hence to deflate the variance and to construct a valid confidence interval. Some of these remedies is to reparameterize the parameter and use the log of the parameter. These remedies can be a blueprint for future studies.

Table (9) expounds the pattern of the variance decay. The slopes in this table are obtained from the empirical variances for the estimated parameters recorded in Tables (5-8). Using the log (variance) at n=500, log (variance) at n=100, log (n=500) and log (n=100), the slope can be calculated as previously explained in section 3, (Monte Carlo Simulation Study). Figure (2) also supports this asymptotic variance decay at a rate (1/n) as n enlarges, where the estimated parameter $(\hat{\theta}_n)$ is asymptotically consistent at large sample size. Table (9) elucidates that the lower and the higher PWMs show better validity for this asymptotic consistency rather than the MLE and MOMs.

Table (7): analysis of the sampling distribution of the estimated $(\hat{\alpha})$ & $(\hat{\beta})$ of MLE

| n | | MLE $\alpha = 0.5$ & $\beta = 0.6$ | | MLE $\alpha = 0.5$ & $\beta = 1.3$ | | MLE $\alpha = 1.1$ & $\beta = 0.6$ | | MLE $\alpha = 1.1$ & $\beta = 1.6$ | |
|---|---|---|---|---|---|---|---|---|---|
| 500 | Var_cov | 2.631e-4 | -2.900e-4 | 5.368e-5 | -9.989e-5 | 0.0031 | 1.683e-4 | 4.522e-4 | -7.028e-5 |
| | | -2.900e-4 | 4.048e-4 | -9.989e-5 | 5.869e-4 | 1.683e-4 | 1.596e-4 | -7.028e-5 | 7.779e-4 |
| | Corr | -0.8887(p=0) | | -0.5628(p=1.26e-84) | | 0.2395(p=1.644e-14) | | -0.1185(p=1.729e-4) | |
| | Cond num. Eigen.val | Cond_num=17.8648 Eigen_val=3.5402e-05, 6.3245e-4 | | Cond_num=17.0047 Eigen_val=3.5578e-05, 6.0499e-4 | | Cond_num=20.6931 Eigen_val=1.5003e-04, 0.0031 | | Cond_num=1.8107 Eigen_val=4.3764e-04, 7.9246e-4 | |
| 100 | Var_cov | 0.0012 | -0.0015 | 2.699e-4 | -6.459e-4 | 0.0149 | -6.407e-4 | 0.0022 | 1.6973e-4 |
| | | -0.0015 | 0.0019 | -6.459e-4 | 0.0019 | -6.407e-4 | 4.676e-4 | 1.6973e-4 | 0.0010 |
| | Corr | -0.9670(p=0) | | -0.8917(p=0) | | -0.2429(p=6.8e-15) | | 0.1128(p=3.7e-4) | |
| | Cond num. Eigen.val | Cond_num=62.3364 Eigen_val=4.9263e-05, 0.0031 | | Cond_num=43.5652 Eigen_val=4.9688e-05, 0.0022 | | Cond_num=33.9381 Eigen_val=4.3920e-04, 0.0149 | | Cond_num=2.3023 Eigen_val=9.8549e-04, 0.0023 | |
| 50 | Var_cov | 0.0028 | -0.0034 | 5.626e-4 | -0.0014 | 0.0315 | -0.0013 | 0.0047 | -2.841e-4 |
| | | -0.0034 | 0.0042 | -0.0014 | 0.0043 | -0.0013 | 7.928e-4 | -2.841e-4 | 0.0016 |
| | Corr | -0.9823(p=0) | | -0.9326(p=0) | | -0.2514(p=7e-16) | | 0.1050(p=0.0009) | |
| | Cond num. Eigen.val | Cond_num=116.3342 Eigen_val=6.0377e-05, 0.0070 | | Cond_num=72.3631 Eigen_val=6.5640e-05, 0.0047 | | Cond_num=42.4821 Eigen_val=7.4153e-04, 0.0315 | | Cond_num=3.1071 Eigen_val=0.0015, 0.0047 | |
| 20 | Var_cov | 0.0071 | -0.0089 | 0.0014 | -0.0035 | 0.0703 | -0.0044 | 0.011 | -0.0013 |
| | | -0.0089 | 0.0116 | -0.0035 | 0.0093 | -0.0044 | 0.0017 | -0.0013 | 0.0030 |
| | Corr | -0.9788(p=0) | | -0.9683(p=0) | | -0.4087(p=0) | | -0.2297(p=0) | |
| | Cond num. Eigen.val | Cond.num=99.0764 Eigen.val=0.0185, 1.8665e-4 | | Cond.num=139.9554 Eigen.val=0.0106, 7.5630e-5 | | Cond.num=51.2278 Eigen.val=0.0014, 0.0705 | | Cond.num=4.0477 Eigen.val=0.0028, 0.0112 | |

Tables (10-13) show the relative efficiency of each method against other methods. The most efficient method among the others is colored. The higher order PWMs $(M_{1,0,2}, M_{1,2,0})$ are slightly more efficient than the lower order PWMs $(M_{1,0,1}, M_{1,1,0})$. This can be confirmed by calculating the relative efficiency of the lower order to the higher order PWMs. Across all the sample sizes, this RE at the pairs $(\alpha = 0.5 \ \& \ \beta = 0.6)$ is in the range from 0.94 to 0.95 for both estimated parameters. While at the pairs $(\alpha = 0.5 \ \& \ \beta = 1.3)$, it is in the range from 0.998 to1. Furthermore, the RE at the pair $(\alpha = 1.1 \ \& \ \beta = 0.6)$ is around 0.8. Additionally, the ER at the pairs $(\alpha = 1.1 \ \& \ \beta = 1.6)$ is in the range from 0.94 to 0.95. So they are more or less close to each other if taking into account the computational time reported in seconds per replicate. At sample size 20, the time taken by the lower PWMs are slightly less than that taken by the higher order PWMs to accomplish the estimation. This is also true for sample size 50 except at the pair $(\alpha = 0.5 \ \& \ \beta = 1.3)$ where the time taken by the lower order PWMs is 0.1289 seconds per replicate while this time is 0.1264 seconds per replicate for the higher order PWMs to complete the task of estimation. At sample size 100, the lower order PWMs take less time than the time of the higher order PWMs. At sample size 500, and at pairs $(\alpha = 0.5 \ \& \ \beta = 0.6)$ and $(\alpha = 0.5 \ \& \ \beta = 1.3)$, the lower order PWMs are faster than the higher order PWMs , in contrast to the pair $(\alpha = 1.1 \ \& \ \beta = 0.6)$ and the pair $(\alpha = 1.1 \ \& \ \beta = 1.6)$, where the higher order PWMs are faster than lower order PWMs. The time taken by both the lower and higher PWMs is linearly increasing across different sample sizes as n increases. The time taken by the MLE is less than that by PWMs of both order. The time taken by MOMs is the least among other methods. The time taken by MLE and MOMs does not vary markedly across sample sizes as seen in PWMs.

Table (8): analysis of the sampling distribution of the estimated $(\hat{\alpha})$ & $(\hat{\beta})$ of MOMs

| n | | MOM $\alpha = 0.5$ & $\beta = 0.6$ | | MOM $\alpha = 0.5$ & $\beta = 1.3$ | | MOM $\alpha = 1.1$ & $\beta = 0.6$ | | MOM $\alpha = 1.1$ & $\beta = 1.6$ | |
|---|---|---|---|---|---|---|---|---|---|
| 500 | Var_cov | 4.565e-4 | -2.535e-4 | 1.385e-5 | -2.408e-5 | 0.0075 | 0.0038 | 0.0021 | -0.0071 |
| | | -2.535e-4 | 1.515e-4 | -2.408e-5 | 4.615e-5 | 0.0038 | 0.0105 | -0.0071 | 0.0555 |
| | Corr | -0.964(p=0) | | -0.3011(p=2.09e-22) | | 0.4323(p=4.0974e-32) | | -0.662(p=4.0381e-127) | |
| | Cond num. Eigen.val | Cond_num=73.6573 Eigen_val=8.1441e-06, 5.9988e-4 | | Cond_num=3.5871 Eigen_val=4.0254e-5, 1.4440e-4 | | Cond_num=2.6805 Eigen_val=0.0049, 0.0131 | | Cond_num=49.5788 Eigen_val=0.0011, 0.0564 | |
| 100 | Var_cov | 0.0023 | -0.0013 | 7.66e-4 | 4.66e-5 | 0.1857 | 0.0101 | 0.0293 | -0.0396 |
| | | -0.0013 | 7.502e-4 | 4.66e-5 | 0.0011 | 0.0101 | 0.0427 | -0.0396 | 0.1557 |
| | Corr | -0.9846(p=0) | | 0.0511(p=0.1065) | | 0.1138(p=0.0214) | | -0.587(p=2.2243e-93) | |
| | Cond num. Eigen.val | Cond. num=176.1722 Eigen_val=1.7454e-5, 0.0031 | | Cond. num=1.4435 Eigen_val=7.5939e-4 , 0.0011 | | Cond. num=4.4403 Eigen_val=0.042, 0.1864 | | Cond. num=9.3439 Eigen_val=0.0179, 0.1671 | |
| 50 | Var_cov | 0.0055 | -0.0030 | 0.0017 | 3.235e-4 | | | | |
| | | -0.0030 | 0.0017 | 3.235e-4 | 0.003 | | | | |
| | Corr | -0.9600(p=0) | | 0.1444(p=4.525e-6) | | | | | |
| | Cond num. Eigen.val | Cond. num=67.5162 Eigen_val=1.0526e-4, 0.0071 | | Cond. num=1.8992 Eigen_val=0.0016, 0.0031 | | | | | |
| 20 | Var_cov | 0.0124 | -0.0041 | 0.0042 | 5.131e-4 | 0.1345 | 0.0016 | | |
| | | -0.0041 | 0.0066 | 5.131e-4 | 0.0052 | 0.0016 | 0.0781 | | |
| | Corr | -0.4506(p=1.619e-50) | | 0.1093(p=-5.333e-4) | | 0.0153(p=0.9255) | | | |
| | Cond num. Eigen.val | Cond.num=3.2243, Eigen.val=0.0045, 0.0145 | | Cond.num=1.3664 Eigen.val=0.0040, 0.0055 | | Cond_num=1.724 Eigen_val=0.0781 , 0.1246 | | | |

Non-significant correlation is colored.

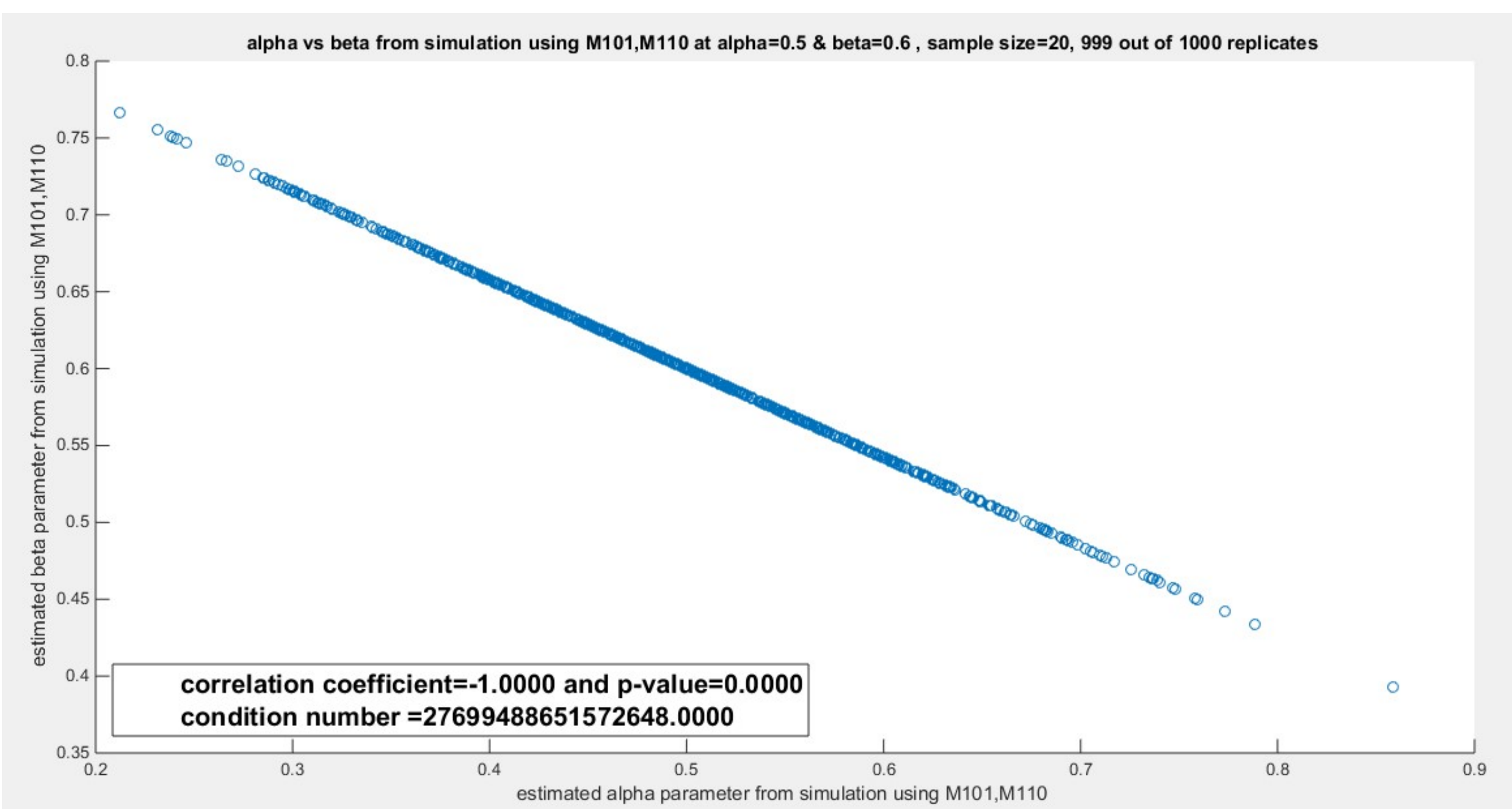

Fig. 1 shows the perfect linear relationship between estimated parameters from the simulation study at n=20 using lower order PWMs.

Table (9): The variance decay of the estimated $(\hat{\alpha})$ & $(\hat{\beta})$ in different simulations

| | PWM: M101,M110 $\alpha = 0.5$ & $\beta = 0.6$ | PWM: M101,M110 $\alpha = 0.5$ & $\beta = 1.3$ | PWM: M101,M110 $\alpha = 1.1$ & $\beta = 0.6$ | PWM: M101,M110 $\alpha = 1.1$ & $\beta = 1.6$ |
|---|---|---|---|---|
| **Slope For $\alpha$** | $-0.9846$ | $-1.0206$ | $-1.0283$ | $-1.0048$ |
| **Slope For $\beta$** | $-0.9882$ | $-1.0206$ | $-1.0277$ | $-0.9933$ |
| | **PWM: M102,M120** $\alpha = 0.5$ & $\beta = 0.6$ | **PWM: M102,M120** $\alpha = 0.5$ & $\beta = 1.3$ | **PWM: M102,M120** $\alpha = 1.1$ & $\beta = 0.6$ | **PWM: M102,M120** $\alpha = 1.1$ & $\beta = 1.6$ |
| **Slope For $\alpha$** | $-0.9876$ | $-1.006$ | $-1.0204$ | $-0.9932$ |
| **Slope For $\beta$** | $-0.9820$ | $-1.0061$ | $-1.0244$ | $-0.9990$ |
| | **MLE** $\alpha = 0.5$ & $\beta = 0.6$ | **MLE** $\alpha = 0.5$ & $\beta = 1.3$ | **MLE** $\alpha = 1.1$ & $\beta = 0.6$ | **MLE** $\alpha = 1.1$ & $\beta = 1.6$ |
| **Slope For α** | $-0.9429$ | $-1.0035$ | $-0.9755$ | $-09830$ |
| **Slope For $\beta$** | $-0.9607$ | $-0.7299$ | $-0.6679$ | $-0.1561$ |
| | **MOM** $\alpha = 0.5$ & $\beta = 0.6$ | **MOM** $\alpha = 0.5$ & $\beta = 1.3$ | **MOM** $\alpha = 1.1$ & $\beta = 0.6$ | **MOM** $\alpha = 1.1$ & $\beta = 1.6$ |
| **Slope For α** | $-1.0047$ | $-1.0627$ | $-1.9940$ | $-1.6376$ |
| **Slope For β** | $-0.9940$ | $-1.9704$ | $-0.8719$ | $-0.6409$ |

The slope that is not approximately equal to (-1) is colored

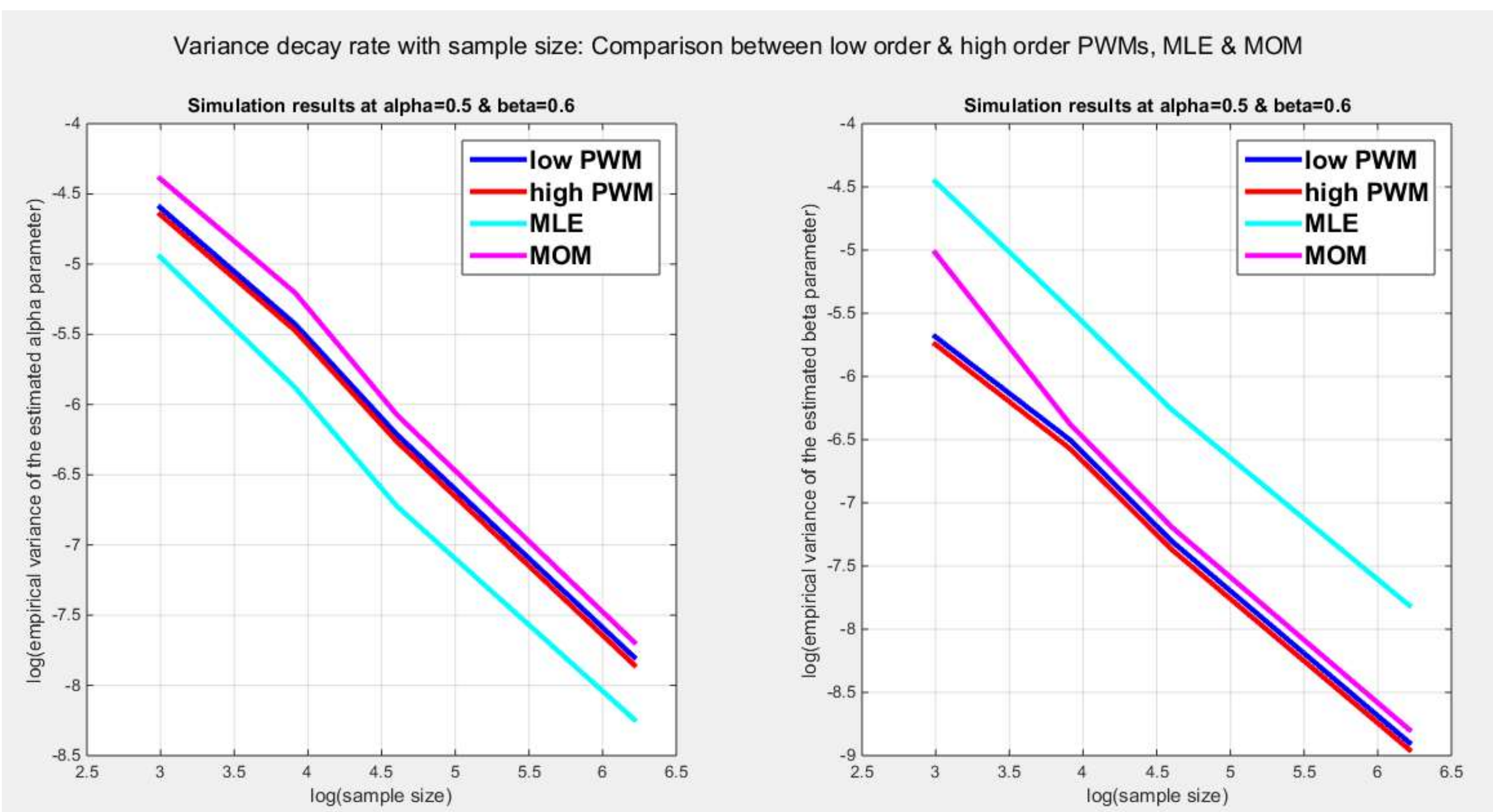

Fig. 2 shows the variance decay emphasizing asymptotic consistency , the slope in a log-log plot of the empirical variance and sample size is approximately -1 as asymptotic consistency of $(\hat{\theta}_n)$ entails.

Table (10) : the relative efficiency of different methods at sample size 20

| | | MLE vs. others | MOM vs. others | M101, M110 vs. others | M102, M120 vs. others |
|---|---|---|---|---|---|
| n=20 | $\alpha = 0.5$ | MLE is more efficient than MOM , M102 & M101 | MOM is less efficient than MLE , M101 & M102 | M101 is more efficient than MOM but is less efficient than MLE & M102 | M102 is less efficient than MLE but is more efficient than MOM & M101 |
| | $\beta = 0.6$ | MLE is less efficient than MOM, M101 &M102 | MOM is more efficient than MLE but less efficient than M101& M102 | M101 is more efficient than MLE, MOM but less efficient than M102 | M102 is more efficient than MLE , MOM & M101 |
| | time | 0.0307s per replicate Total time=30.7250s | 0.0220s per replicate Total time=22.0016 | 0.0423s per replicate Total time=42.2866 s | 0.0430s per replicate Total time=42.9643s |
| n=20 | $\alpha = 0.5$ | MLE is more efficient than MOM , M102 & M101 | MOM is less efficient than MLE, M101 & M102 | M101 is less efficient than MLE but is more efficient than MOM and is likely as M102 | M102 is less efficient than MLE but more efficient than MOM & is likely as M101 |
| | $\beta = 1.3$ | MLE is more efficient than MOM but is less efficient than both M101 & M102 | MOM is less efficient than MLE, M101 & M102 | M101 is more efficient than MLE & MOM but is less efficient than M102 | M102 is more efficient than MLE, MOM & M101 |
| | time | 0.0307s per replicate Total time=30.6922s | 0.0224 s per replicate Total time=22.4416s | 0.0430s per replicate Total time=43.0307s | 0.0442s per replicate Total time=44.2048s |
| n=20 | $\alpha = 1.1$ | MLE is more efficient than MOM & M101 & is less efficient than M102 | MOM is less efficient than MLE, M101 & M102 | M101 is more efficient than MOM but is less efficient than MLE & M102 | M102 is more efficient than MLE, MOM & M101 |
| | $\beta = 0.6$ | MLE is more efficient than MOM but is less efficient than M101 & M102 | MOM is less efficient than MLE, M101 & M102 | M101 is more efficient than MLE & MOM but is less efficient than M102 | M102 is more efficient than MLE, MOM & M101 |
| | time | 0.0305s per replicate Total time=30.4843s | 0.0071s per replicate Total time=7.1177s | 0.0443s per replicate Total time=44.3011s | 0.0443 s per replicate Total time=44.2731s |
| n=20 | $\alpha = 1.1$ | MLE is more efficient than M102 & M101 | | M101 is less efficient than MLE & M102 | M102 is less efficient than MLE but is more efficient than M101. |
| | $\beta = 1.6$ | MLE is markedly less efficient than M101 & M102 | | M101 is more efficient than MLE but less efficient than M102 | M102 is more efficient than MLE & M101 |
| | time | 0.0329s per replicate Total time=32.8729s | | 0.0450s per replicate Total time=45.0361s | 0.0454s per replicate Total time=45.4295s |

Table (11) : the relative efficiency of different methods at sample size 50

| | | MLE vs. others | MOM vs. others | M101, M110 vs. others | M102, M120 vs. others |
|---|---|---|---|---|---|
| n=50 | $\alpha = 0.5$ | **MLE is more efficient than MOM , M102 & M101** | **MOM is less efficient than MLE , M101 & M102** | **M101 is more efficient than MOM but is less efficient than MLE & M102** | **M102 is more efficient than MOM & M101 but is less efficient than MLE** |
| | $\beta = 0.6$ | **MLE is less efficient than MOM,M101 &M102** | **MOM is more efficient than MLE but is less efficient than M101 & M102** | **M101 is more efficient than MLE, MOM but is less efficient than M102** | **M102 is more efficient than MOM , M101 & MLE** |
| | time | **0.0342s per replicate Total time=34.1581s** | **0.0248s per replicate Total time=24.8382s** | **0.1227s per replicate Total time=122.7369s** | **0.1285 s per replicate Total time=128.5091s** |
| n=50 | $\alpha = 0.5$ | **MLE is more efficient than MOM , M102 & M101** | **MOM is less efficient than MLE, M101 & M102** | **M101 is more efficient than MOM but is less efficient than MLE and is likely as M102** | **M102 is more efficient than MOM but is less efficient than MLE& is likely as M101** |
| | $\beta = 1.3$ | **MLE is more efficient than MOM but less efficient than both M101 & M102** | **MOM is less efficient than MLE, M101 & M102** | **M101 is more efficient than MLE, MOM & M102** | **M102 is more efficient than MLE and MOM but is less efficient than M101** |
| | time | **0.0342s per replicate Total time=34.2283s** | **0.0255s per replicate Total time=25.5316s** | **0.1289s per replicate Total time=128.9309s** | **0.1264s per replicate Total time=126.4414s** |
| n=50 | $\alpha = 1.1$ | **MLE is slightly more efficient than M101 but is less efficient than M102** | | **M101 is less efficient than MLE, & M102** | **M102 is more efficient than MLE & M101** |
| | $\beta = 0.6$ | **MLE is less efficient than M101 & M102** | | **M101 is more efficient than MLE but is less efficient than M102** | **M102 is more efficient than MLE & M101** |
| | time | **0.0354s per replicate Total time=35.353s** | | **0.1266s per replicate Total time=126.6163s** | **0.1281s per replicate Total time=128.0973s** |
| n=50 | $\alpha = 1.1$ | **MLE is more efficient than M102 & M101** | | **M101 is less efficient than MLE & M102** | **M102 is less efficient than MLE but is more efficient than M101** |
| | $\beta = 1.6$ | **MLE is markedly less efficient than M101 & M102** | | **M101 is more efficient than MLE but is less efficient than M102** | **M102 is more efficient than MLE & M101** |
| | time | **0.0340s per replicate Total time=34.0159s** | | **0.1276s per replicate Total time=127.5708s** | **0.1344s per replicate Total time=134.377s** |

Tables (10-13) almost depict that the descending order of the efficient methods from the most efficient to the least efficient in estimating ($\alpha$) parameter across different sample sizes and across different pairs is MLE, $(M_{1,0,2}, M_{1,2,0})$, $(M_{1,0,1}, M_{1,1,0})$ then MOM. The exception for this is the pair ($\alpha = 1.1$ & $\beta = 0.6$ ), which shows the descending order is $(M_{1,0,2}, M_{1,2,0})$, MLE, $(M_{1,0,1}, M_{1,1,0})$ and finally MOM across different sample sizes excluding n=500 & for this pair ($\alpha = 1.1$ & $\beta = 0.6$ ) the order is $(M_{1,0,2}, M_{1,2,0})$, $(M_{1,0,1}, M_{1,1,0})$ , MLE then MOM. While the descending order of the efficient methods in estimating ($\beta$) across different sample sizes and different pairs is $(M_{1,0,2}, M_{1,2,0})$, $(M_{1,0,1}, M_{1,1,0})$, MLE then MOM. The exception for this is the pair ($\alpha = 0.5$ & $\beta = 0.6$ ), which shows

the descending order is $(M_{1,0,2}, M_{1,2,0})$, $(M_{1,0,1}, M_{1,1,0})$, MOM and lastly MLE across different sample sizes. This last order is also true when n=500 and the pair is ($\alpha = 0.5$ & $\beta = 1.3$ ).

Table (12) : the relative efficiency of different methods at sample size 100

| | | MLE vs. others | MOM vs. others | M101, M110 vs. others | M102, M120 vs. others |
|---|---|---|---|---|---|
| n=100 | $\alpha = 0.5$ | MLE is more efficient than MOM , M102 & M101 | MOM is less efficient than MLE , M101 & M102 | M101 is more efficient than MOM but is less efficient than MLE & M102 | M102 is more efficient than MOM & M101 but is less efficient than MLE |
| | $\beta = 0.6$ | MLE is less efficient than MOM,M101 & M102 | MOM is more efficient than MLE but is less efficient than M102 & M101 | M101 is more efficient than MLE & MOM but is less efficient than M102 | M102 is more efficient than MLE , MOM & M101 |
| | time | 0.0377s per replicate Total time=37.7212s | 0.0262s per replicate Total time=26.2368s | 0.3503s per replicate Total time=350.3439s | 0.3536s per replicate Total time=353.56s |
| n=100 | $\alpha = 0.5$ | MLE is more efficient than MOM , M102 & M101 | MOM is less efficient than MLE, M101 & M102 | M101 is more efficient than MOM but is less efficient than MLE & M102 | M102 is more efficient than MOM & M101 but is less efficient MLE |
| | $\beta = 1.3$ | ME is more efficient than MOM but less efficient than both M101 & M102 | MOM is less efficient than MLE, M101 & M102 | M101 is more efficient than MLE, MOM but is slightly less efficient than M102 | M102 is more efficient than MLE, MOM & M101 |
| | time | 0.0364s per replicate Total time=36.4423s | 0.0270s per replicate Total time=27.0006s | 0.3519s per replicate Total time=351.9213s | 0.3612s per replicate Total time=361.189s |
| n=100 | $\alpha = 1.1$ | MLE is more efficient than MOM & M101 but is less efficient than M102 | MOM is less efficient than MLE,M101 & M102 | M101 is more efficient than MOM but is less efficient than MLE & M102 | M102 is more efficient than MOM, MLE & M101 |
| | $\beta = 0.6$ | MLE is more efficient than MOM but is less efficient than M101, M102 | MOM is less efficient than MLE,M101 & M102 | M101 is more efficient than MOM, MLE but is less efficient than M102 | M102 is more efficient than MOM, MLE & M101 |
| | time | 0.0356s per replicate Total time=35.5829s | 0.0126s per replicate Total time=12.6141s | 0.3094s per replicate Total time=309.4498s | 0.3746s per replicate Total time=374.639s |
| n=100 | $\alpha = 1.1$ | MLE is more efficient than M102 , MOM & M101 | MOM is less efficient than MLE, M101, & M102. | M101 is more efficient than MOM but is less efficient than MLE & M102 | M102 is more efficient than MOM & M101 but is less efficient than MLE |
| | $\beta = 1.6$ | MLE is more efficient than MOM but is less efficient than M101 & M102 | MOM is less efficient than MLE, M101, & M102. | M101 is more efficient than MOM, MLE but is less efficient than M102 | M102 is more efficient than MOM, MLE & M101 |
| | time | 0.0384s per replicate Total time=38.3641s | 0.0124s per replicate Total time=12.4405s | 0.3546s per replicate Total time=354.597s | 0.3627s per replicate Total time=362.7022s |

Table (13) : the relative efficiency of different methods at sample size 500

| | | MLE vs. others | MOM vs. others | M101, M110 vs. others | M102, M120 vs. others |
|---|---|---|---|---|---|
| n= 500 | $\alpha = 0.5$ | MLE is more efficient than MOM , M102 & M101 | MOM is less efficient than MLE , M101 & M102 | M101 is more efficient than MOM but is less efficient than MLE & M102 | M102 is more efficient than MOM & M101 but is less efficient than MLE |
| | $\beta = 0.6$ | MLE is less efficient than MOM,M101 &M102 | MOM is more efficient than MLE but is less efficient than M102 & M101 | M101 is more efficient than MLE, MOM but is less efficient than M102 | M102 is more efficient than MLE , MOM & M101 |
| | time | 0.0569s per replicate Total time=56.8912s | 0.0296s per replicate Total time=29.5591s | 4.4784s per replicate Total time=4.4784e+3s | 5.5616s per replicate Total time=5.5616e+3s |
| n= 500 | $\alpha = 0.5$ | MLE is more efficient than MOM , M102 & M101 | MOM is less efficient than MLE, M101 & M102 | M101 is more efficient than MOM but is less efficient than MLE & M102 | M102 is more efficient than MOM & M101 but is less efficient than MLE |
| | $\beta = 1.3$ | MLE is less efficient than MOM, M101 & M102 | MOM is more efficient than MLE but is less efficient than M101 & M102 | M101 is more efficient than MLE, MOM but is less efficient than M102 | M102 is more efficient than MLE, MOM & M101 |
| | time | 0.0608s per replicate Total time=60.8357 | 0.032s per replicate Total time=31.9762s | 4.2734s per replicate Total time=4.2734e+3s | 4.323s per replicate Total time=4.323e+3s |
| n= 500 | $\alpha = 1.1$ | MLE is less efficient than M102 & M101 but is more efficient than MOM | MOM is less efficient than MLE,M101 & M102 | M101 is more efficient than MOM, MLE but is less efficient than M102 | M102 is more efficient than MOM , MLE & M101 |
| | $\beta = 0.6$ | MLE is less efficient than M102 & M101 but is more efficient than MOM | MOM is less efficient than MLE, M101 & M102 | M101 is more efficient than MOM, MLE but is less efficient than M102 | M102 is more efficient than MOM , MLE & M101 |
| | time | 0.0542s per replicate Total time=54.1649s | 0.0404s per replicate Total time=27.245s | 4.486s per replicate Total time=4.486+3s | 4.267s per replicate Total time=4.267e+3s |
| n= 500 | $\alpha = 1.1$ | MLE is more efficient than M102 , MOM & M101 | MOM is less efficient than MLE, M101, & M102. | M101 is more efficient than MOM but is less efficient than MLE & M102 | M102 is more efficient than MOM & M101 but is less efficient than MLE |
| | $\beta = 1.6$ | MLE is more efficient than MOM but is less efficient than M101 & M102 | MOM is less efficient than MLE, M101, & M102. | M101 is more efficient than MOM, MLE but is less efficient than M102 | M102 is more efficient than MOM, MLE &M101 |
| | time | 0.0588s per replicate Total time=58.7812s | 0.0355s per replicate Total time=35.5128s | 4.519s per replicate Total time=4.519e+3s | 4.353s per replicate Total time=4.353e+3s |

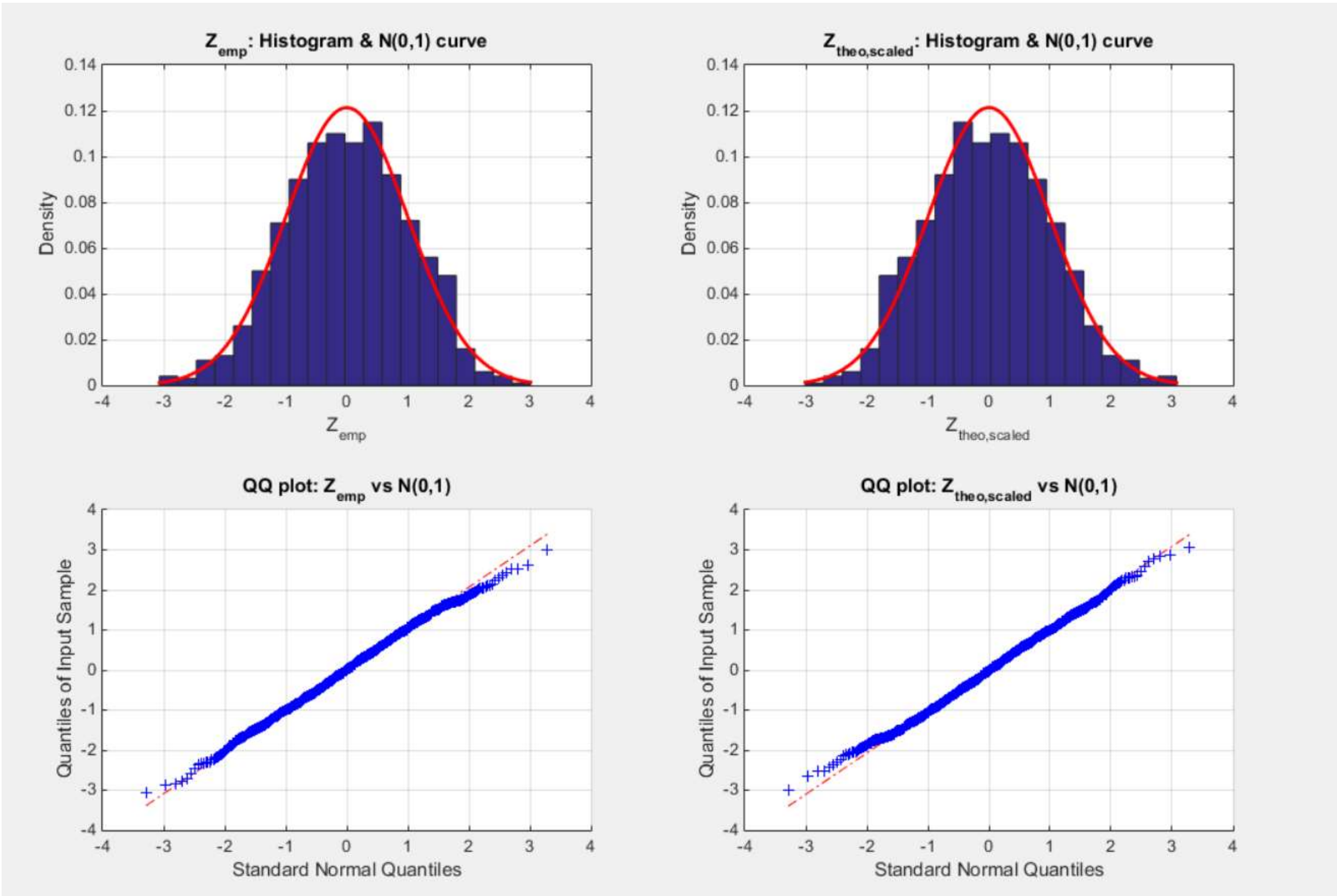

Fig. 3 shows the histogram, fitted standard normal curve and QQ plot for the error derived from the difference between the function of the parameter at the estimated parameters and the true parameters; $g(\hat{\theta}) - g(\theta_0)$. The error standardized by the empirical covariance matrix obtained from the empirical distribution is a standard normal variable (left upper and lower panels). The same error standardized by the covariance matrix obtained from rescaling the theoretical covariance matrix obtained via applying the delta method on each replicate is also a standard normal variable. (right upper and lower panels). This is the results of the simulation using n=500 , N=1000 , $\alpha = 0.5$ and $\beta = 1.3$ and the lower order PWMs (M101,M110)

Figure (3) shows the validation of asymptotic normality of the delta method. The error obtained from the difference between $g(\hat{\theta})$ and the $g(\theta_0)$ where $\theta_0$ are the pair of parameter ($\alpha = 0.5$ & $\beta = 1.3$ ), the sample size is 500 and the function is the lower order PWMs , is standardized by the empirical covariance matrix . The same error is also standardized by the theoretical covarinace matrix obtained via applying the delta method. However, both matrices the empirical and the theoretical matrices are nearly singular whose one of their eigenvalues are nearly zero. So both matrices are subjected to singular value decomposition (SVD) so as to project the error along the principle vector with high variability (the vector corresponding to the large non-zero singular value). And to match the empirical with the theoretical matrix, the last one is multiplied by a small scalar that is the ratio between the trace of the empirical covariance and the trace of the theoretical covariance. Both standardized error should be a standard normal variable as shown in figure (3). This procedure was repeated for the higher order PWMs with different sample sizes and different pairs of the parameters. The same results were obtained which confirms the asymptotic normality of the delta method. The regularity conditions essential to apply the delta method is that the function should be a continuously differentiable function which also holds for the PWMs of both orders.

# 5. Real data analysis

The OECD platform is available at https://stats.oecd.org/index.aspx?DataSetCode=BLI

In OECD platform, many indicators are recorded. The author transformed these indicators into unit ratios and a conducted distributional fit for those indicators. In this paper, the author discussed three indicators; the water quality, the educational attainment and the self-reported health. Table (14) shows the descriptive analysis of the indicators. Figure (4) shows the boxplot of each indicator.

Water quality: is the percentage of the people who report being satisfied with the quality of their water. This reflects the proportion of people who have access to clean water supply. The values of this indicator are 0.92, 0.92, 0.79, 0.90, 0.62, 0.82, 0.87, 0.89, 0.93, 0.86, 0.97, 0.78, 0.91, 0.67, 0.81, 0.97, 0.8, 0.77, 0.77, 0.87, 0.82, 0.83, 0.83, 0.85, 0.75, 0.91, 0.85, 0.98, 0.82, 0.89, 0.81, 0.93, 0.76, 0.97 ,0.96, 0.62, 0.82, 0.88, 0.7, 0.62, 0.72.

Educational attainment: is presented as the percentage of a given population who has completed a specific level of education mainly the tertiary education. The values of this indicator are 0.84, 0.86, 0.8, 0.92, 0.67, 0.59, 0.43, 0.94, 82, 0.91, 0.91, 0.81, 0.86, 0.76, 0.86, 0.76, 0.85, 0.88, 0.63, 0.89, 0.89, 0.94, 0.74, 0.42, 0.81, 0.81, 0.82, 0.93, 0.55, 0.92, 0.90, 0.63, 0.84, 0.89, 0.42, 0.82, 0.92, 0.57, 0.95, 0.48

Self-reported health is presented as the percentage of the population that rates their health as good or very good and this reflects access to health services, healthy living conditions, and better life-style choices (diets and exercises). The values of this indicator are 0.85, 0.71, 0.74, 0.89, 0.60, 0.8, 0.73, 0.62, 0.7, 0.57, 0.68, 0.67, 0.66, 0.79, 0.58, 0.77, 0.84, 0.74, 0.73, 0.37, 0.34, 0.47, 0.46, 0.72, 0.66, 0.75, 0.86, 0.75, 0.6, 0.5, 0.65, 0.67, 0.75, 0.76, 0.81, 0.67, 0.73, 0.88, 0.43

Table 14 : Summary of descriptive statistics for each indicator.

| | Water quality, n=41 | Educational attainment, n=40 | Self-reported Health, n=39 |
|---|---|---|---|
| Mean | 0.8332 | 0.7810 | 0.6795 |
| Standard deviation | 0.0972 | 0.1568 | 0.1358 |
| Skewness | -0.6059 | -1.1480 | -0.8026 |
| Kurtosis | 2.9144 | 3.2077 | 3.2807 |
| Min | 0.62 | 0.42 | 0.3400 |
| Max | 0.98 | 0.95 | 0.8900 |
| 25$^{th}$ percentile | 0.7775 | 0.705 | 0.6050 |
| Median | 0.83 | 0.83 | 0.7100 |
| 75$^{th}$ percentile | 0.91 | 0.895 | 0.7575 |

The variables exhibit left skewness with different degrees. Educational attainment shows significant left skewness then the self-reported health and lastly the water quality indicator. Water quality shows mesokurtic shape while the educational attainment and the self-reported health show slightly more than excess kurtosis (leptokurtic, the kurtosis coefficient is more than 3). So the data exhibit different shapes.

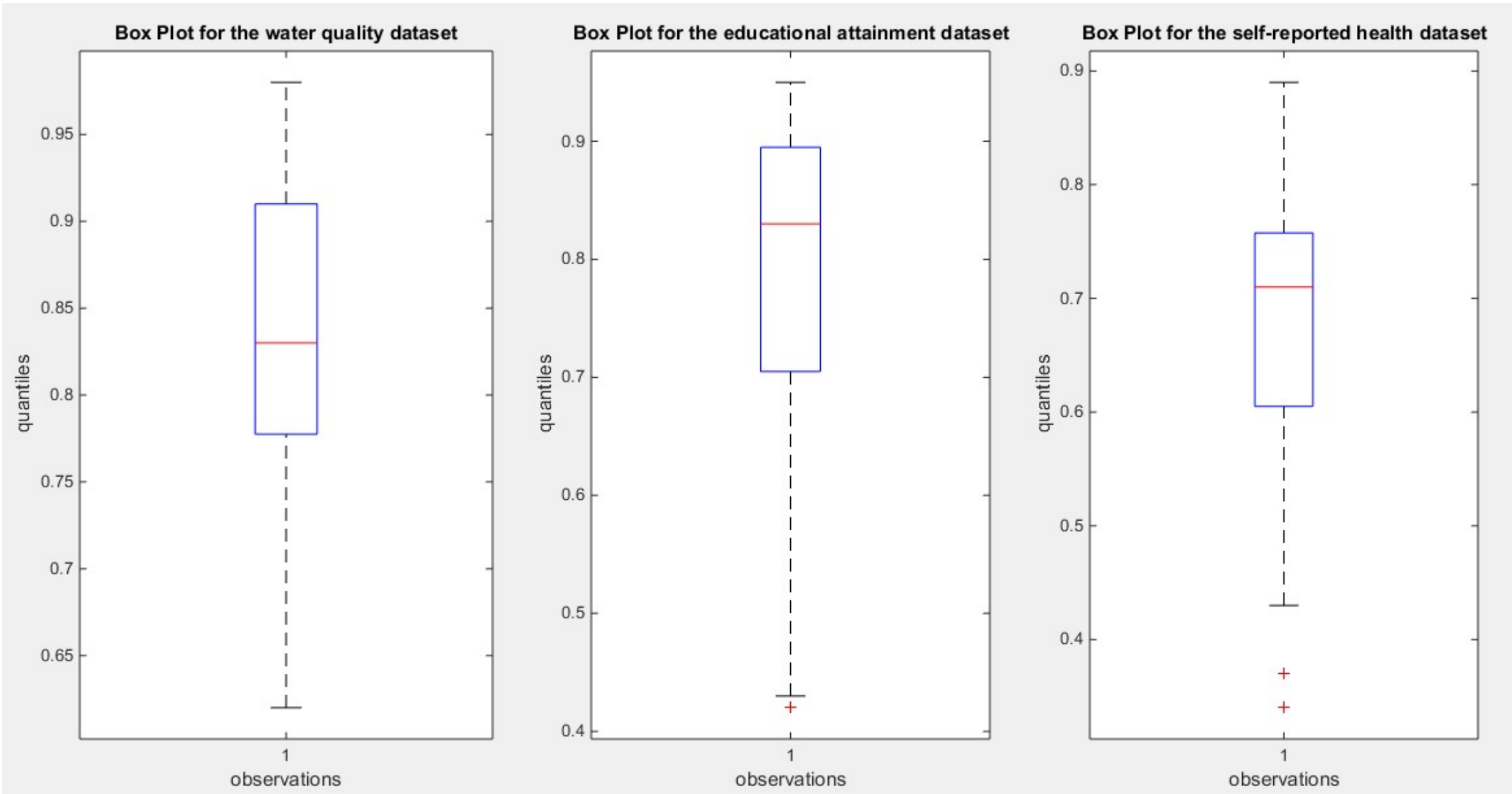

Fig. 4 shows the box plot of the indicators. The variables are negatively skewed.

The different unit distributions tested was Beta, Kumaraswamy, unit Lindley, Topp Leone and the new MBUW distribution. MLE method was applied using the Nelder Mead optimizer in MATLAB. For each distribution, the estimated parameters, variance, Log-Likelihood, AIC, CAIC, BIC, and HQIC are recorded. Also the KS-test, AD, CVM test are reported. Figures for the fitted CDFs and fitted PDFs are shown.

Beta Distribution: $f(y;\alpha,\beta) = \frac{\Gamma(\alpha+\beta)}{\Gamma(\alpha)\Gamma(\beta)} y^{\alpha-1}(1-y)^{\beta-1}, 0<y<1, \alpha>0, \ \beta>0$

Kumaraswamy Distribution: $f(y;\alpha,\beta) = \alpha\beta y^{\alpha-1}(1-y^{\alpha})^{\beta-1}, 0<y<1, \ \alpha>0, \ \beta>0$

Topp-Leone Distribution: $f(y;\theta) = \theta(2-2y)\,(2y-y^2)^{\theta-1}, 0<y<1, \ \theta>0$

Unit-Lindley: $f(y;\theta) = \frac{\theta^2}{1+\theta}(1-y)^3 \exp\left(\frac{-\theta}{1-y}\right), \ 0<y<1, \ \theta>0$

Table (15): Estimators and validation indices for the water quality dataset

| | Beta | | Kumaraswamy | | MBUW | | Topp-Leone | Unit-Lindley |
|---|---|---|---|---|---|---|---|---|
| **theta** | $\alpha = 10.8716$ | | $\alpha = 8.4271$ | | $\alpha = 0.3559$ | | **25.8989** | **0.2001** |
| | $\beta = 2.1667$ | | $\beta = 2.2817$ | | $\beta = 1.4308$ | | | |
| **Var-cov** | **8.1698** | **1.2164** | **1.9713** | **0.5680** | | | **16.3599** | **0.000495** |
| | **1.2164** | **0.2274** | **0.5680** | **0.2906** | | | | |
| **SE** | **2.858** | | **1.4040** | | | | **4.0447** | **0.0222** |
| | **0.4769** | | **0.5391** | | | | | |
| **AIC** | **-77.3397** | | **-76.8933** | | **-76.9952** | | **-78.5197** | **-61.8161** |
| **CAIC** | **-77.0239** | | **-76.5775** | | **-76.6794** | | **-78.4171** | **-61.7135** |
| **BIC** | **-73.9125** | | **-73.4662** | | **-73.5680** | | **-76.8061** | **-60.1025** |
| **HQIC** | **-76.0917** | | **-75.6453** | | **-75.7472** | | **-77.8957** | **-61.1921** |
| **LL** | **40.6698** | | **40.4467** | | **40.4976** | | **40.2599** | **31.9080** |
| **K-S Value** | **0.0929** | | **0.0996** | | **0.0991** | | **0.0569** | **0.1952** |
| **$H_0$** | **Fail to reject** | | **Fail to reject** | | **Fail to reject** | | **Fail to reject** | **Fail to reject** |
| **P-value** | **0.8387** | | **0.7741** | | **0.7789** | | **0.9288** | **0.0765** |
| **AD** | **0.3114** | | **0.3499** | | **0.3463** | | **0.5916** | **2.5329** |
| **CVM** | **0.0398** | | **0.0483** | | **0.0543** | | **0.0567** | **0.4423** |

Table 15 shows that all the unit distributions fit the water quality data well. The variance-covariance obtained after fitting the MBUW distribution is not identified, although the other statistical indices like AIC, CAIC, BIC, HQIC, LL are comparable to other distributions. Figures (5-6) show that the fitted CDF and the fitted PDF of the MBUW align with the fitted CDFs and fitted PDFs of both the Beta distribution and the Kumaraswamy distribution. But the problem is mainly the variance obtained from employing the MLE method utilizing Nelder Mead optimizer in MATLAB. Using the values of the estimated parameter as initial guesses to substitute in the LM algorithm and PWMs (lower and higher orders) gives the results shown in Table (18) and in Figures (11-12).

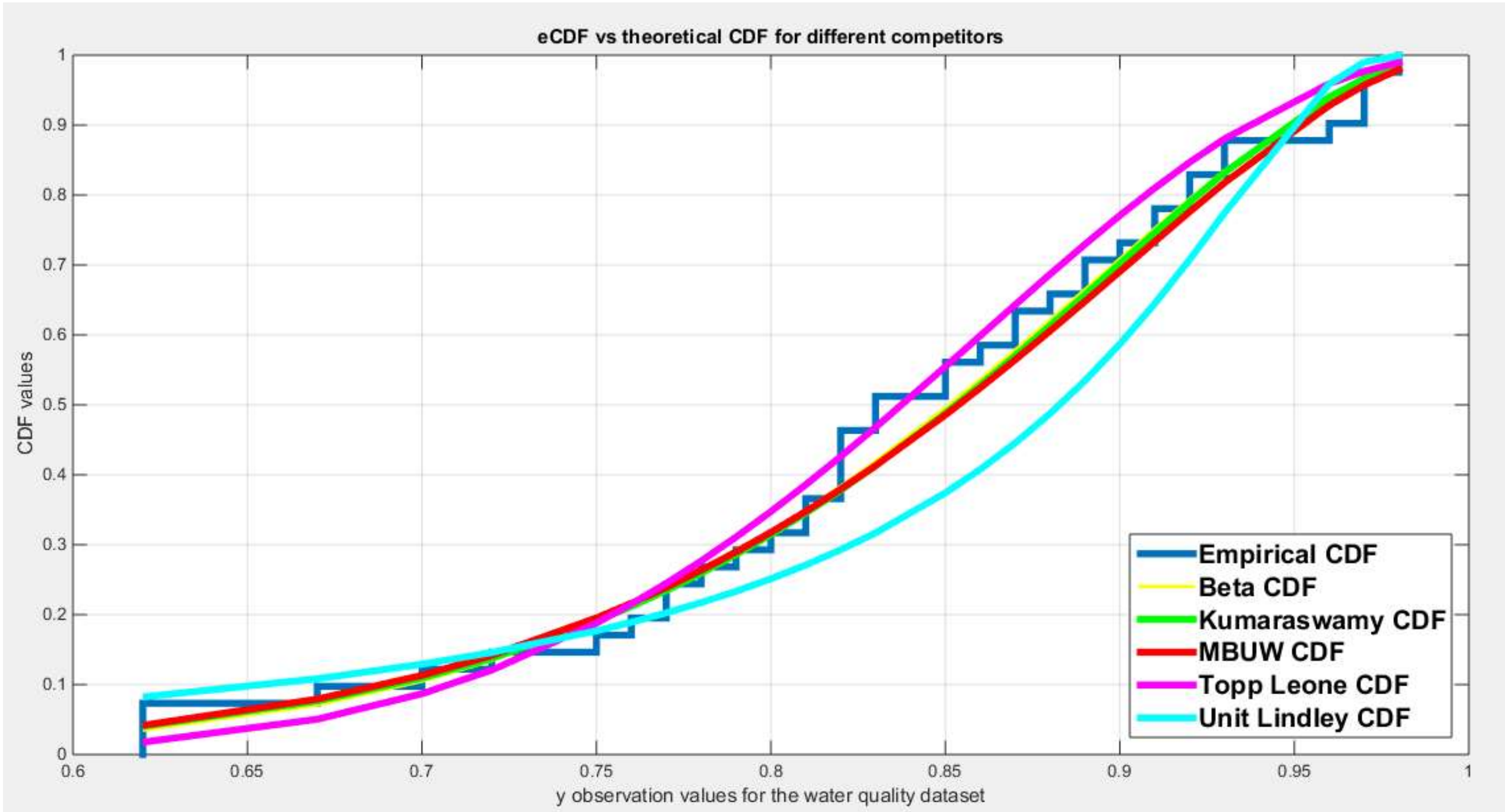

Fig. 5 shows the empirical CDF vs. fitted CDF for the different competitors of the water quality dataset

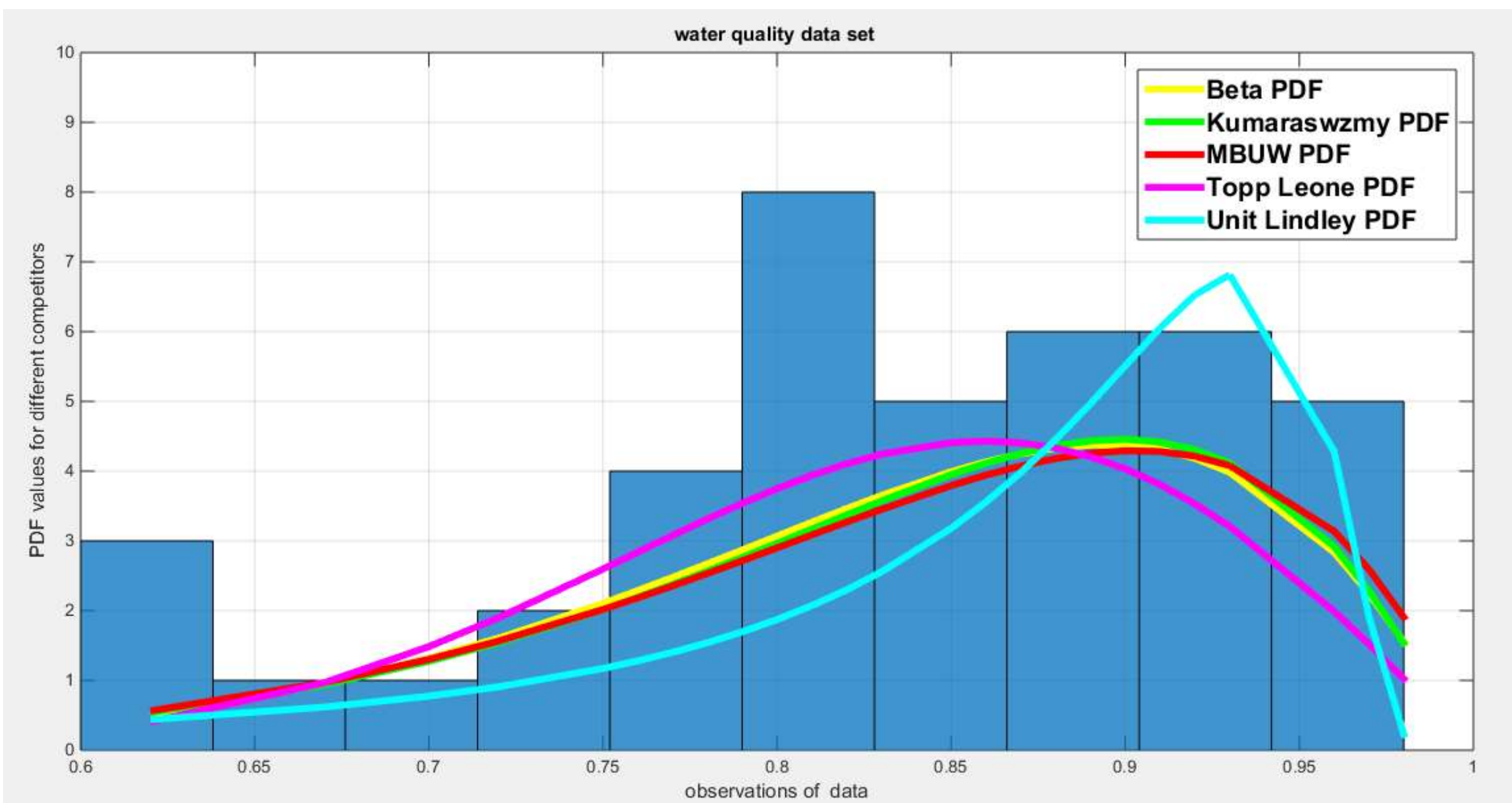

Fig. 6 shows the histogram and the fitted PDFs for the different competitors of the water quality dataset

Table (16): Estimators and validation indices for the educational attainment dataset

| | Beta | | Kumaraswamy | | MBUW | | Topp-Leone | Unit-Lindley |
|---|---|---|---|---|---|---|---|---|
| **theta** | $\alpha = 6.0373$ | | $\alpha = 5.4735$ | | $\alpha = 0.4333$ | | **12.3585** | **0.2891** |
| | $\beta = 1.7313$ | | $\beta = 1.9272$ | | $\beta = 1.3477$ | | | |
| **Var-cov** | **2.9116** | **0.9141** | **1.0268** | **0.3651** | | | **3.8183** | **0.0011** |
| | **0.9141** | **0.3410** | **0.3651** | **0.2227** | | | | |
| **SE** | **1.7063** | | **1.0133** | | | | **1.9540** | **0.0332** |
| | **0.5839** | | **0.4719** | | | | | |
| **AIC** | **-48.1413** | | **-48.8385** | | **-48.0574** | | **-40.9642** | **-57.3738** |
| **CAIC** | **-47.817** | | **-48.5142** | | **-47.7331** | | **-40.8589** | **-57.2685** |
| **BIC** | **-44.7636** | | **-45.4607** | | **-44.6796** | | **-39.2753** | **-55.6849** |
| **HQIC** | **-46.92** | | **-47.6172** | | **-46.8361** | | **-40.3535** | **-56.7631** |
| **LL** | **26.0707** | | **26.4193** | | **26.0287** | | **21.4821** | **29.6869** |
| **K-S Value** | **0.1481** | | **0.1434** | | **0.1577** | | **0.2598** | **0.0697** |
| **$H_0$** | **Fail to reject** | | **Fail to reject** | | **Fail to reject** | | **Reject** | **Fail to reject** |
| **P-value** | **0.1613** | | **0.1841** | | **0.1219** | | **0.0023** | **0.9442** |
| **AD** | **1.1861** | | **1.1575** | | **1.4386** | | **4.5709** | **0.2596** |
| **CVM** | **0.2068** | | **0.1985** | | **0.2546** | | **0.8485** | **0.0344** |

Table 16 shows that all the unit distributions fit the educational attainment data well except the Topp-Leone distribution. The same trouble of the variance-covariance obtained after fitting the MBUW distribution is shown, even though the other statistical indices like AIC, CAIC, BIC, HQIC, LL are more or less analogous to other fitting distributions. Figures (7-8) show that the fitted CDF and the fitted PDF of the MBUW is side by side with the fitted CDFs and fitted PDFs of both the Beta distribution and the Kumaraswamy distribution. But then again; the trouble is essentially the variance gained from deploying the MLE method consuming Nelder Mead optimizer in MATLAB. Treating the values of the estimated parameters as initial guesses to replace in the LM algorithm and PWMs (lower and higher orders) produces the outcomes shown in Table (19) and in Figures (13-14).

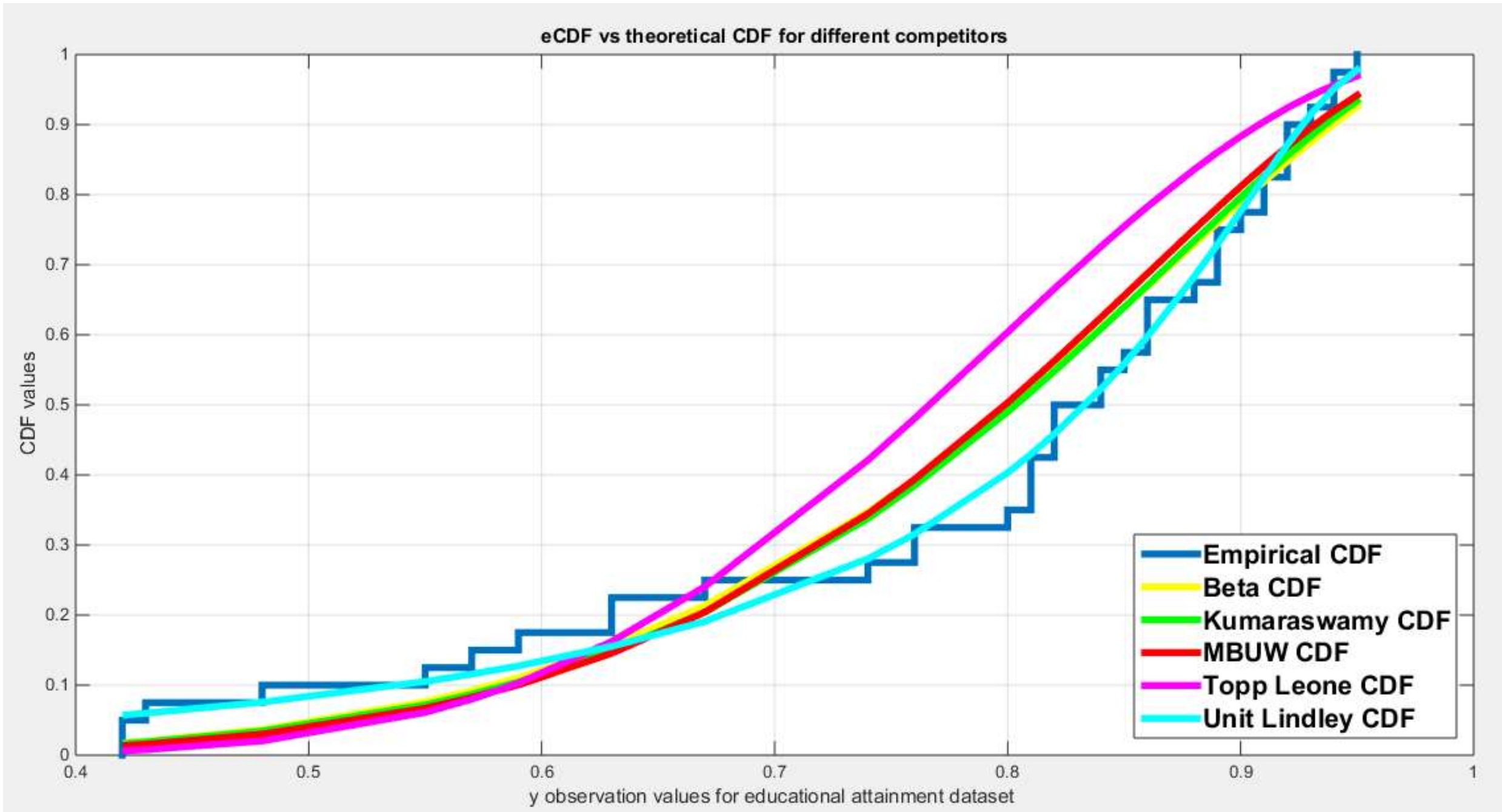

Fig. 7 shows empirical CDF vs. fitted CDFs for different competitors for the educational attainment dataset

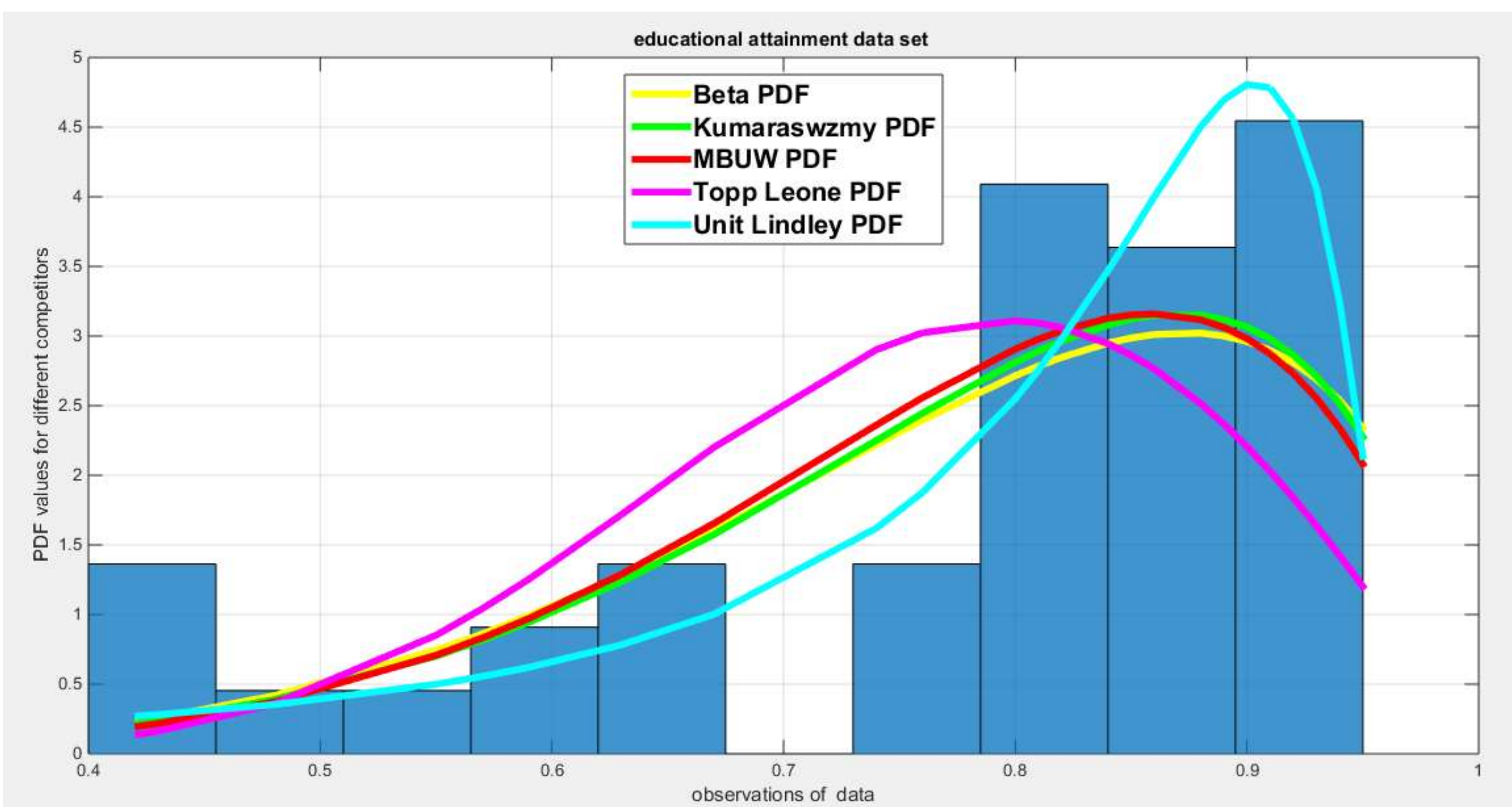

Fig. 8 shows the histogram and the fitted PDFs for different competitors for the educational attainment dataset

Table (17): Estimators and validation indices for the self-reported health dataset

| | Beta | | Kumaraswamy | | MBUW | | Topp-Leone | Unit-Lindley |
|---|---|---|---|---|---|---|---|---|
| **theta** | $\alpha = 8.0698$ | | $\alpha = 5.7073$ | | $\alpha = 0.5771$ | | **7.3278** | **0.5929** |
| | $\beta = 3.8357$ | | $\beta = 5.2250$ | | $\beta = 1.2824$ | | | |
| **Var** | **3.3978** | **1.6515** | **0.7936** | **1.2015** | | | **1.3768** | **0.0058** |
| | **1.6515** | **0.9425** | **1.2015** | **2.5192** | | | | |
| **SE** | **1.8433** | | **0.8908** | | | | **1.1734** | **0.0693** |
| | **0.9708** | | **1.5872** | | | | | |
| **AIC** | **-46.2684** | | **-47.3865** | | **-38.5536** | | **-44.4298** | **-40.09617** |
| **CAIC** | **-45.9351** | | **-47.0531** | | **-38.2202** | | **-44.3217** | **-39.9617** |
| **BIC** | **-42.9413** | | **-44.0594** | | **-35.2264** | | **-42.7662** | **-38.4062** |
| **HQIC** | **-45.0746** | | **-46.1927** | | **-37.3598** | | **-43.8329** | **-39.4729** |
| **LL** | **25.1342** | | **25.6932** | | **21.2768** | | **23.2149** | **21.0349** |
| **K-S Value** | **0.0867** | | **0.0684** | | **0.1625** | | **0.1234** | **0.1400** |
| **$H_0$** | **Fail to reject** | | **Fail to reject** | | **Fail to reject** | | **Fail to reject** | **Fail to reject** |
| **P-value** | **0.6674** | | **0.8577** | | **0.2284** | | **0.5512** | **0.2102** |
| **AD** | **0.4674** | | **0.3414** | | **1.4935** | | **0.8744** | **1.6862** |
| **CVM** | **0.0807** | | **0.0547** | | **0.2492** | | **0.1397** | **0.2886** |

Table 17 shows that all the unit distributions fit the self-reported health data well. The same dilemma of the variance-covariance obtained after fitting the MBUW distribution is displayed, albeit the other statistical indices like AIC, CAIC, BIC, HQIC, LL are slightly less than that of the Beta, Kumaraswamy, and the Topp –Leone distributions. Figures (9-10) show that the fitted CDF and the fitted PDF of the MBUW is lined up with the Beta and Kumaraswamy at the lower tail and is aligned with the Topp-Leone and the Unit Lindley distribution at the upper tail. Nevertheless, the misfortune is fundamentally the variance returned from implementing the MLE method executing Nelder Mead optimizer in MATLAB. Employing the values of the estimated parameters as initial guesses to insert in the LM algorithm and PWMs (lower and higher orders) generates the sequels shown in Table (20) and in Figures (15-16).

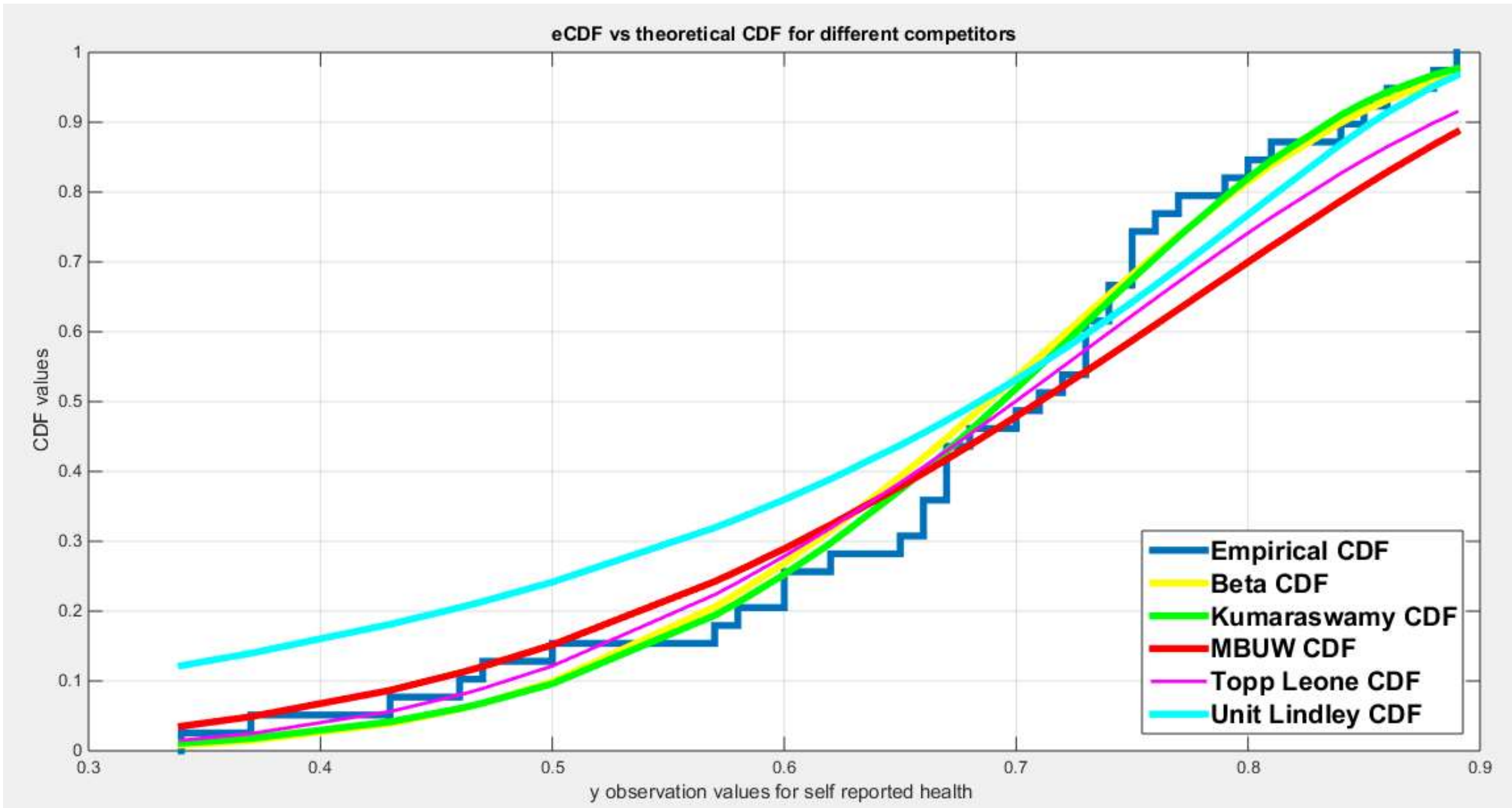

Fig. 9 shows the empirical CDF vs the fitted CDFs for different competitors for self-reported health

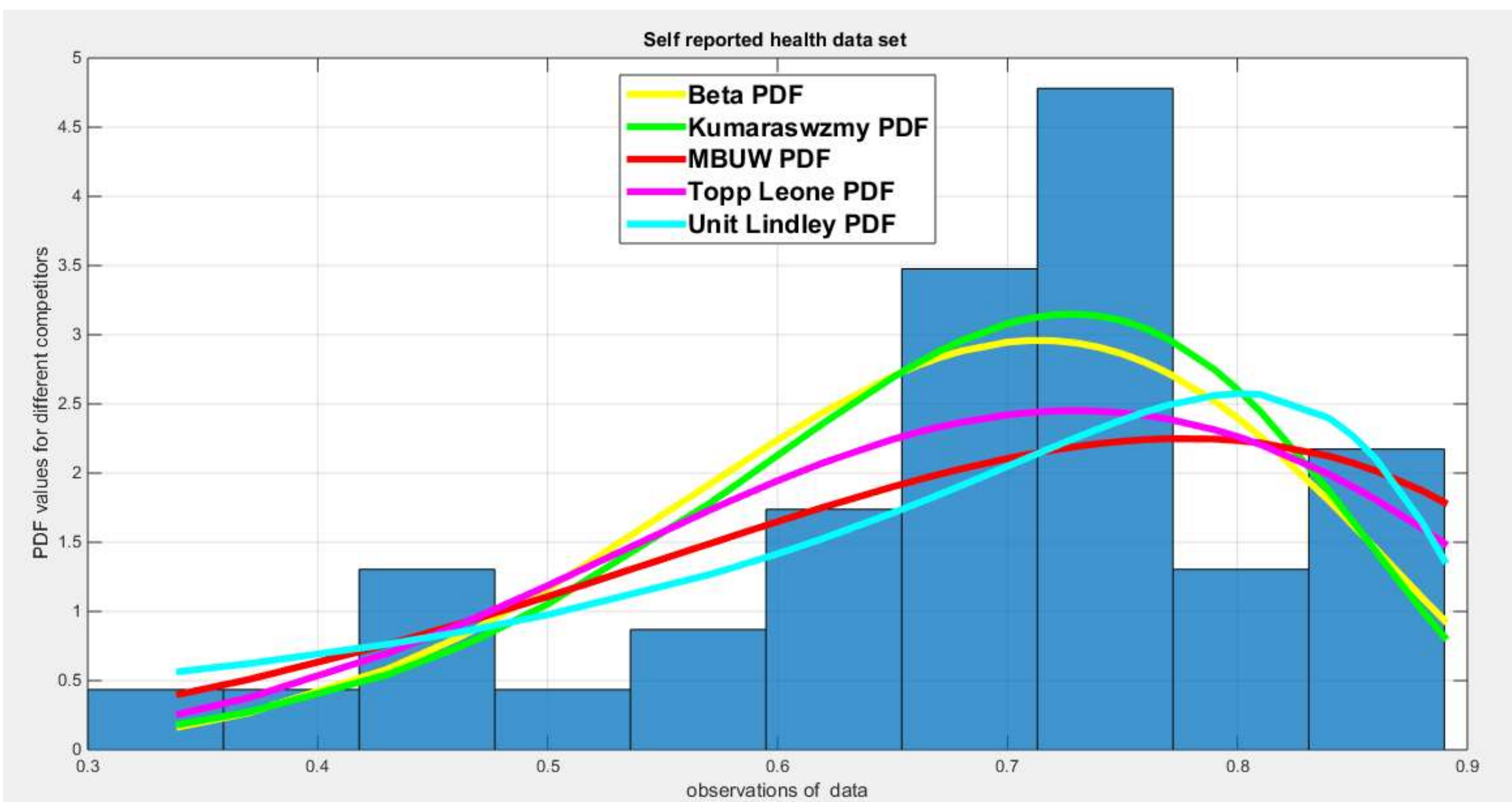

Fig. 10 shows the histogram vs the fitted PDFs for different competitors for self-reported health

Table (18): comparisons of the results of the PWMs method for parameter estimation of the water quality dataset using $M_{1,0,1}$ & $M_{1,1,0}$ and $M_{1,0,2}$ & $M_{1,2,0}$

| | | Using unbiased Sample estimator for $M_{1,0,1}$ *and* $M_{1,1,0}$ | | Using unbiased Sample estimator for $M_{1,0,2}$ *and* $M_{1,2,0}$ | |
|---|---|---|---|---|---|
| thetas | $\alpha$ | 0.3564 | | 0.3529 | |
| | $\beta$ | 1.4307 | | 1.4316 | |
| Var-cov matrix of parameter | | 10.7370 | 22.9418 | 627.8954 | 2.4064e+3 |
| | | 22.9418 | 94.1037 | 2.4064e+3 | 9.3821e+3 |
| | | Eigenvalues=4.8406, 100 | | Eigenvalue=10.0408 , 1e+4 | |
| AD | | 0.3418 ( p=0.8810) | | 0.3820 (p=0.865) | |
| CVM | | 0.0531 (p=0.8420) | | 0.0637 (p=0.785) | |
| KS & p-value | | 0.0980 (p=0.7901) | | 0.1061 (p=0.7051 | |
| H0 | | Fail to reject | | Fail to reject | |
| SSE | | 1.8169e-19 | | 1.7110e-19 | |
| $\hat{\alpha}_r$, unbiased estimator | | 0.3891 | | 0.2491 | |
| $\hat{\beta}_s$, unbiased estimator | | 0.4441 | | 0.3041 | |
| Sig.of $\alpha$ parameter | | 1.648(p=0.1072) | | 0.5950(p=0.5552) | |
| Sig.of $\beta$ parameter | | 3.3898(p=0.0016) | | 1.2365(p=0.2235) | |
| Variance of the function of the parmaeter, after the delta method application. Determinant and trace of this matrix | | 0.0181 | 0.0096 | 0.0214 | 0.0080 |
| | | 0.0096 | 0.0051 | 0.0080 | 0.0030 |
| | | Eigenvalues=0.0232, 8.6736e-19<br>Determinant=0<br>Trace=0.0232 | | Eigenvalues=0.0244 , -4.3368e-19<br>Determinant=0<br>Trace=0.0244 | |
| Var-cov between M101&M110 | | 8.8987e-4 | 0.0042 | | |
| | | 0.0042 | 0.0020 | | |
| Var-cov between M102 & M120 | | | | 9.5798e-4 | 0.0197 |
| | | | | 0.0197 | 0.002 |
| $var(\hat{\theta})$ and associated $\xi$ | | 0.0468 | 0.0361 | 0.3518 | 0.2717 |
| | | 0.0361 | 0.1781 | 0.2717 | 1.3403 |
| | | $\xi =$ 5.3356 to achieved condition number=5 | | $\xi =$ 0.7092 to achieved condition number=5 | |
| Jacobian matrix | | -0.3796 | 0.0977 | -0.2863 | 0.0735 |
| | | -1.4718 | 0.3786 | -0.1067 | 0.0274 |

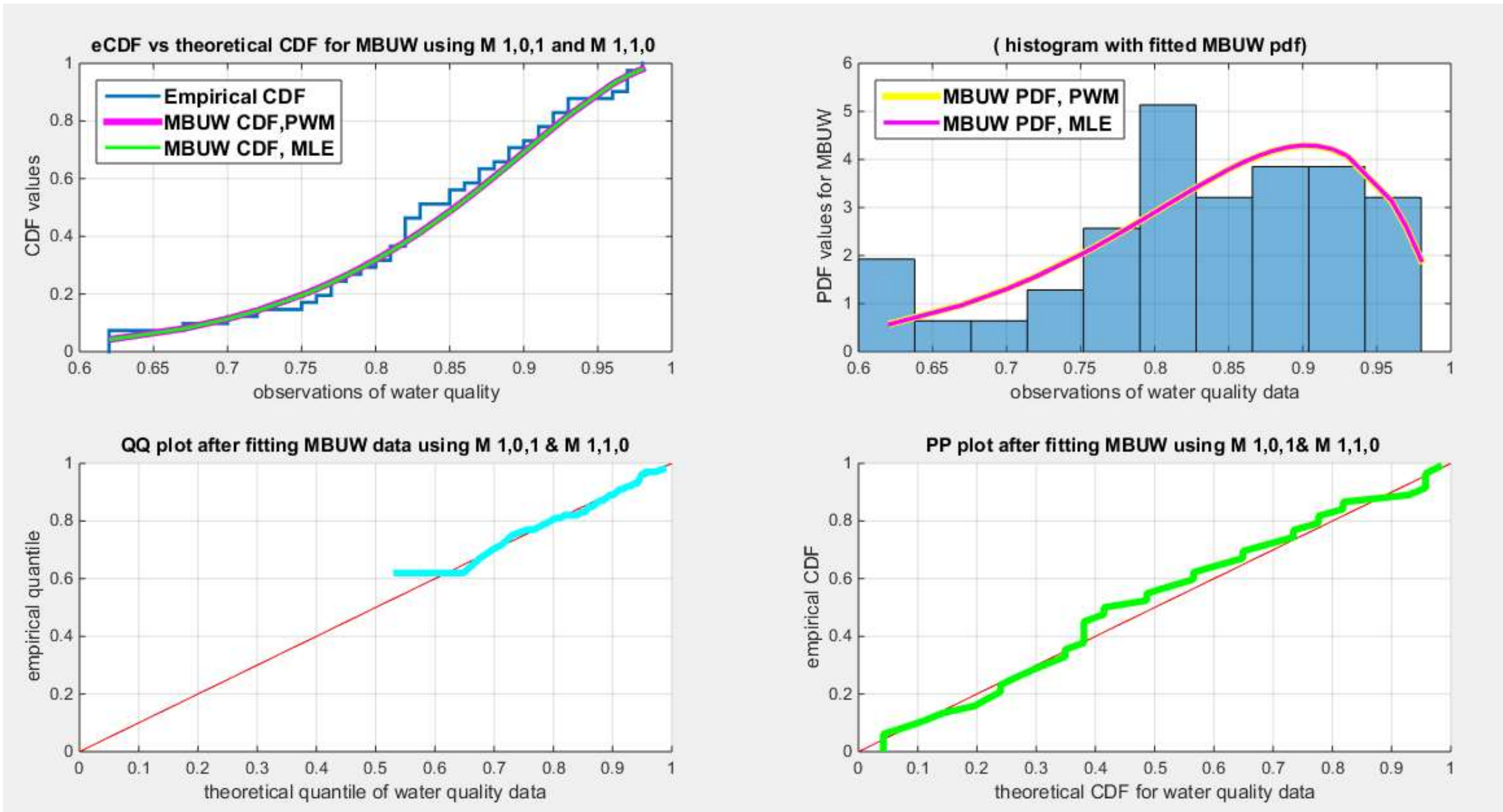

Fig. 11 shows the eCDF vs. fitted CDF for MLE & low order PWMs, M101, M110 (upper left subplot), histogram & fitted PDFs ( upper right subplot), QQ plot (lower left) and PP plot (lower right) for water quality dataset.

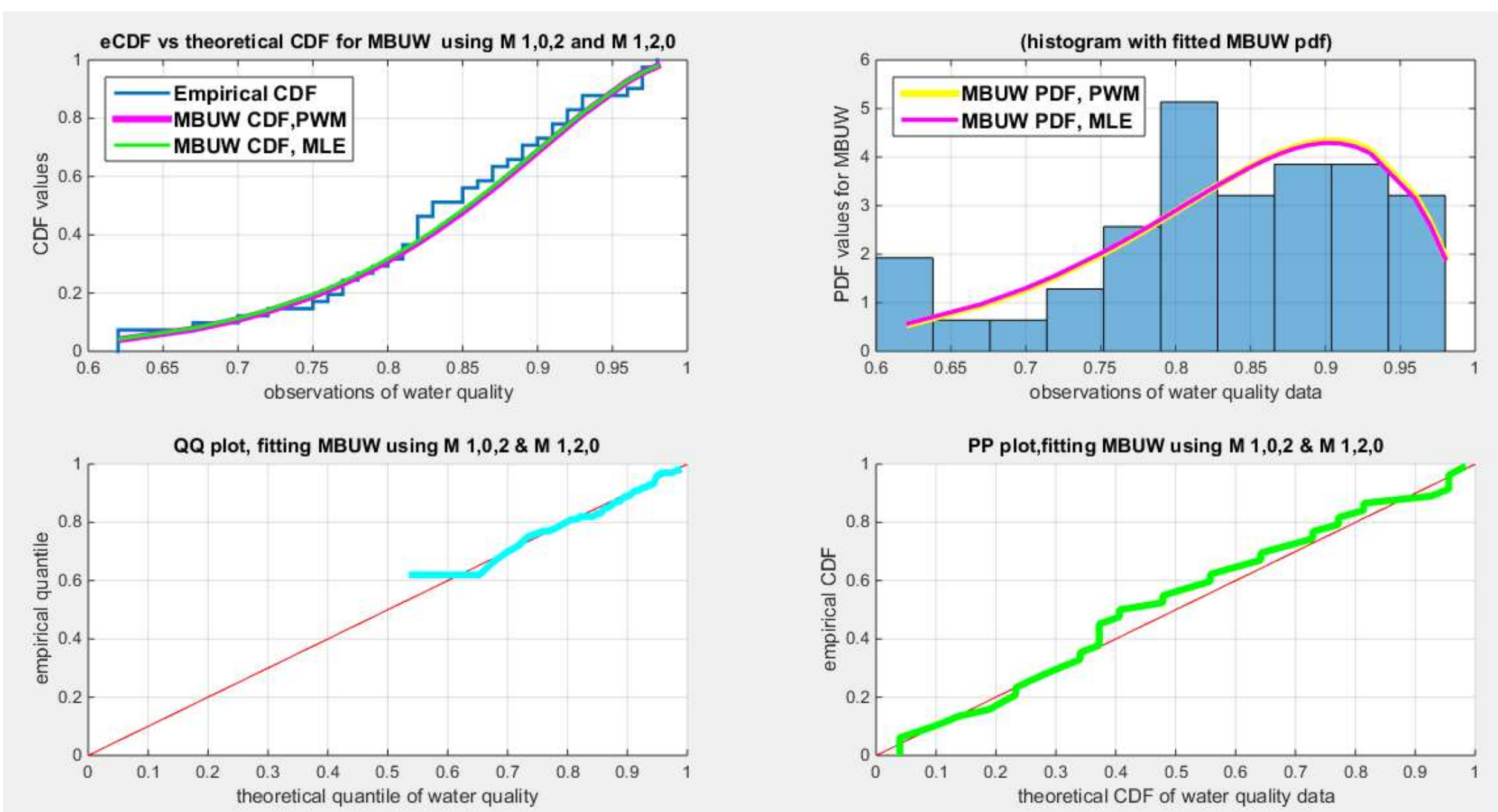

Fig. 12 shows the eCDF vs. fitted CDF for MLE & high order PWMs, M102, M120 (upper left subplot), histogram & fitted PDFs ( upper right subplot), QQ plot (lower left) and PP plot (lower right) for water quality dataset.

Table (19): comparisons of the results of the PWMs method for parameter estimation of the educational attainment dataset using $M_{1,0,1}$ & $M_{1,1,0}$ and $M_{1,0,2}$ & $M_{1,2,0}$

| | | Using unbiased Sample estimator for $M_{1,0,1}$ *and* $M_{1,1,0}$ | | Using unbiased Sample estimator for $M_{1,0,2}$ *and* $M_{1,2,0}$ | |
|---|---|---|---|---|---|
| thetas | $\alpha$ | 0.4310 | | 0.4462 | |
| | $\beta$ | 1.3483 | | 1.3442 | |
| Var-cov matrix of parameter | | 679.8845 | 2.5075e+3 | 679.7206 | 2.4966e+3 |
| | | 2.5075e+3 | 9.3254e+3 | 2.4966e+3 | 9.3313e+3 |
| | | Eigenvalues=5.2672, 10000 | | Eigenvalue=10.9762 , 1e+4 | |
| AD | | 1.4027 ( p=0.2090) | | 1.6993( p=0.1330) | |
| CVM | | 0.2420 (p=0.1990) | | 0.3333(p=0.1110) | |
| KS & p-value | | 0.1787 (p=0.1371) | | 0.2041(p=0.0615) | |
| H0 | | Fail to reject | | Fail to reject | |
| SSE | | 2.2362e-19 | | 2.4395e-19 | |
| $\hat{\alpha}_r$, unbiased estimator | | 0.3485 | | 0.2139 | |
| $\hat{\beta}_s$, unbiased estimator | | 0.4325 | | 0.2979 | |
| Sig.of $\alpha$ parameter | | 2.0345(p=0.0487) | | 0.6592(p=0.5136) | |
| Sig.of $\beta$ parameter | | 3.2983(p=0.0021) | | 1.0279(p=0.3104) | |
| Variance of the function of the parmaeter, after the delta method application. Determinant and trace of this matrix | | 0.0190 | 0.0107 | 0.0213 | 0.0089 |
| | | 0.0107 | 0.0060 | 0.0089 | 0.0037 |
| | | Eigenvalues=0.0250, 2.6021e-18<br>Determinant=0<br>Trace=0.0250 | | Eigenvalues=0.0250 , 4.3368e-19<br>Determinant=0<br>Trace=0.0250 | |
| Var-cov between M101&M110 | | 6.7399e-4 | 0.0038 | | |
| | | 0.0038 | 0.0020 | | |
| Var-cov between M102 & M120 | | | | 6.7302e-4 | 0.0264 |
| | | | | 0.0264 | 0.0020 |
| $var(\hat{\theta})$ and associated $\xi$ | | 0.0449 | 0.0355 | 0.4582 | 0.3613 |
| | | 0.0355 | 0.1671 | 0.3613 | 1.7103 |
| | | $\xi =$ 5.6607 to achieve a condition number=5 | | $\xi =$ 0.5534 to achieve a condition number=5 | |
| Jacobian matrix | | -0.3668 | 0.0987 | -0.2691 | 0.0721 |
| | | -0.2059 | 0.0554 | -0.1119 | 0.0300 |

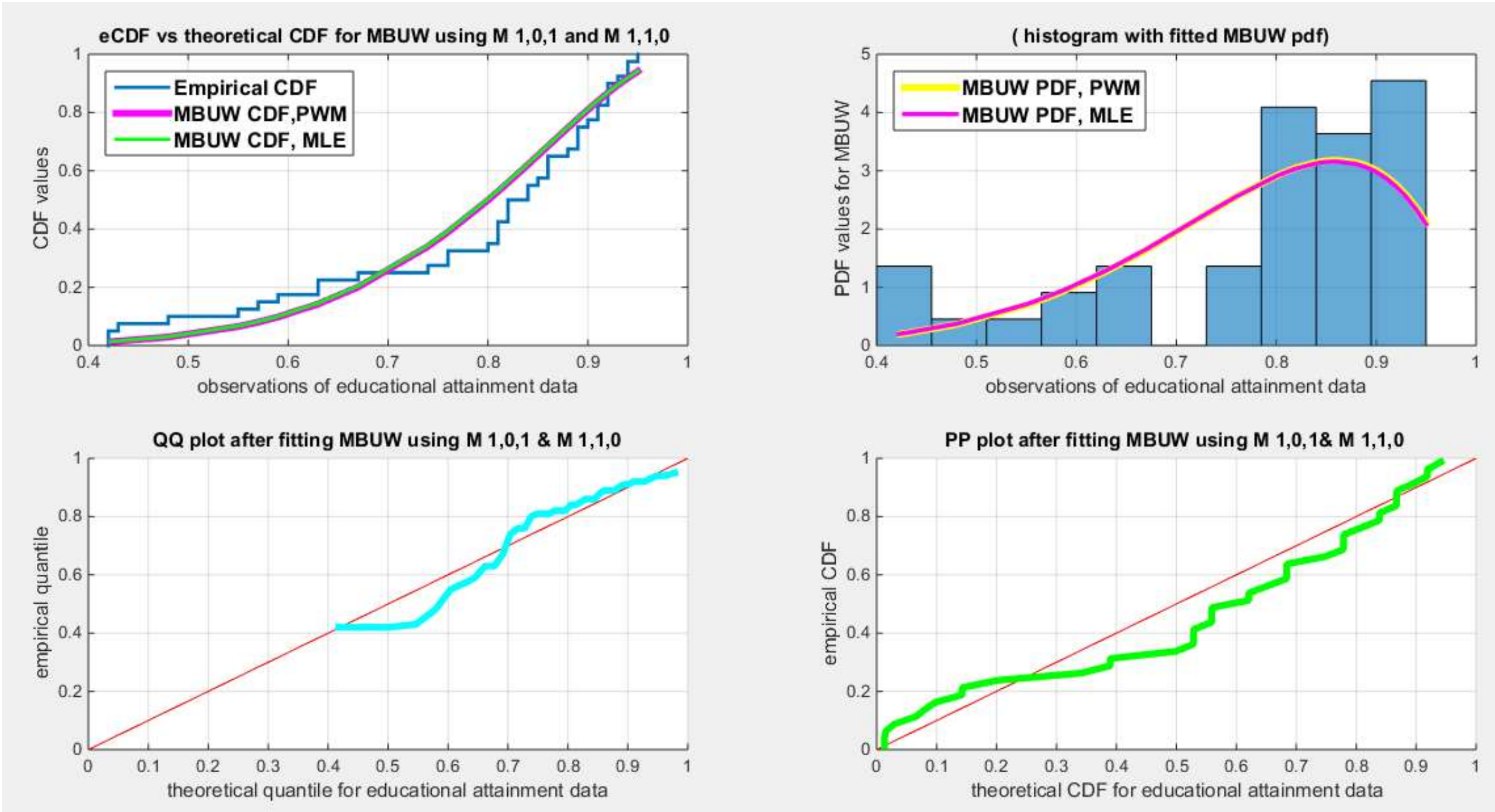

Fig. 13 shows the eCDF vs. fitted CDF for MLE & low order PWMs, M101, M110 (upper left subplot), histogram & fitted PDFs ( upper right subplot), QQ plot (lower left) and PP plot (lower right) for educational attainment dataset.

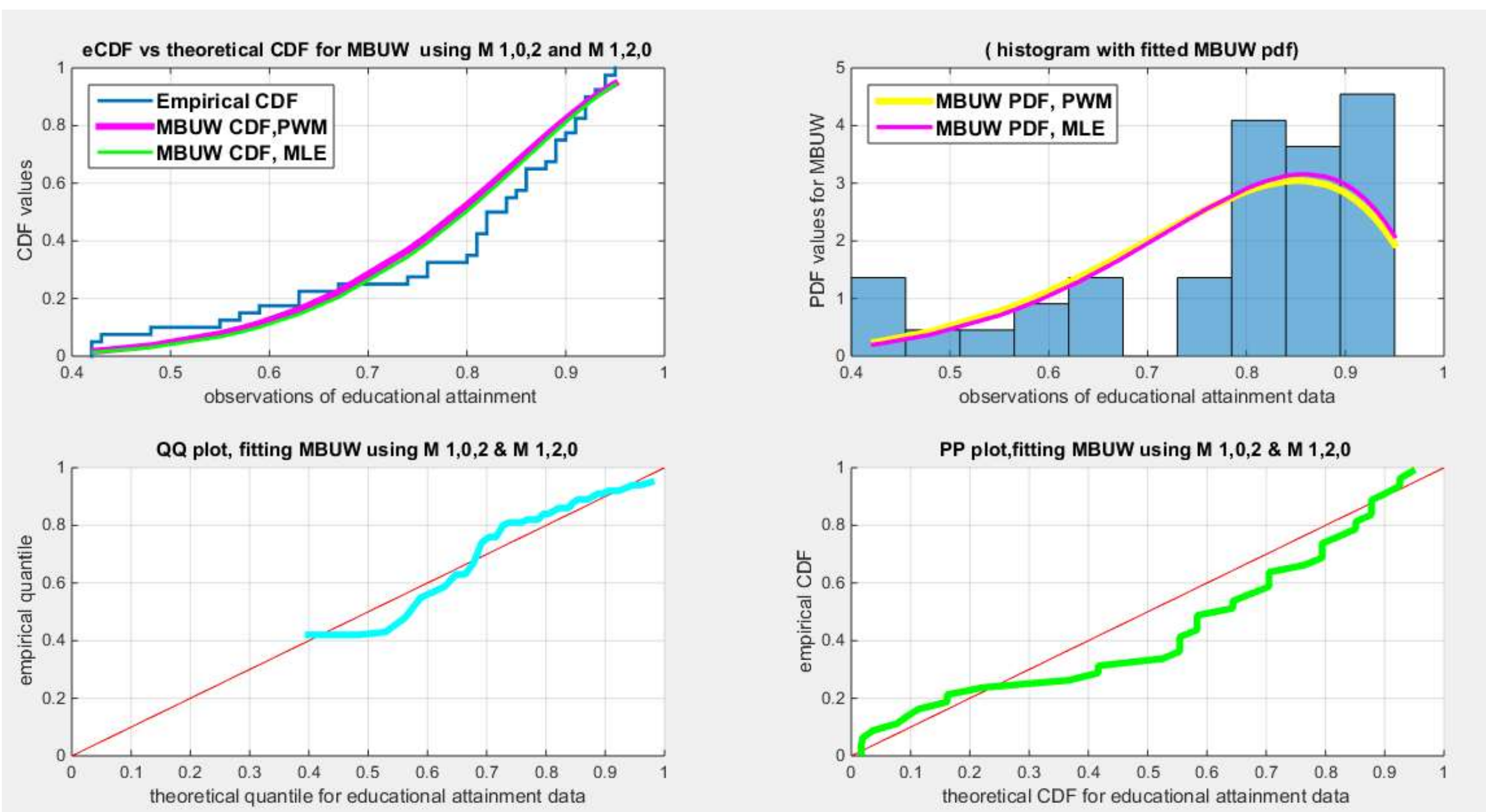

Fig. 14 shows the eCDF vs. fitted CDF for MLE & high order PWMs, M102, M120 (upper left subplot), histogram & fitted PDFs ( upper right subplot), QQ plot (lower left) and PP plot (lower right) for educational attainment dataset.

Table (20): comparisons of the results of the PWMs method for parameter estimation of the self-reported health dataset using $M_{1,0,1}$ & $M_{1,1,0}$ and $M_{1,0,2}$ & $M_{1,2,0}$

<table>
<tr><th colspan="2"></th><th colspan="2">Using unbiased Sample estimator for $M_{1,0,1}\ and\ M_{1,1,0}$</th><th colspan="2">Using unbiased Sample estimator for $M_{1,0,2}\ and\ M_{1,2,0}$</th></tr>
<tr><td rowspan="2">thetas</td><td>$\alpha$</td><td colspan="2">0.5843</td><td colspan="2">0.5752</td></tr>
<tr><td>$\beta$</td><td colspan="2">1.2806</td><td colspan="2">1.2829</td></tr>
<tr><td colspan="2" rowspan="3">Var-cov matrix of parameter</td><td>11.610</td><td>21.6717</td><td>71.1767</td><td>230.3148</td></tr>
<tr><td>21.6717</td><td>94.6865</td><td>230.3148</td><td>942.8902</td></tr>
<tr><td colspan="2">Eigenvalues=6.2964 , 100</td><td colspan="2">Eigenvalue=14.0669 , 1000</td></tr>
<tr><td colspan="2">AD</td><td colspan="2">1.4443 ( p=0.2010)</td><td colspan="2">1.5100 ( p=0.1780)</td></tr>
<tr><td colspan="2">CVM</td><td colspan="2">0.2382 (p=0.2100)</td><td colspan="2">0.2529 (p=0.1900)</td></tr>
<tr><td colspan="2">KS & p-value</td><td colspan="2">0.1549 (p=0.2768)</td><td colspan="2">0.1645 (p=0.2167)</td></tr>
<tr><td colspan="2">H0</td><td colspan="2">Fail to reject</td><td colspan="2">Fail to reject</td></tr>
<tr><td colspan="2">SSE</td><td colspan="2">5.9631e-19</td><td colspan="2">0</td></tr>
<tr><td colspan="2">$\hat{\alpha}_r$, unbiased estimator</td><td colspan="2">0.3019</td><td colspan="2">0.1865</td></tr>
<tr><td colspan="2">$\hat{\beta}_s$, unbiased estimator</td><td colspan="2">0.3775</td><td colspan="2">0.2621</td></tr>
<tr><td colspan="2">Sig.of $\alpha$ parameter</td><td colspan="2">2.7038(p=0.0102)</td><td colspan="2">0.7048 (p=0.4852)</td></tr>
<tr><td colspan="2">Sig.of $\beta$ parameter</td><td colspan="2">3.0045(p=0.0047)</td><td colspan="2">0.7988 (p=0.4293)</td></tr>
<tr><td colspan="2" rowspan="3">Variance of the function of the parmaeter, after the delta method application. Determinant and trace of this matrix</td><td>0.0172</td><td>0.0108</td><td>0.0205</td><td>0.0099</td></tr>
<tr><td>0.0108</td><td>0.0068</td><td>0.0099</td><td>0.0047</td></tr>
<tr><td colspan="2">Eigenvalues=0.0240, 1.7347e-18<br>Determinant=0<br>Trace=0.0240</td><td colspan="2">Eigenvalues=0.0253 , 0<br>Determinant=0<br>Trace=0.0253</td></tr>
<tr><td colspan="2" rowspan="2">Var-cov between M101& M110</td><td>4.0448e-4</td><td>0.0029</td><td colspan="2" rowspan="2"></td></tr>
<tr><td>0.0029</td><td>0.0020</td></tr>
<tr><td colspan="2" rowspan="2">Var-cov between M102 & M120</td><td colspan="2" rowspan="2"></td><td>4.2498e-4</td><td>0.0347</td></tr>
<tr><td>0.0347</td><td>0.0020</td></tr>
<tr><td colspan="2" rowspan="3">$var(\hat{\theta})$ and associated $\xi$</td><td>0.0467</td><td>0.0352</td><td>0.6661</td><td>0.5053</td></tr>
<tr><td>0.0352</td><td>0.1817</td><td>0.5053</td><td>2.5787</td></tr>
<tr><td colspan="2">$\xi =$ 5.2546 to achieve a condition number=5</td><td colspan="2">$\xi =$ 0.3698 to achieve a condition number=5</td></tr>
<tr><td colspan="2" rowspan="2">Jacobian matrix</td><td>-0.3062</td><td>0.0717</td><td>-0.2317</td><td>0.0574</td></tr>
<tr><td>-1.5296</td><td>0.3579</td><td>-0.1112</td><td>0.0276</td></tr>
</table>

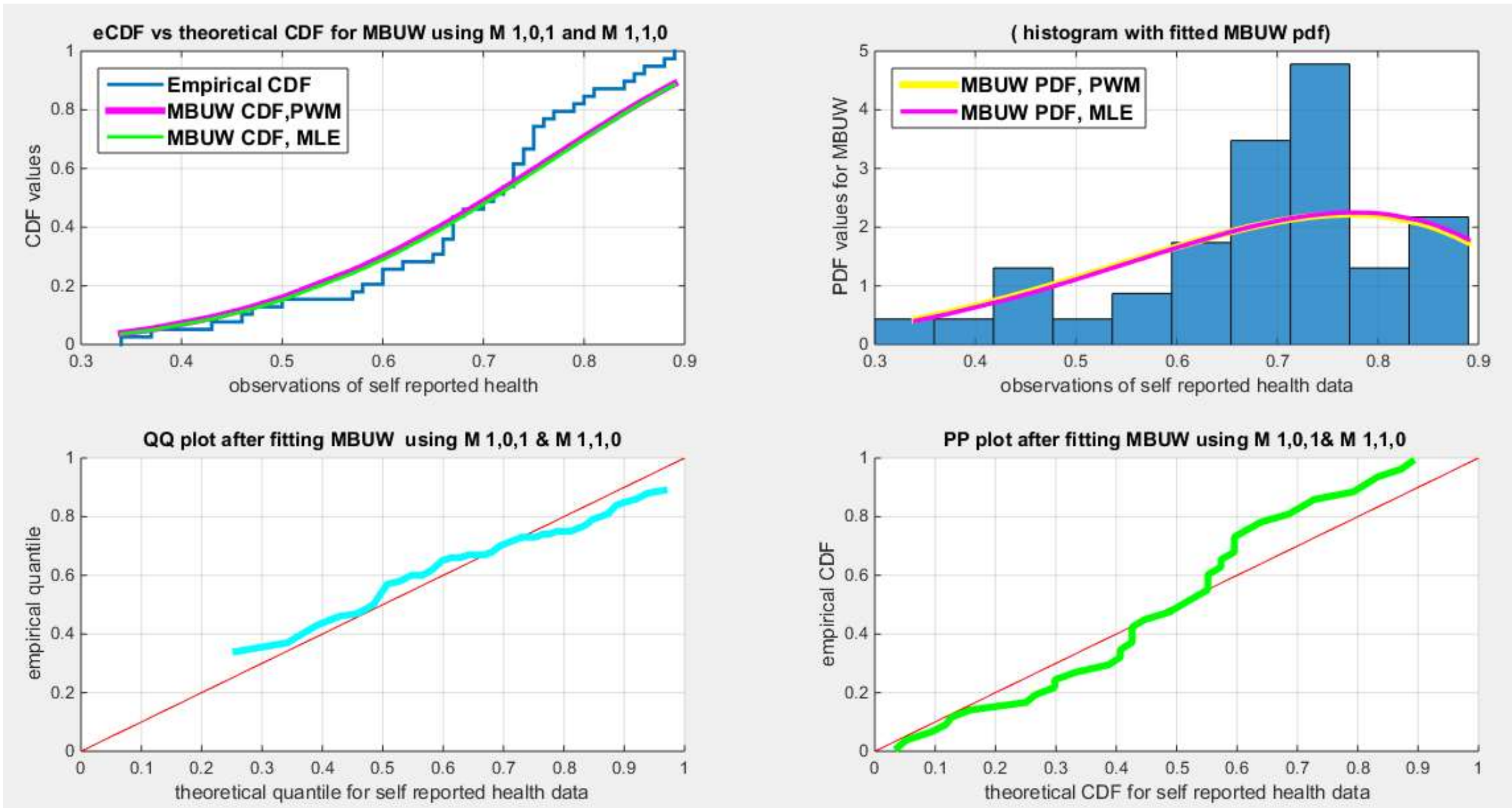

Fig. 15 shows the eCDF vs. fitted CDF for MLE & low order PWMs, M101, M110 (upper left subplot), histogram & fitted PDFs ( upper right subplot), QQ plot (lower left) and PP plot (lower right) for self-reported health dataset.

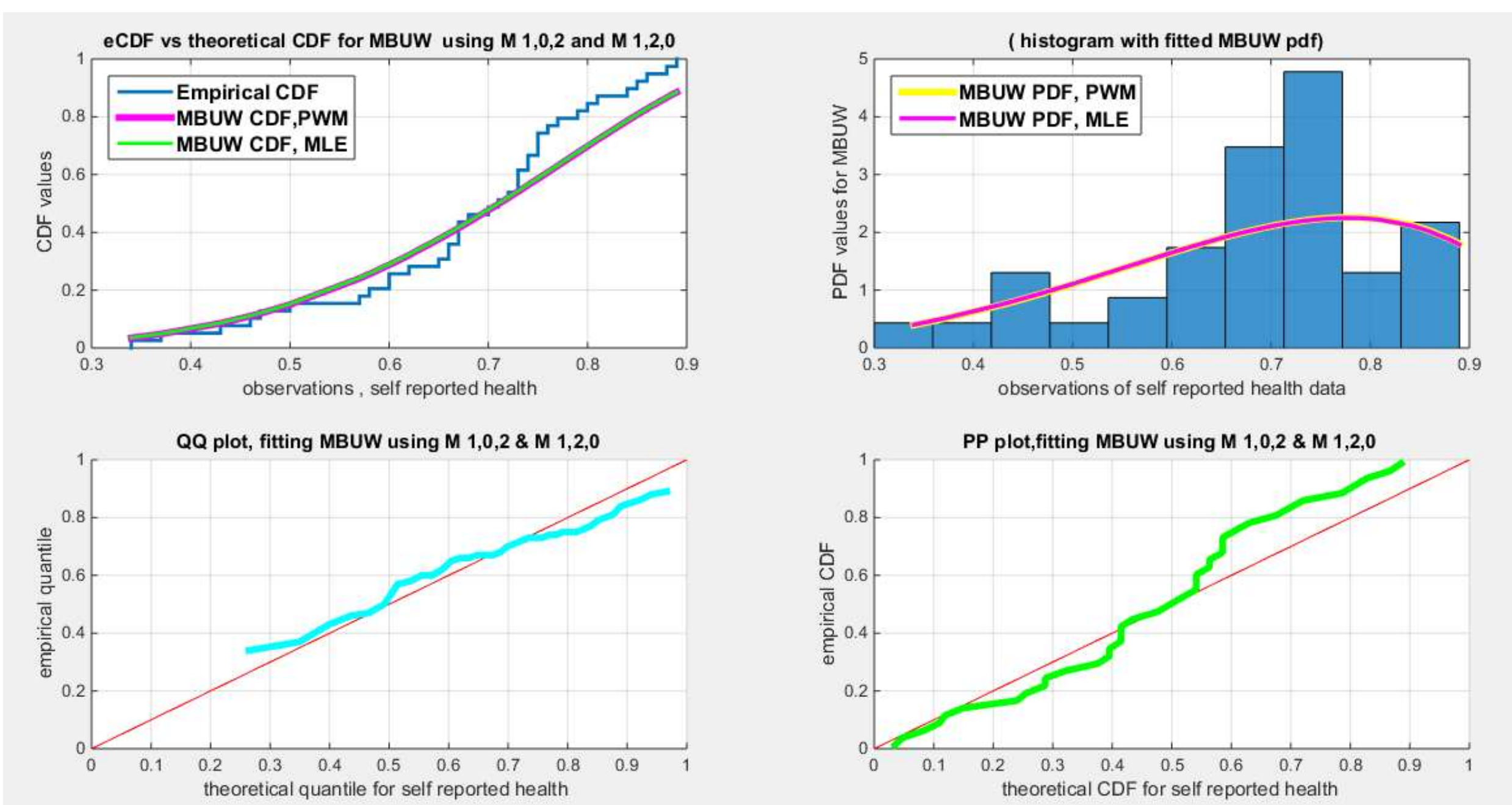

Fig. 16 shows the eCDF vs. fitted CDF for MLE & high order PWMs, M102, M120 (upper left subplot), histogram & fitted PDFs ( upper right subplot), QQ plot (lower left) and PP plot (lower right) for self-reported health dataset.

Table 18 shows the results of the water quality dataset. The values of the estimated parameters using the lower PMWs and higher PWMs are nearly equal. The approximated variance-covariance obtained from the LM algorithm using the lower order PWMs is less than that obtained from using the higher order PWMs. Applying the delta method on both lower and higher PWMs gives variances comparable to each other. The determinant for the matrix of the lower order (0.0232) is less than that of the higher order (0.0244). The Jacobian matrix is an ill condition matrix for both PWMs. Using a regularization factor, which is lesser for the higher order PWMs, enhances the estimated variance of the estimated parameters. This regularized factor is chosen so the condition number of this matrix $\left([g'(\theta_0)]^T \left(var\left[g(\hat{\theta})\right]\right)^{-1}[g'(\theta_0)]\right)$ is 5. This is because the estimated $var(\hat{\theta}) = \left[[g'(\theta_0)]^T \left(var\left[g(\hat{\theta})\right]\right)^{-1}[g'(\theta_0)] + \xi I\right]^{-1}$ requires this factor ($\xi$). The standard errors of the estimated parameters are obtained by taking the square root of the diagonal of this last matrix. The estimated variance; $var(\hat{\theta})$ ; obtained from the lower order PWMs is less than that from the higher PWMS. The covariance between the samples PWMs is also reported; $var\left[g(\hat{\theta})\right] = \begin{bmatrix} var(M_{101}) & cov(M_{101}, M_{110}) \\ cov(M_{101}, M_{110}) & var(M_{110}) \end{bmatrix}$ and it is less for the lower order PWMs than the covariance for the higher order PWMs. The AD, CVM, and KS tests are statistically significant for both order PWMs. The estimated $(\hat{\beta})$ obtained from lower PWMs is statistically significant. Figures (11-12) show the fitted CDFs and the fitted PDFs for the lower and higher order PWMs respectively. The curves of the fitted CDF and the fitted PDF obtained from MLE are perfectly aligned with the curves obtained from using both orders of PWMs. The QQ plot and PP plot are almost perfectly aligned with the diagonal.

Table 19 clarifies the same type of the results for the educational attainment dataset. The values of the estimated parameters are nearly alike. The approximated variance yielded from the LM algorithm is nearly comparable between the two orders of the PWMs. After applying the delta method, the determinant of the covariance matrix obtained from both orders of the PWMs is identical (0.025). The variance of the estimated parameter, $var(\hat{\theta})$ , is less for lower order PWMs than that for the higher order PWMs. The AD, CVM, KS tests are statistically significant. The estimated $(\hat{\alpha})$ &$(\hat{\beta})$ obtained from lower PWMs are statistically significant. Figures (13-14) disclose the fitted CDFs and the fitted PDFs for the lower and higher order PWMs respectively. The fitted CDF and the fitted PDF exhibit curves gained from MLE that are seamlessly aligned with those curves attained from using both orders of PWMs. The QQ plot and PP plot display fair alignment with the diagonal at the upper tail and at the lower tail respectively.

Table 20 expresses similar type of the outcomes for the self-reported health dataset. The values of the estimated parameters are nearly indistinguishable. The approximated variance gained from the LM algorithm is nearly comparable between the two orders of the PWMs. After applying the delta method, the determinant of the covariance matrix obtained from lower order PWMs (0.0240) is less than that produced from the higher order PWMs (0.0253). The variance of the estimated parameter, $var(\hat{\theta})$ , is less for lower order PWMs than that for the higher order PWMs. The AD, CVM, KS tests are statistically significant. The estimated $(\hat{\alpha})$ &$(\hat{\beta})$ obtained from lower PWMs are statistically significant. Figures (15-16) unveil the fitted CDFs and the fitted PDFs for the lower and higher order PWMs respectively. The curves of the fitted CDF and the fitted PDF obtained from MLE are perfectly aligned with the curves obtained from using both orders of PWMs. The QQ plot and PP plot illustrate reasonable alignment with the diagonal.

## 6. Conclusion

The classic (PWMs) method can be effectively utilized for the parameter estimation of the new Median Based Unit Weibull (MBUW) distribution. This method is robust against outliers and is more straightforward to obtain than the maximum likelihood estimator. In the context of the fitting datasets employed in this study, the unbiased estimators for both order of PWMs yielded comparable results. The PWMs method offers several advantages over alternative estimation techniques. It is both rapid and easy to compute, consistently producing feasible values for the estimated parameters and the estimated variances. Although the higher order PWMs take much longer time to execute at lager sample sizes than MLE, they are more efficient in estimating $(\hat{\beta})$ . Furthermore, PWMs estimators exhibit asymptotic normal distributions. It was observed that higher-order moments like $(M_{1,0,2}\ and\ M_{1,2,0})$ did not significantly contribute more information than the commonly used moments $(M_{1,0,1}\ and\ M_{1,1,0})$. The relative efficiency of the lower order PWMs to the higher order PWMs is in the range from 0.8 to 1 depending on the values of the parameters. The Average Absolute Biases for both orders are more or less comparable and nearly equal so there is no big difference between them. The computational time is less for lower order PWMs than the time for the higher PWMs. The estimated parameters from both orders of the PWMs are asymptotically consistent. The delta method used to estimate the variance of the function of the parameters is asymptotically normal. Using regularization factor to estimate the variance of the estimated parameters, after accounting for the covariance between the sample PWMs, can be a solution to mitigate the dependency between the parameters. The main root cause of this dependency between the parameters is the distribution itself. The method may configure the shape of the dependency, being linear or non-linear and monotonic or non-monotonic.

In this paper, the formula employed for defining PWMs is primarily based on the binomial expansion of the cumulative distribution function and the survival function, thus integrating with respect to the variable (dy) rather than integrating with respect to the cumulative distribution function (dF). The quantile function of the MBUW is difficult to integrate in a sense to obtain a system of equations that can be solved numerically.

## 7. Future work

PWMs serve as a powerful foundation for L-moments, paving the way for a wealth of analytical possibilities. In future research, we can leverage the estimation of various L-moments, including L-skewness and L-kurtosis, alongside the L-moment method for precise parameter estimation. Furthermore, by extending the $(\beta)$ parameter of the Median Based unit Weibull (MBUW) distribution to accommodate negative values, we can broaden our analytical horizons. In such scenarios, the Generalized Probability-Weighted Moments (GPWMs) method stands out as an essential tool. Additionally, we can effectively apply Partial GPWMs to handle censored data, ensuring a comprehensive approach to our analyses. The Bayesian methods with different loss functions can be consumed in the future to estimate the parameters. The correlation between the two parameters is a proposal for future studies to alleviate this dependency and hence to enhance the estimation and inference procedures. The tail indices can be further assessed to see if they can benefit from higher order PWMs. These directions not only enhance our statistical methodologies but also significantly enrich our understanding of the underlying data distributions.

## Appendix A

### A code for the PWMs method using the LM algorithm is available at:

https://doi.org/10.5281/zenodo.17616890

### The Jacobian for the M101 & M110:

$$M_{1,0,1} = \frac{360 + 108\alpha^{\beta}}{1044\alpha^{\beta} + 580\alpha^{2\beta} + 155\alpha^{3\beta} + 20\alpha^{4\beta} + \alpha^{5\beta} + 720} = \frac{p}{q}$$

Let us call numerator for M101 p and the denominator q

$$\frac{\partial\, p}{\partial\, \alpha} = 108\beta\alpha^{\beta-1} \quad \& \qquad \frac{\partial\, p}{\partial\, \beta} = 108\alpha^{\beta}\ln\alpha$$

$$\frac{\partial\, q}{\partial\, \alpha} = 1044\beta\alpha^{\beta-1} + 580(2\beta)\alpha^{2\beta-1} + 155(3\beta)\alpha^{3\beta-1} + 20\alpha(4\beta)^{\,4\beta-1} + (5\beta)\alpha^{5\beta-1}$$

$$\frac{\partial\, q}{\partial\, \beta} = 1044\alpha^{\beta}\ln\alpha + 580(2)\alpha^{2\beta}\ln\alpha + 155(3)\alpha^{3\beta}\ln\alpha + 20\,(4)\alpha^{4\beta}\ln\alpha + (5)\alpha^{5\beta}\ln\alpha$$

$$\frac{\partial M_{1,0,1}}{\partial\alpha} = \frac{\frac{\partial\, p}{\partial\, \alpha}\,(q) - \frac{\partial\, q}{\partial\, \alpha}(p)}{q^2} \qquad \& \qquad \frac{\partial M_{1,0,1}}{\partial\beta} = \frac{\frac{\partial\, p}{\partial\, \beta}\,(q) - \frac{\partial\, q}{\partial\, \beta}(p)}{q^2}$$

$$M_{1,1,0} = \frac{60 + 6\alpha^{\beta}}{74\alpha^{\beta} + 15\alpha^{2\beta} + \alpha^{3\beta} + 120} = \frac{R}{T}$$

Let us call numerator for M110 R and the denominator T

$$\frac{\partial R}{\partial\, \alpha} = 6\beta\alpha^{\beta-1} \quad \& \qquad \frac{\partial\, R}{\partial\, \beta} = 6\alpha^{\beta}\ln\alpha$$

$$\frac{\partial\, T}{\partial\, \alpha} = 74\,\beta\alpha^{\beta-1} + 15(2\beta)\alpha^{2\beta-1} + (3\beta)\alpha^{3\beta-1}$$

$$\frac{\partial\, T}{\partial\, \beta} = 74\,\alpha^{\beta}\ln\alpha + 15(2)\alpha^{2\beta}\ln\alpha + (3)\alpha^{3\beta}\ln\alpha$$

$$\frac{\partial M_{1,1,0}}{\partial \alpha} = \frac{\frac{\partial R}{\partial \alpha}\ (T) - \frac{\partial T}{\partial \alpha}(R)}{T^2} \quad \& \quad \frac{\partial M_{1,1,0}}{\partial \beta} = \frac{\frac{\partial R}{\partial \beta}\ (T) - \frac{\partial T}{\partial \beta}(R)}{T^2}$$

**The Jacobian for the M102 & M120:**

$M_{1,0,2} = \frac{K}{L}$ *where* $K = 58320\ \alpha^{\beta} + 6480\, \alpha^{2\beta} + 120960$

$$L = 663696\, \alpha^{\beta} + 509004\, \alpha^{2\beta} + 214676\, \alpha^{3\beta} + 54649\alpha^{4\beta} + 8624\, \alpha^{5\beta} + 826\, \alpha^{6\beta} + 44\, \alpha^{7\beta} + \alpha^{8\beta} + 362880$$

$$\frac{\partial K}{\partial \alpha} = 58320\, \beta\, \alpha^{\beta-1} + 6480(2\beta)\, \alpha^{2\beta-1}$$

$$\frac{\partial K}{\partial \beta} = 58320\ \alpha^{\beta} \ln \alpha + 6480(2)\, \alpha^{2\beta} \ln \alpha$$

$$\frac{\partial L}{\partial \alpha} = 663696\beta\, \alpha^{\beta-1} + 509004(2\beta)\, \alpha^{2\beta-1} + 214676(3\beta)\, \alpha^{3\beta-1} + 54649\, (4\beta)\alpha^{4\beta-1} + 8624\, (5\beta)\, \alpha^{5\beta-1} + 826\, (6\beta)\, \alpha^{6\beta-1} + 44(7\beta)\, \alpha^{7\beta-1} + (8\beta)\, \alpha^{8\beta-1}$$

$$\frac{\partial L}{\partial \beta} = 663696\, \alpha^{\beta} \ln \alpha + 509004(2)\, \alpha^{2\beta} \ln \alpha + 214676(3)\, \alpha^{3\beta} \ln \alpha + 54649\, (4)\alpha^{4\beta} \ln \alpha + 8624\, (5)\alpha^{5\beta} \ln \alpha + 826(6)\, \alpha^{6\beta} \ln \alpha + 44\, (7)\alpha^{7\beta} \ln \alpha + (8)\, \alpha^{8\beta} \ln \alpha$$

$$\frac{\partial M_{1,0,2}}{\partial \alpha} = \frac{\frac{\partial K}{\partial \alpha}\ (L) - \frac{\partial L}{\partial \alpha}(K)}{L^2} \quad \& \quad \frac{\partial M_{1,0,2}}{\partial \beta} = \frac{\frac{\partial K}{\partial \beta}\ (L) - \frac{\partial L}{\partial \beta}(K)}{L^2}$$

$$M_{1,2,0} = \frac{150\, \alpha^{\beta} + 6\, \alpha^{2\beta} + 1008}{1650\, \alpha^{\beta} + 335\, \alpha^{2\beta} + 30\ \alpha^{3\beta} + \alpha^{4\beta} + 3024} = \frac{H}{G}$$

$$\frac{\partial H}{\partial \alpha} = 150\, \beta\, \alpha^{\beta-1} + 6(2\beta)\, \alpha^{2\beta-1} \quad \& \quad \frac{\partial \mathrm{H}}{\partial \mathrm{\beta}} = 150\alpha^{\beta} \ln \alpha + 6\, (2)\alpha^{2\beta} \ln \alpha$$

$$\frac{\partial G}{\partial \alpha} = 1650\, \beta\, \alpha^{\beta-1} + 335\, (2\beta)\, \alpha^{2\beta-1} + 30\, (3\beta)\, \alpha^{3\beta-1} + (4\beta)\ \alpha^{4\beta-1}$$

$$\frac{\partial G}{\partial \beta} = 1650\, \alpha^{\beta} \ln \alpha + 335(2)\, \alpha^{2\beta} \ln \alpha + 30\, (3)\, \alpha^{3\beta} \ln \alpha + (4)\ \alpha^{4\beta} \ln \alpha$$

$$\frac{\partial M_{1,2,0}}{\partial \alpha} = \frac{\frac{\partial H}{\partial \alpha}\ (G) - \frac{\partial G}{\partial \alpha}(H)}{G^2} \quad \& \quad \frac{\partial M_{1,2,0}}{\partial \beta} = \frac{\frac{\partial H}{\partial \beta}\ (G) - \frac{\partial G}{\partial \beta}(H)}{G^2}$$

## Maximum Likelihood Estimation

Let $y_1, y_2, \dots, y_n$ an observed random sample from MBUW distribution with parameters $\alpha, \beta$. The likelihood function and the log-likelihood function are defined in

$$L(\alpha,\beta;y) = \left(\frac{6}{\alpha^{\beta}}\right)^{n} \prod_{i=1}^{n}\left[1 - y_i^{\frac{1}{\alpha^{\beta}}}\right] \prod_{i=1}^{n} y_i^{\left(\frac{2}{\alpha^{\beta}}-1\right)} \quad , \ 0 < y < 1 , \qquad \alpha \,\&\beta > 0$$

$$l(\alpha,\beta;y) = nln(6) - n\beta ln(\alpha) + \sum_{i=1}^{n} ln\left[1 - y_i^{\frac{1}{\alpha^{\beta}}}\right] + \left(\frac{2}{\alpha^{\beta}} - 1\right)\sum_{i=1}^{n} ln(y_i)$$

The Maximum Likelihood (ML) equations are the first derivative of the log-likelihood function concerning each parameter as defined in

$$\frac{dl(\alpha,\beta;y)}{d\alpha} = \frac{-\beta n}{\alpha} + \sum_{i=1}^{n} \frac{-y_i^{\frac{1}{\alpha^{\beta}}}(ln[y_i])\left(-\beta\alpha^{-\beta-1}\right)}{1 - y_i^{\frac{1}{\alpha^{\beta}}}} - 2\beta\alpha^{-\beta-1}\sum_{i=1}^{n} ln(y_i)$$

$$\frac{dl(\alpha,\beta;y)}{d\beta} = -nln\alpha + \alpha^{-\beta}\sum_{i=1}^{n} \frac{y_i^{\frac{1}{\alpha^{\beta}}} ln[\alpha] ln[y]}{1 - y_i^{\frac{1}{\alpha^{\beta}}}} - 2\alpha^{-\beta}[ln\alpha]\sum_{i=1}^{n} ln(y_i)$$

# Appendix B

### The variance of the M101:

$$var_{M101} = E\{y\,[F_Y(y)]^0\,[1 - F_Y(y)]^1\}^2 - \{E(y\,[F_Y(y)]^0\,[1 - F_Y(y)]^1)\}^2$$

Let us call this, $E\{y\,[F_Y(y)]^0\,[1 - F_Y(y)]^1\}^2$ , the integral $I_{M101}$

$$E\{y^2\ [1 - F_Y(y)]^2\} = \int_0^1 y^2\ [1 - F_Y(y)]^2 \frac{6}{\alpha^{\beta}}\left[1 - y^{\frac{1}{\alpha^{\beta}}}\right] y^{\frac{2}{\alpha^{\beta}}-1}\, \mathrm{dy}$$

$$I_{M101} = \int_0^1 y^2\ \left[1 - 3y^{\frac{2}{\alpha^{\beta}}} + 2y^{\frac{3}{\alpha^{\beta}}}\right]^2 \frac{6}{\alpha^{\beta}}\left[1 - y^{\frac{1}{\alpha^{\beta}}}\right] y^{\frac{2}{\alpha^{\beta}}-1}\, \mathrm{dy}$$

After expansion of the squared quantity and algebraic manipulations this integral $(I_{M101})$ is

$$I_{M101} = 6\left(\frac{1}{2+2\alpha^{\beta}} - \frac{1}{3+2\alpha^{\beta}} - \frac{6}{4+2\alpha^{\beta}} + \frac{10}{5+2\alpha^{\beta}} + \frac{5}{6+2\alpha^{\beta}} - \frac{21}{7+2\alpha^{\beta}} + \frac{16}{8+2\alpha^{\beta}} - \frac{4}{9+2\alpha^{\beta}}\right)$$

$E(y\,[F_Y(y)]^0\,[1 - F_Y(y)]^1)$ is calculated in the text of the paper.

### The variance of the M110:

$$var_{M110} = E\{y\,[F_Y(y)]^1\,[1 - F_Y(y)]^0\}^2 - \{E(y\,[F_Y(y)]^1\,[1 - F_Y(y)]^0)\}^2$$

Let us call this $E\{y\ [F_Y(y)]^1\ [1-F_Y(y)]^0\}^2$ , the integral $I_{M110}$

$$E\{y^2\ \ [F_Y(y)]^2\} = \int_0^1 y^2\ \ [F_Y(y)]^2 \frac{6}{\alpha^\beta}\left[1-y^{\frac{1}{\alpha^\beta}}\right] y^{\frac{2}{\alpha^\beta}-1}\ \mathrm{dy}$$

$$I_{M11}\ \ = \int_0^1 y^2\ \ \left[3y^{\frac{2}{\alpha^\beta}} - 2y^{\frac{3}{\alpha^\beta}}\right]^2 \frac{6}{\alpha^\beta}\left[1-y^{\frac{1}{\alpha^\beta}}\right] y^{\frac{2}{\alpha^\beta}-1}\ \mathrm{dy}$$

After expansion of the squared quantity and algebraic manipulations this integral $(I_{M110})$ is

$$I_{M110} = 6\left(\frac{9}{6+2\alpha^\beta} - \frac{21}{7+2\alpha^\beta} + \frac{16}{8+2\alpha^\beta} - \frac{4}{9+2\alpha^\beta}\right)$$

$E(y\ [F_Y(y)]^1\ [1-F_Y(y)]^0)$ is calculated in the text of the paper.

## The variance of the M102:

$$var_{M102} = E\{y\ [F_Y(y)]^0\ [1-F_Y(y)]^2\}^2 - \{E(y\ [F_Y(y)]^0\ [1-F_Y(y)]^2)\}^2$$

Let us call this $E\{y\ [F_Y(y)]^0\ [1-F_Y(y)]^2\}^2$ , the integral $I_{M102}$

$$E\{y^2\ \ [1-F_Y(y)]^4\} = \int_0^1 y^2\ \ [1-F_Y(y)]^4 \frac{6}{\alpha^\beta}\left[1-y^{\frac{1}{\alpha^\beta}}\right] y^{\frac{2}{\alpha^\beta}-1}\ \mathrm{dy}$$

$$I_{M102} = \int_0^1 y^2\ \ \left[1-3y^{\frac{2}{\alpha^\beta}} + 2y^{\frac{3}{\alpha^\beta}}\right]^4 \frac{6}{\alpha^\beta}\left[1-y^{\frac{1}{\alpha^\beta}}\right] y^{\frac{2}{\alpha^\beta}-1}\ \mathrm{dy}$$

After algebraic manipulations this integral $(I_{M102})$ is

$$I_{M102} = 6\left(\frac{1}{2+2\alpha^\beta} - \frac{1}{3+2\alpha^\beta} - \frac{12}{4+2\alpha^\beta} + \frac{20}{5+2\alpha^\beta} + \frac{46}{6+2\alpha^\beta} - \frac{126}{7+2\alpha^\beta} - \frac{12}{8+2\alpha^\beta} + \frac{300}{9+2\alpha^\beta} - \frac{279}{10+2\alpha^\beta} - \frac{121}{11+2\alpha^\beta} + \frac{400}{12+2\alpha^\beta} - \frac{312}{13+2\alpha^\beta} + \frac{112}{14+2\alpha^\beta} - \frac{16}{15+2\alpha^\beta}\right)$$

$E(y\ [F_Y(y)]^0\ [1-F_Y(y)]^2)$ is calculated in the text of the paper.

## The variance of the M120:

$$var_{M120} = E\{y\ [F_Y(y)]^2\ [1-F_Y(y)]^0\}^2 - \{E(y\ [F_Y(y)]^2\ [1-F_Y(y)]^0)\}^2$$

Let us call this $E\{y\ [F_Y(y)]^2\ [1-F_Y(y)]^0\}^2$ , the integral $I_{M12}$

$$E\{y^2\ \ [F_Y(y)]^4\} = \int_0^1 y^2\ \ [F_Y(y)]^4 \frac{6}{\alpha^\beta}\left[1-y^{\frac{1}{\alpha^\beta}}\right] y^{\frac{2}{\alpha^\beta}-1}\ \mathrm{dy}$$

$$I_{M120} = \int_0^1 y^2\ \ \left[3y^{\frac{2}{\alpha^\beta}} - 2y^{\frac{3}{\alpha^\beta}}\right]^4 \frac{6}{\alpha^\beta}\left[1-y^{\frac{1}{\alpha^\beta}}\right] y^{\frac{2}{\alpha^\beta}-1}\ \mathrm{dy}$$

After algebraic manipulations this integral $(I_{M120})$ is

$$I_{M120} = 6\left(\frac{81}{10+2\alpha^{\beta}} - \frac{297}{11+2\alpha^{\beta}} + \frac{432}{12+2\alpha^{\beta}} - \frac{312}{13+2\alpha^{\beta}} + \frac{112}{14+2\alpha^{\beta}} - \frac{16}{15+2\alpha^{\beta}}\right)$$

$E(y\ [F_Y(y)]^2\ [1-F_Y(y)]^0)$ is calculated in the text of the paper.

## The covariance between M101, M110

$$COV_{(M101,M110)} = COV\{X\ [F_X(x)]^0\ [1-F_X(x)]^1\ , Y\ [F_Y(y)]^1\ [1-F_Y(y)]^0\ \}$$

where $x_{j:n} = \ x_{1:2} = M_{1,0,1}$ & $y_{k:n} = \ x_{2:2} = M_{1,1,0}$

The joint pdf of M101 and M110 is the pdf of order statistics; $f_{j,k:n}(x_j, y_k)$

$$= \frac{n!}{(j-1)!\,(k-j-1)!\,(n-k)!}\left[[F_X(x)]\right]^{j-1}[F_Y(y)-F_X(x)]^{k-j-1}[1-F_Y(y)]^{n-k} f_X(x) f_Y(y)$$

$$= \frac{2!}{(1-1)!\,(2-1-1)!\,(2-2)!}\left[[F_X(x)]\right]^{1-1}[F_Y(y)-F_X(x)]^{2-1-1}[1-F_Y(y)]^{2-2} f_X(x) f_Y(y)$$

$$f_{j,k:n}(x_j, y_k) = 2\ f_X(x) f_Y(y)$$

$$COV_{(M101,M110)} = E\{X\ \ [1-F_X(x)]^1\ \ Y\ [F_Y(y)]^1\ \ \}$$

$$= \iint_0^1 X\ \ [1-F_X(x)]^1\ \ Y\ [F_Y(y)]^1 * 2 * f_X(x) f_Y(y) dx dy$$

$$E\{X\ \ [1-F_X(x)]^1\ \ Y\ [F_Y(y)]^1\ \ \} = 2\int_0^1 Y\ [F_Y(y)]^1 f_Y(y) \left\{\int_0^1 X\ \ [1-F_X(x)]^1 f_X(x) dx\right\} dy$$

$\left\{\int_0^1 X\ \ [1-F_X(x)]^1 f_X(x)\ dx\right\}$ , the inner integral can be integrated by parts

$$\int_0^1 X\ \ [1-F_X(x)]^1 f_X(x)\ dx =$$

$Let\ u = x\ so\ du = dx\ \ \&\ \ dv = [1-F_X(x)]^1 f_X(x)\ dx\ \ \ so\ \int_0^1 [1-F_X(x)]^1 f_X(x)\ dx = \frac{-[1-F_X(x)]^2}{2}$

$$\int_0^1 X\ \ [1-F_X(x)]^1 f_X(x)\ dx = \frac{1}{2}\left\{1 + \left(\frac{-6\alpha^{\beta}}{2+\alpha^{\beta}} + \frac{4\alpha^{\beta}}{3+\alpha^{\beta}} + \frac{9\alpha^{\beta}}{4+\alpha^{\beta}} - \frac{12\alpha^{\beta}}{5+\alpha^{\beta}} + \frac{4\alpha^{\beta}}{6+\alpha^{\beta}}\right)\right\} = \frac{1}{2}c$$

Then the outer integral is $\int_0^1 Y\ [F_Y(y)]^1 f_Y(y) dy$ can also be integrated by parts

$Let\ u = y\ so\ du = dy\ \ \&\ \ dv = [F_Y(y)]^1 f_Y(y)\ dy\ \ so\ \int_0^1 [F_Y(y)]^1 f_Y(y)\ dy = \frac{[F_X(x)]^2}{2}$

$$\int_0^1 Y\ [F_Y(y)]^1 f_Y(y)dy = \frac{1}{2} - \frac{1}{2}\left\{\frac{9\alpha^\beta}{4+\alpha^\beta} - \frac{12\alpha^\beta}{5+\alpha^\beta} + \frac{4\alpha^\beta}{6+\alpha^\beta}\right\} = \frac{1}{2} - \frac{1}{2}d$$

$$COV_{(M101,M110)} = E\{X\ [1-F_X(x)]^1\ Y\ [F_Y(y)]^1\ \} = 2\left(\frac{c}{2}\right)\left(\frac{1}{2} - \frac{1}{2}d\right) = \frac{c}{2}(1-d)$$

## The covariance between M102, M120

$$COV_{(M102,M12\ \ )} = COV\{X\ [F_X(x)]^0\ [1-F_X(x)]^2\ , Y\ [F_Y(y)]^2\ [1-F_Y(y)]^0\ \}$$

where $x_{j:n} = x_{1:3} = M_{1,0,2}$ & $y_{k:n} = x_{3:3} = M_{1,2,0}$

The joint pdf of M102 and M120 is the pdf of the order statistics; $f_{j,k:n}(x_j, y_k)$

$$= \frac{n!}{(j-1)!\,(k-j-1)!\,(n-k)!}\big[[F_X(x)]\big]^{j-1}[F_Y(y)-F_X(x)]^{k-j-1}[1-F_Y(y)]^{n-k} f_X(x) f_Y(y)$$

$$= \frac{3!}{(1-1)!\,(3-1-1)!\,(3-3)!}\big[[F_X(x)]\big]^{1-1}[F_Y(y)-F_X(x)]^{3-1-1}[1-F_Y(y)]^{3-3} f_X(x) f_Y(y)$$

$$f_{j,k:n}(x_j, y_k) = 6[F_Y(y)-F_X(x)]\ f_X(x) f_Y(y)$$

$$COV_{(M10\ \ ,M12\ \ )} = E\{X\ [1-F_X(x)]^2\ Y\ [F_Y(y)]^2\ \}$$

$$= \iint_0^1 X\ [1-F_X(x)]^2\ Y\ [F_Y(y)]^2 * 6 * f_X(x) f_Y(y)\ [F_Y(y)-F_X(x)]\ dxdy$$

$$= 6\left(\int_0^1 Y\ [F_Y(y)]^3 f_Y(y)\left\{\int_0^1 X\ [1-F_X(x)]^2 f_X(x)dx\right\}dy\right.$$

$$\left. - \int_0^1 Y\ [F_Y(y)]^2 f_Y(y)\left\{\int_0^1 X\ F_X(x)\ [1-F_X(x)]^2 f_X(x)dx\right\}dy\right)$$

$$= \int_0^1 Y\ [F_Y(y)]^3 f_Y(y)\left\{\int_0^1 X\ [1-F_X(x)]^2 f_X(x)dx\right\}dy = \frac{c_1}{3}\left(\frac{1}{4} - \frac{1}{4}d_1\right) = \frac{c_1}{12}(1-d_1)$$

The above integration can be solved by integration by parts where

$$c_1 = 1 - \frac{9\alpha^\beta}{2+\alpha^\beta} + \frac{6\alpha^\beta}{3+\alpha^\beta} + \frac{27\alpha^\beta}{4+\alpha^\beta} - \frac{36\alpha^\beta}{5+\alpha^\beta} - \frac{15\alpha^\beta}{6+\alpha^\beta} + \frac{54\alpha^\beta}{7+\alpha^\beta} - \frac{36\alpha^\beta}{8+\alpha^\beta} + \frac{8\alpha^\beta}{9+\alpha^\beta}$$

$$d_1 = \frac{81\alpha^\beta}{8+\alpha^\beta} - \frac{216\alpha^\beta}{9+\alpha^\beta} + \frac{216\alpha^\beta}{10+\alpha^\beta} - \frac{96\alpha^\beta}{11+\alpha^\beta} + \frac{16\alpha^\beta}{12+\alpha^\beta}$$

$$\int_0^1 Y\ [F_Y(y)]^2 f_Y(y)\left\{\int_0^1 X\ F_X(x)\ [1-F_X(x)]^2 f_X(x)dx\right\}dy\ =\ c_2\left(\frac{1}{3} - \frac{1}{3}d_2\right) =\ \frac{c_2}{3}(1-d_2)$$

This integration can be solved by parts and expansion of the $[1-F_X(x)]^2$ where

$$c_2=\frac{18}{4+\alpha^\beta}-\frac{30}{5+\alpha^\beta}-\frac{96}{6+\alpha^\beta}+\frac{252}{7+\alpha^\beta}-\frac{30}{8+\alpha^\beta}-\frac{438}{9+\alpha^\beta}+\frac{540}{10+\alpha^\beta}-\frac{264}{11+\alpha^\beta}+\frac{48}{12+\alpha^\beta}$$

$$d_2=\frac{27\alpha^\beta}{6+\alpha^\beta}-\frac{54\alpha^\beta}{7+\alpha^\beta}+\frac{36\alpha^\beta}{8+\alpha^\beta}-\frac{8\alpha^\beta}{9+\alpha^\beta}$$

$$COV_{(M102,M12\ )}=E\{X\ [1-F_X(x)]^2\ Y\ [F_Y(y)]^2\ \}=6\left(\frac{c_1}{12}(1-d_1)-\frac{c_2}{3}(1-d_2)\right)$$

$$COV_{(M102,M120)}=\frac{c_1}{2}(1-d_1)-2\,c_2(1-d_2)$$

**Declarations:**
**Ethics approval and consent to participate**
Not applicable.
**Consent for publication**
Not applicable
**Availability of data and material**
Not applicable. Data sharing does not apply to this article as no new datasets were generated or analyzed during the current study.
**Competing interests**
The author declares no competing interests of any type.
**Funding**
No funding resources. No funding roles in the design of the study and collection, analysis, and interpretation of data and in writing the manuscript are declared.
**Authors' contribution**
AI (Attia Iman) carried the conceptualization by formulating the goals, and aims of the research article, formal analysis by applying the statistical, mathematical, and computational techniques to synthesize and analyze the hypothetical data, carried the methodology by creating the model, software programming and implementation, supervision, writing, drafting, editing, preparation, and creation of the presenting work.
**Acknowledgment**
Not applicable

## Appendix C

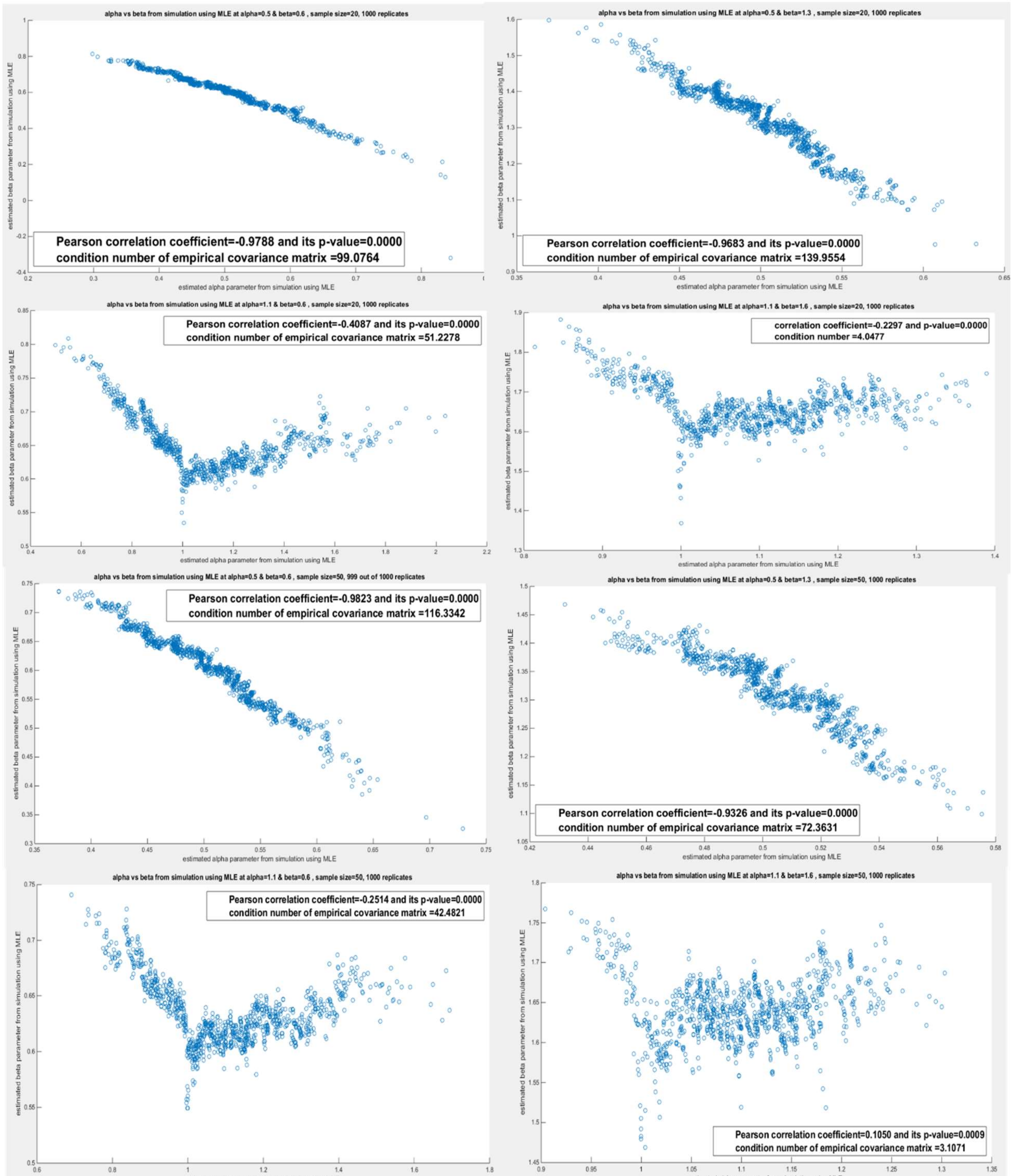

The first picture contains 8 figures with the following description: The first four figures are obtained from the replicates of sample size n=20, while the last four figures are from the replicates of sample size, n=50. The figures for each sample size is arranged as the pairs mentioned in the text, the first pair is $(a = 0.5 \, \& \, b = 0.6)$ , the second pair is $(a = 0.5 \, \& \, b = 1.3)$ , the third pair $(a = 1.1 \, \& \, b = 0.6)$ , and the fourth pair $(a = 1.1 \, \& \, b = 1.6)$. These are the runs obtained from the MLE method.

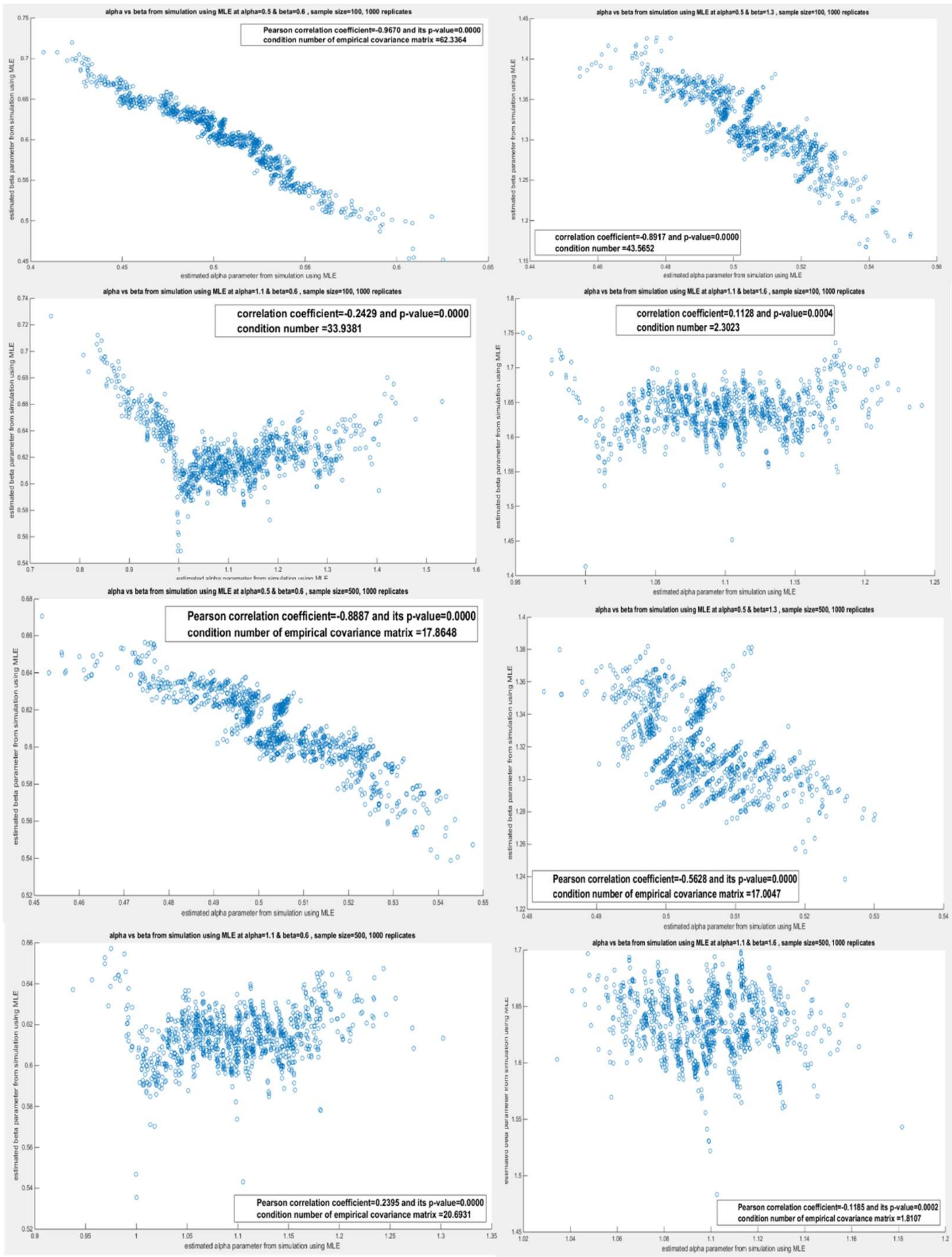

The second picture contains 8 figures with the following discription: The first four figures are obtained from the replicates of sample size n=100, while the last four figures are from the replicates of sample size, n=500. The figures for each sample size is arranged as the pairs mentioned in the text, the first pair is $(a = 0.5 \,\&\, b = 0.6)$ , the second pair is $(a = 0.5 \,\&\, b = 1.3)$ , the third pair $(a = 1.1 \,\&\, b = 0.6)$ , and the fourth pair $(a = 1.1 \,\&\, b = 1.6)$. These are the runs obtained from the MLE method

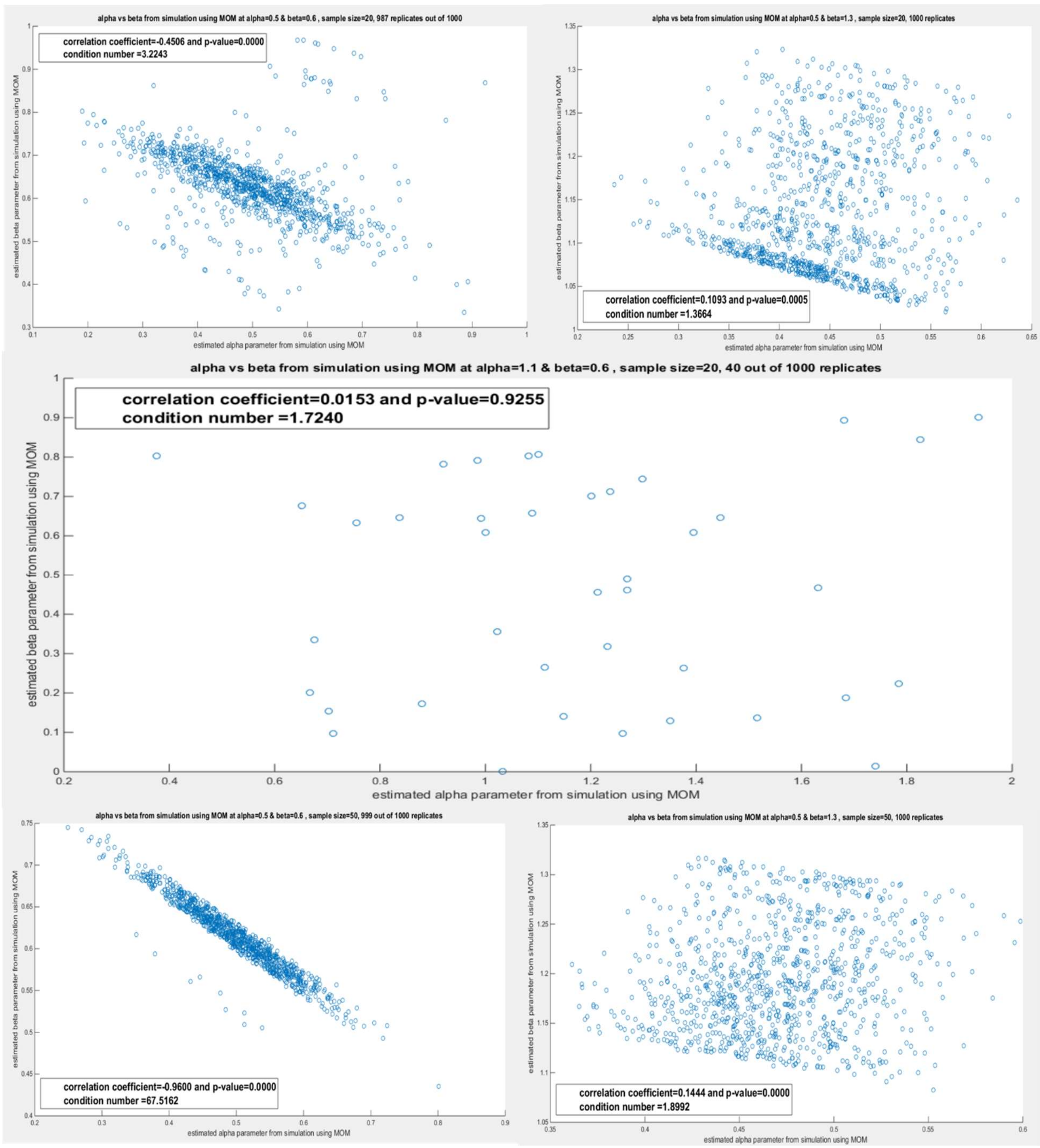

The third picture contains 5 figures with the following discription: The first three figures are obtained from the replicates of sample size n=20, while the last two figures are from the replicates of sample size, n=50. The figures for each sample size is arranged as the pairs mentioned in the text, the first pair is ($a = 0.5$ & $b = 0.6$) , the second pair is ($a = 0.5$ & $b = 1.3$) , the third pair ($a = 1.1$ & $b = 0.6$) , and the fourth pair ($a = 1.1$ & $b = 1.6$). These are the runs obtained from the MOM method

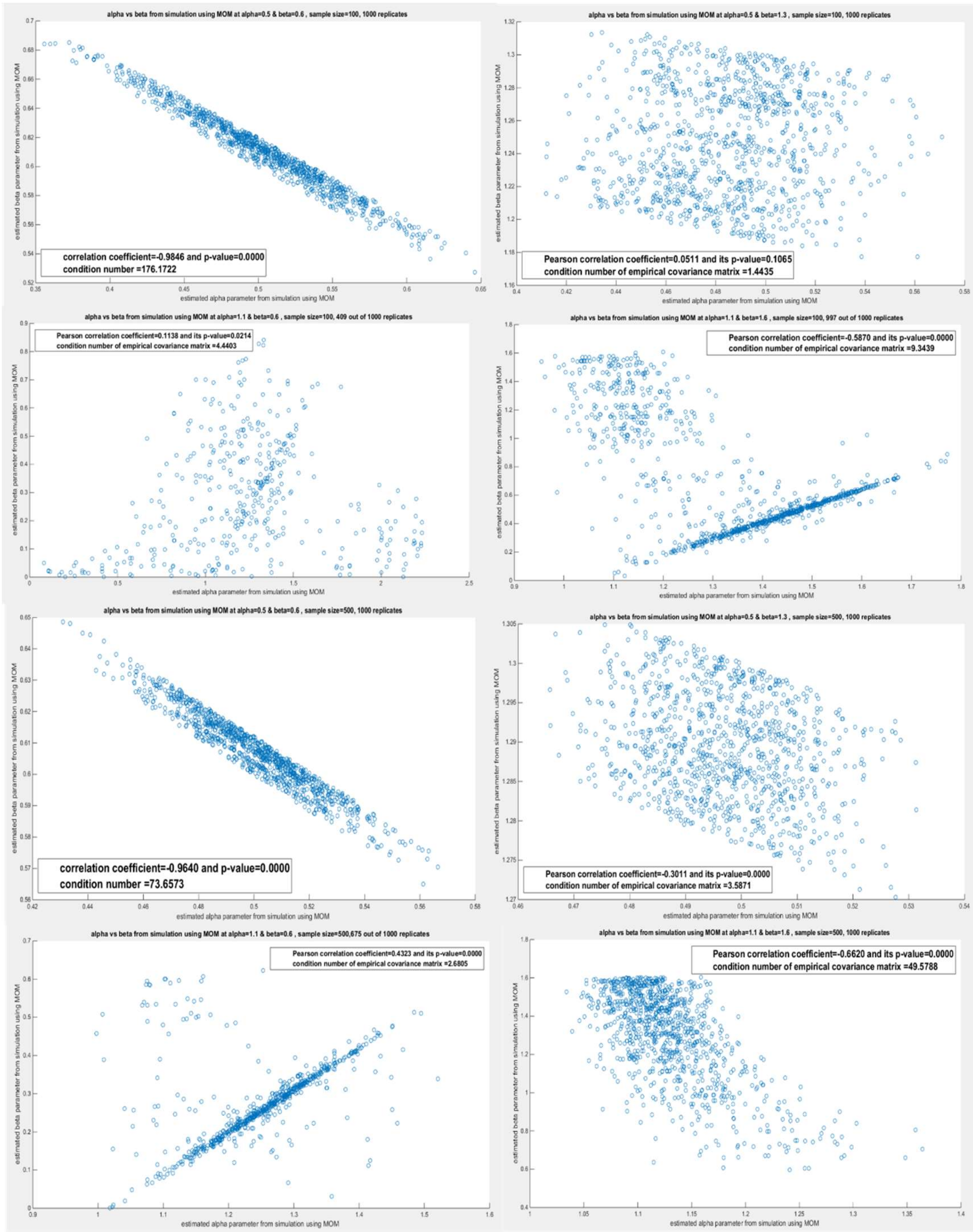

The fourth picture contains 8 figures with the following discription: The first four figures are obtained from the replicates of sample size n=100, while the last four figures are from the replicates of sample size, n=500. The figures for each sample size is arranged as the pairs mentioned in the text, the first pair is ($a = 0.5$ & $b = 0.6$), the second pair is ($a = 0.5$ & $b = 1.3$), the third pair ($a = 1.1$ & $b = 0.6$), and the fourth pair ($a = 1.1$ & $b = 1.6$). These are the runs obtained from the MOM method.